\documentclass[preprint,authoryear,times,12pt]{elsarticle}

\usepackage[latin1]{inputenc}
\usepackage[german,american]{babel}
\usepackage{babelbib}

\usepackage[T1]{fontenc}
\usepackage{latexsym} 

\usepackage{amsmath} 
\usepackage{amssymb} 
\usepackage{amsxtra} 
\usepackage{bm} 
\usepackage{mathtools}

\usepackage[OMLmathsfit]{isomath}

\def\ssans#1{\mathsfbfit{#1}}

\usepackage{enumitem}

\usepackage[all]{xy}

\usepackage{graphicx} 
\usepackage{color} 
\usepackage{pdfpages}
\usepackage[percent]{overpic}
\usepackage{pgf,tikz}

\usepackage{float}
\floatstyle{ruled}
\newfloat{algorithmT}{htb}{loa}
\floatname{algorithmT}{Algorithm}
\newfloat{algorithmP}{htb}{loa}
\floatname{algorithmP}{Algorithm}
\newfloat{algorithmD}{htb}{loa}
\floatname{algorithmD}{Algorithm}

\usepackage{epic,eepic}

\definecolor{darkgreen}{rgb}{0.0,0.5,0.0}
\definecolor{darkblue}{rgb}{0.0,0.0,0.85}

\def\clap#1{\hbox to 0pt{\hss#1\hss}}

\journal{...}
\begin{document} 

\begin{frontmatter}

\title{Development and comparison of spectral algorithms for numerical 
modeling of the quasi-static mechanical behavior of inhomogeneous materials}

\author{M.~Khorrami${}^{1}$, 
J.~R.~Mianroodi${}^{1,2}$, 
P.~Shanthraj${}^{2,3}$, 
B.~Svendsen${}^{1,2}$\corref{cor1}}
\cortext[cor1]{Corresponding author}
\address{${}^{1}$Material Mechanics, RWTH Aachen University, 
Aachen, Germany\\[1mm] 
${}^{2}$Microstructure Physics and Alloy Design, 
Max-Planck Institute for Iron Research,
D\"usseldorf, Germany\\[1mm]
${}^{3}$The School of Materials,  
University of Manchester, 
Manchester, U.K.}

\begin{abstract} 

In the current work, a number of algorithms are developed and compared for the numerical solution of periodic (quasi-static) linear elastic mechanical boundary-value problems (BVPs) based on two different discretizations of Fourier series. The first is standard and based on the trapezoidal approximation of the Fourier mode integral, resulting in trapezoidal discretization (TD) of the truncated Fourier series. Less standard is the second discretization based on piecewise-constant approximation of the Fourier mode integrand, yielding a piecewise-constant discretization (PCD) of this series. Employing these, fixed-point algorithms are formulated via Green-function preconditioning (GFP) and finite-difference discretization (of differential operators; FDD). In particular, in the context of PCD, this includes an algorithm based on the so-called "discrete Green operator" (DGO) recently introduced by \citet[][]{Eloh2019}, which employs GFP, but not FDD. For computational comparisons, the (classic) benchmark case of a cubic inclusion embedded in a matrix \citep[e.g.,][]{Suq97,Willot2015} is employed. Both discontinuous and smooth transitions in elastic stiffness at the matrix-inclusion (MI) interface are considered. In the context of both TD and PCD, a number of GFP- and FDD-based algorithms are developed. Among these, one based on so-called averaged-forward-backward-differencing (AFB) is shown to result in the greatest improvement in convergence rate. As it turns out, AFB is equivalent to the "rotated scheme" (R) of \citet[][]{Willot2015} in the context of TD. 
In the context of PCD, comparison of the performance and convergence behavior of AFB/R- and DGO-based algorithms shows that the former is more efficient than the latter for larger phase contrasts. 

\end{abstract}


\end{frontmatter}

\section{Introduction}
\label{sec:Int}

Numerical modeling of the mechanical behaviour of inhomogeneous 
materials is often based on the assumption that their microstructure 
is periodic. In this case, a well-known alternative to finite difference, 
finite element, or isogeometric, methods for the numerical solution 
of the corresponding boundary value problems (BVPs) is offered by 
spectral methods \cite[e.g.,][]{Got77,Can88,Trefethen2000,Pre07,Kopriva2009}. 
Besides the possibility of exponential convergence, spectral methods 
generally offer much higher spatial resolution than finite-difference-  
or finite-element-based ones. Indeed, as discussed for example by 
\citet[][\S 20.7]{Pre07}, when applicable, these are the methods of 
choice for this purpose. Another difference here is that, in contrast to 
finite-difference and finite-element methods, which approximate / discretize 
the model field relations, spectral methods approximate / discretize 
their solution. In particular, in the Fourier case, this latter is based on 
discretization of truncated Fourier series, referred to as Fourier 
series discretization in what follows. 

For the current case of numerical solution of the (quasi-static) momentum 
balance and corresponding mechanical BVPs, 
Fourier series discretization has traditionally been 
combined with differential-operator preconditioning \cite[e.g.,][]{Mou94,Suq97}, 
formally analogous to differential-operator inversion for analytic solution of 
BVPs in mathematical physics based on Green functions (e.g., Lippmann-Schwinger). 
This is referred to as "Green-function preconditioning" (GFP) in what follows. 
Together with continuous and discrete Fourier transformation, such methods 
have a long history of application in material science and in particular 
in continuum micromechanics \cite[e.g.,][]{Kha83,Mur87,Mou94,Suq97,Che02} 
to the modeling of material microstructure. The corresponding mechanical 
BVP is formulated at the level of a unit cell or representative volume element. 
Traditionally, these are strain-based, geometrically linear, and 
based on fixed-point iterative solution 
\cite[e.g.,][]{Suq97,Ey99,Mich00,Mich01,Leb01,Eis13,Lebensohn2020}. 
More recently, these have been extended 
conjugate-gradient- and Newton-Krylov-based numerical solution 
and geometric nonlinearity \cite[e.g.,][]{Kabel2014,Shanthraj2015,Mishra2016}. 

As well-known, in the case of material heterogeneity due to discontinuous 
material properties, Fourier series discretizationn 
suffers from oscillations due to the Gibbs 
phenomena and to aliasing \cite[e.g.,][]{Trefethen2000,Kopriva2009}. 
In general, both have an adverse effect on convergence behavior, 
resulting in a significant loss of algorithmic efficiency and robustness. 
To address these issues, a number of algorithms going beyond the original 
"basic scheme" of \cite{Suq97} have been developed. One class of such 
algorithms combines Fourier series discretization 
with finite-difference discretization (FDD) of differential 
operators \cite[e.g.,][]{Dre00,Bro02,Mou14,Willot2015}, resulting in a 
reduction of the effect of oscillations on convergence behavior due to 
modes at wavelengths shorter than the nodal or grid spacing. As shown in 
particular by \cite{Willot2015}, among such "accelerated schemes" 
\cite[e.g.,][]{Mou14}, 
his "rotated scheme" is most effective in improving algorithmic efficiency 
and robustness. This scheme has been employed for numerical solution 
of BVPs for heterogeneous materials in a number of works 
\cite[e.g., for field dislocation mechanics in][]{Djaka2017}. More recently, 
\cite{Schneider2017b} showed that finite-element discretization based on 
linear hexahedral elements with reduced integration is equivalent to the 
rotated scheme of \cite{Willot2015}. Quite recently, \cite{Lucarini2019} 
developed a displacement-based approach via direct Fourier transformation 
and real-function-based Fourier transform symmetry reduction, yielding an 
algorithmic system based on a full-rank associated Hermitian matrix 
amenable to preconditioning. For the latter purpose, they worked with a 
Green function based on the average elastic stiffness. In contrast to 
\cite{Willot2015} and the current work, FDD was not employed. 

All algorithms or "schemes" discussed up to this point employ Fourier series 
discretization  
in "standard" form, i.e., based on trapezoidal approximation / discretization (TD) 
of the integral for Fourier modes with respect to the unit cell. TD 
is the basis of Fourier interpolation and the standard form of the discrete 
Fourier transform \cite[e.g.,][]{Trefethen2000,Pre07,Kopriva2009}. 
Alternatively, \cite{Brisard2010} and \cite{Anglin2014} discretize 
by assuming the solution is piecewise-constant, resulting in piecewise-constant 
discretization (PCD). This approach has been exploited 
recently by \cite{Eloh2019}, who combine PCD with GFP to obtain an 
algorithm depending on a so-called discrete Green operator (DGO) for 
numerical solution of periodic mechanical BVPs for materially 
heterogeneous elastostatics via fixed-point interation. 

One purpose of the current work is to develop additional algorithms in the 
context of Fourier series discretization 
based on TD and PCD. All of these exploit GFP, and some 
of them FDD. A second purpose is to carry out detailed theoretical and 
computational comparisons of these with existing algorithms, in particular 
with the "rotated" (R) scheme of \cite{Willot2015}, and with the DGO-based 
scheme of \cite{Eloh2019}. 
The corresponding computational comparisons are based on the classic 
benchmark case of a periodic matrix-inclusion microstructure with cubic 
inclusions having sharp corners. The material inhomogeneity takes the 
form of (both discontinuous and smooth) phase contrast in elastic 
stiffness \cite[e.g.,][]{Suq97,NematNasser1999,Willot2015} across the 
matrix-inclusion (MI) interface. 
In contrast, \cite{Eloh2019} focus mainly on material 
heterogeneity due to eigenstrains \cite[see also e.g.,][]{Anglin2014}; in 
one example, however, spherical inclusions are considered. 
Likewise, \cite{Lucarini2019} focus (solely) on spherical inclusions. 

To establish relevant basic concepts and issues, the work begins in Section 
\ref{sec:ConAna1D} with basic considerations in one dimension (1D). These 
include analytic solution of the periodic matrix-inclusion mechanical BVP and 
its truncated Fourier series representation. Attention is focused next on the 
development of algorithms in 1D based on TD and PCD of truncated Fourier 
series in Section \ref{sec:AlgStr1D}. As discussed above, all of these employ 
GFP, and some of them FDD. 
Computational comparison of these is then carried out in Section \ref{sec:Res1D} 
for the matrix-inclusion benchmark case. In particular, material heterogeneity 
in the form of both (i) discontinuous and (ii) smooth compliance / stiffness 
distributions are considered. Multidimensional forms of the 1D algorithms 
are obtained via direct tensor product generalization in Section \ref{sec:AlgStr3D}. 
Corresponding computational comparsions are presented in Section \ref{sec:Res3D}. 
The work ends with a summary and discussion in Section \ref{sec:DisSum}.

In this work, (three-dimensional) Euclidean vectors are represented by 
lower-case bold italic characters \(\bm{a},\bm{b},\ldots,\). 
In particular, let \(\bm{i}_{1},\bm{i}_{2},\bm{i}_{3}\) 
represent the Cartesian basis vectors. 
Second-order tensors are represented by upper-case bold italic characters 
\(\bm{A},\bm{B},\ldots\), with \(\bm{I}\) the second-order identity. 
By definition, any \(\bm{A}\) maps any \(\bm{b}\) linearly into a vector 
\(\bm{A}\bm{b}\). 
Fourth-order Euclidean tensors \(\ssans{A},\ssans{B},\ldots,\) are 
denoted by upper-case slanted sans-serif characters. By definition, 
any \(\ssans{A}\) maps any \(\bm{B}\) linearly into a second-order 
tensor \(\ssans{A}\bm{B}\). Let  
\(\mathcal{A}\cdot\mathcal{B}=A_{ijk\ldots}B_{ijk\ldots}\) 
(summation convention) 
represent the scalar product of two arbitrary-order tensors 
\(\mathcal{A}\) and \(\mathcal{B}\). Using this, 
\(
\bm{A}^{\!\mathrm{T}}\bm{b}\cdot\bm{c}
:=\bm{b}\cdot\bm{A}\bm{c}
\) 
defines the transpose \(\bm{A}^{\!\mathrm{T}}\) of \(\bm{A}\). 
Let 
\(
\mathrm{sym}\,\bm{A}
:=\frac{1}{2}(\bm{A}+\bm{A}^{\!\mathrm{T}})
\) 
represent the symmetric part of \(\bm{A}\). The tensor products 
\((\bm{A}\,\square\,\bm{B})\,\bm{C}:=\bm{A}\bm{C}\bm{B}\) 
and 
\(
(\bm{A}\,\triangle\,\bm{B})\,\bm{C}
:=\bm{A}\bm{C}^{\mathrm{T}}\bm{B}
\) 
will also be employed. 
Additional notation will be introduced as needed in what follows. 

\section{Basic considerations in 1D}
\label{sec:ConAna1D}

\subsection{Analytic solution of periodic boundary-value problem} 
\label{sec:EquMec1D}

Let the interval \(\lbrack 0,l\rbrack\) of length \(l>0\) represent the 
unit cell of the periodic matrix-inclusion (MI) microstructure. In what 
follows, we work with the split 
\begin{equation}
f=\bar{f}+\tilde{f}
\,,\quad
\bar{f}:=\frac{1}{l}\int_{0}^{\,l}f(x)\ dx
\,,\quad
\tilde{f}:=f-\bar{f}
\,,
\label{equ:FieMeaCelUni1D}
\end{equation} 
of any (integrable) \(f\) into mean \(\bar{f}\) and "fluctuating" \(\tilde{f}\) 
parts. 

Let \(u\) represent the displacement field on the unit cell and \(H=\nabla u\) 
the corresponding distortion. Integration of \(\nabla u=\bar{H}+\tilde{H}\) 
yields the split 
\begin{equation} 
u=u^{\mathrm{h}}+u^{\mathrm{p}}
\,,\quad
u^{\smash{\mathrm{h}}}(x):=c+\bar{H}x
\,,\quad 
u^{\smash{\mathrm{p}}}(x):=\int\tilde{H}(x)\ dx
\,,
\label{equ:DisFluParHom1D}
\end{equation} 
of \(u\) into "homogeneous" \(u^{\smash{\mathrm{h}}}\) 
and "particular" \(u^{\smash{\mathrm{p}}}\) parts, with \(c\) 
the constant of integration. In what follows, \(c\) and \(\bar{H}\) 
are assumed given or known. Then \(u^{\smash{\mathrm{h}}}(x)\) 
is determined, and \(u^{\smash{\mathrm{p}}}(x)\) 
represents the primary unknown field. 

Excluding cracking, kinematic compatibility requires \(u\), and so 
\(u^{\smash{\mathrm{p}}}\), to be continuous everywhere, in 
particular at MI interfaces. In addition, quasi-static mechanical 
equilibrium \(\mathop{\mathrm{div}}T=0\) requires the stress \(T\) to be 
continuous everywhere and in fact constant in 1D; then 
\(\tilde{T}=0\) and \(T=\bar{T}\). In this case, the linear elastic relation 
\(E=ST=C^{-1}T\) for the 1D strain \(E\equiv H\) in terms of the 
compliance \(S=C^{-1}\) and stiffness \(C\) results in the equilibrium 
relations 
\(
\bar{E}=\bar{S}\bar{T}
\) 
and 
\(
\tilde{E}=\tilde{S}\bar{T}
\) 
for \(E=\bar{E}+\tilde{E}\) in 1D. In turn, 
\begin{equation}
\frac{\tilde{E}(x)}{\bar{E}}
=\frac{\tilde{S}\!(x)}{\bar{S}}
\,,\quad
\frac{u^{\smash{\mathrm{p}}}(x)}{\bar{E}}
=\int\frac{\tilde{S}\!(x)}{\bar{S}}\ dx
\,,
\label{equ:EquMecStnFluStsDer1D}
\end{equation} 
then hold for \(\tilde{E}(x)\) and \(u^{\smash{\mathrm{p}}}(x)\), 
the latter via \eqref{equ:DisFluParHom1D}${}_{3}$. 

\begin{figure}[H]
\centering
\includegraphics[width=0.45\textwidth]{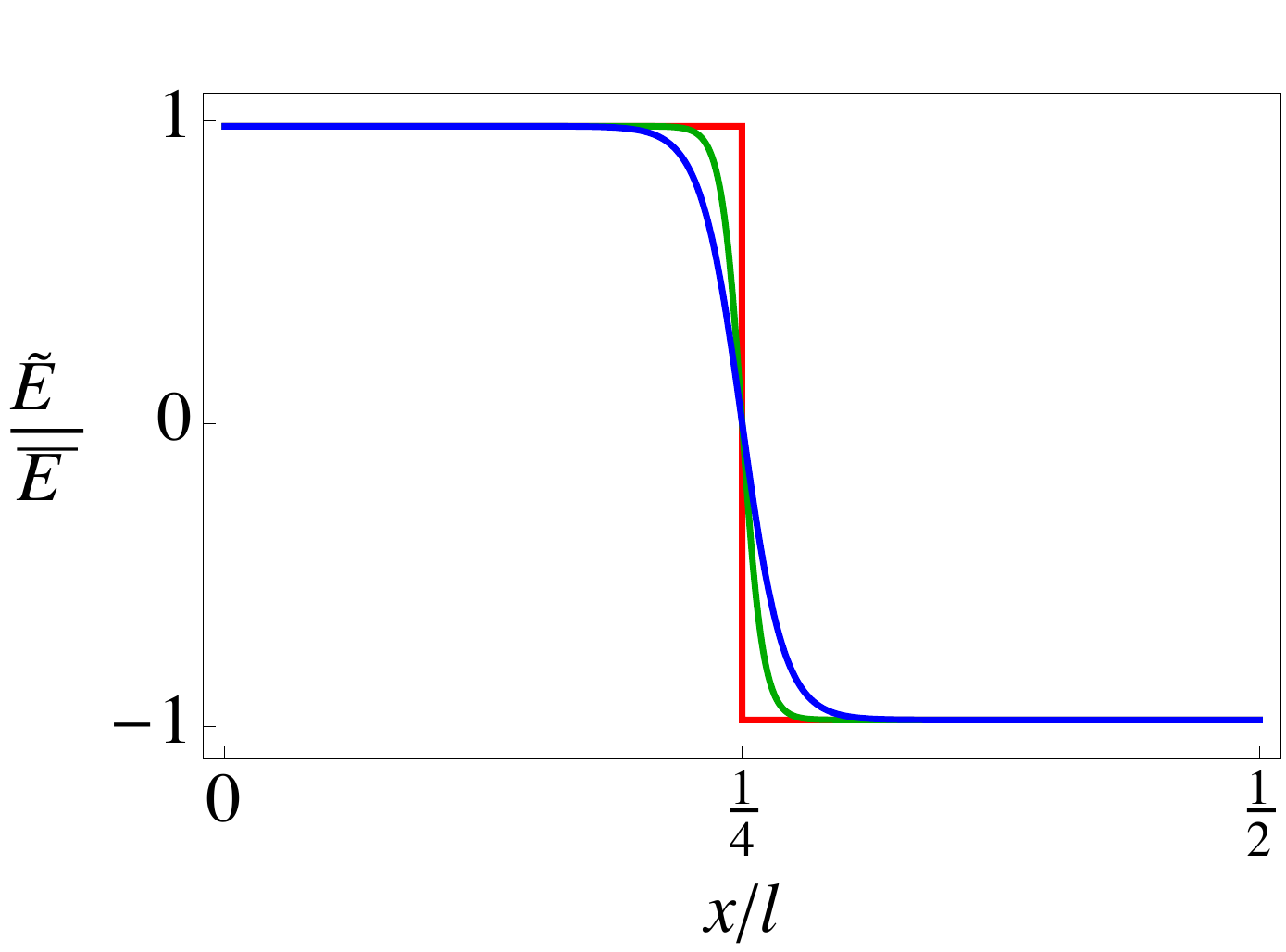}
\hspace{5mm}
\includegraphics[width=0.45\textwidth]{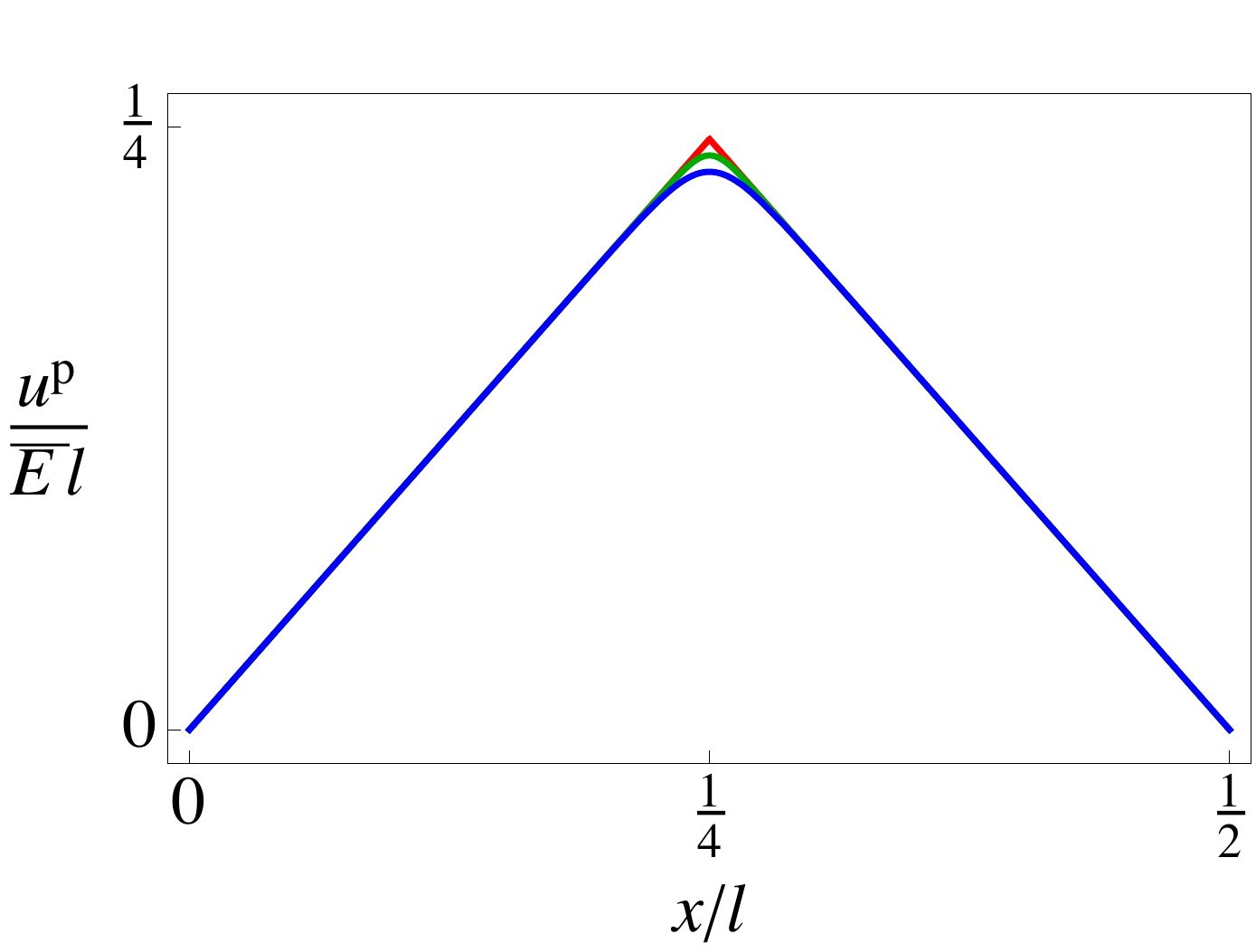}
\caption{\(\tilde{E}/\bar{E}\) (left) and 
\(u^{\smash{\mathrm{p}}}/\bar{E}l\) (right) 
from \eqref{equ:FieStnDisIncMat1D} on the left half 
\(\lbrack 0,l/2\rbrack\) of the unit cell \(\lbrack 0,l\rbrack\) for sharp 
(red:~\(\epsilon/l=0\)) and smooth 
(green:~\(\epsilon/l=1/100\); blue:~\(\epsilon/l=1/50\)) 
MI interfaces. These and all 1D results to follow are based on 
\(c_{\mathrm{l}}=l/4\), 
\(c_{\mathrm{r}}=3l/4\), 
and \(s_{\mathrm{IM}}=-99/100\) 
(i.e., 
\(C_{\mathrm{I}}/C_{\mathrm{M}}
=S_{\!\mathrm{M}\,}/S_{\!\mathrm{I}\,}
=100\)). 
See text for details.} 
\label{fig:StnDisIncMat}
\end{figure}

In the classic composite case \cite[e.g.,][]{Suq97}, the interface between 
two bulk phases is idealized as "sharp" in the sense that material properties 
like \(S\!\) are assumed to vary discontinuously across the interface. On the 
other hand, phase field models idealize such interfaces as a mixture of the 
bulk phases in which properties like \(S\!\) vary smoothly from one bulk 
phase to the other. 
Assuming the classic case of a double well potential and 
"flat" interface region of half-width \(\epsilon\) centered at \(x=c\), solution 
of the equilibrium Ginzburg-Landau relation at zero stress yields 
\begin{equation}
\phi_{\epsilon}(x;c)
=\tfrac{1}{2}
+\tfrac{1}{2}\tanh((x-c)/\epsilon)
\label{equ:FiePhaIncMat}
\end{equation}
for the phase field of the MI system varying between 
\(\phi_{\epsilon}=0\) (matrix) and \(\phi_{\epsilon}=1\) (inclusion) 
in the MI interface region. The "sharp interface" limit of this is given by 
\(
\phi_{0}(x;c)
:=\lim_{\epsilon\to 0}\phi_{\epsilon}(x;c)
=\theta(x-c)
\) 
in terms of the modified Heaviside step function \(\theta(x)\) 
\cite[i.e., \(\theta(x<0)=0\), 
\(\theta(x=0):=1/2\), 
\(\theta(x>0)=1\); e.g.,][Chapter 4]{Bra00}. 
Assuming that the left (right) MI interface is at \(x=c_{\mathrm{l}}\) 
(\(x=c_{\mathrm{r}}\)), 
\begin{equation}
\nu_{\!\epsilon}(x)
:=\phi(x;c_{\mathrm{l}},\epsilon)-\phi(x;c_{\mathrm{r}},\epsilon)
\label{equ:DenVolIncMat}
\end{equation}
represents the 1D "volume density" of the inclusion. In terms of this, 
\begin{equation}
S=(1-\nu_{\!\epsilon})\,S_{\!\mathrm{M}\,}
+\nu_{\!\epsilon}\,S_{\!\mathrm{I}\,}
=S_{\!\mathrm{M}\,}+S_{\!\mathrm{IM}\,}\nu_{\!\epsilon}
\,,\quad
S_{\!\mathrm{IM}\,}
:=S_{\!\mathrm{I}\,}-S_{\!\mathrm{M}\,}
\,,
\label{equ:FieComIncMat1D}
\end{equation} 
holds for the elastic compliance of the MI unit cell 
(note 
\(
S_{\!\mathrm{M}\,}/S_{\!\mathrm{I}\,}
=C_{\mathrm{I}}/C_{\mathrm{M}}
\)). 
Substituting \eqref{equ:FieComIncMat1D} into \eqref{equ:EquMecStnFluStsDer1D} 
yields
\begin{equation}
\tilde{E}(x)
=s_{\mathrm{IM}}\skew1\tilde{\nu}_{\!\epsilon}(x)\,\bar{E}
\,,\quad 
u^{\smash{\mathrm{p}}}(x)
=s_{\mathrm{IM}}
\lbrack
\lambda_{\epsilon}(x)
-\lambda_{\epsilon}(0)
-\skew1\bar{\nu}_{\!\epsilon}\,x/l
\rbrack
\,\bar{E}l
\,,\quad
s_{\mathrm{IM}}:=S_{\!\mathrm{IM}\,}/\bar{S}
\,,
\label{equ:FieStnDisIncMat1D}
\end{equation} 
with 
\(
\lambda_{\epsilon}(x)
:=\lbrack
\ell_{\epsilon}(x;c_{\mathrm{l}})
-\ell_{\epsilon}(x;c_{\mathrm{r}})
\rbrack/l
\) 
and 
\(
\ell_{\epsilon}(x;c)
:=\int\phi_{\epsilon}(x;c)\ dx
\). 
Note that \(u^{\smash{\mathrm{p}}}(0)=0\) has been assumed 
in \eqref{equ:FieStnDisIncMat1D}${}_{2}$ without loss of physical 
generality. \(\tilde{E}(x)\) and \(u^{\smash{\mathrm{p}}}(x)\) 
are displayed in Figure \ref{fig:StnDisIncMat}. 
\vspace{-3mm}
As expected from \eqref{equ:EquMecStnFluStsDer1D}, 
\(\tilde{S}\) determines a continuous and smooth \(\tilde{E}\) 
for \(\epsilon>0\), and a discontinuous \(\tilde{E}\) for \(\epsilon=0\). 
Likewise, \(u^{\smash{\mathrm{p}}}\) is continuous and 
smooth for \(\epsilon>0\), but only continuous for \(\epsilon=0\). 

\subsection{Approximation based on truncated Fourier series} 
\label{sec:SerFouTru}

\begin{figure}[H]
\centering
\includegraphics[width=0.45\textwidth]{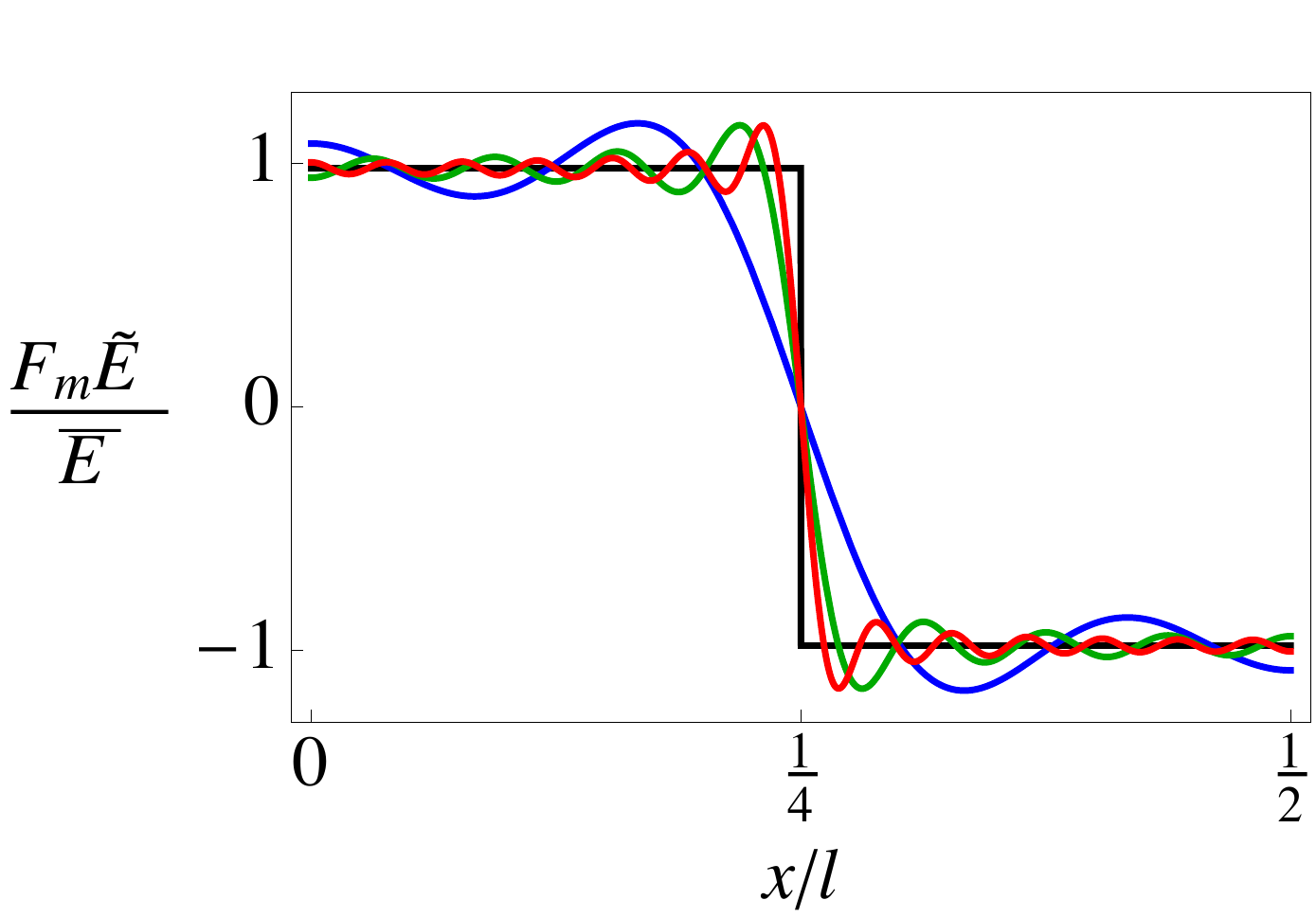}
\hspace{5mm}
\includegraphics[width=0.45\textwidth]{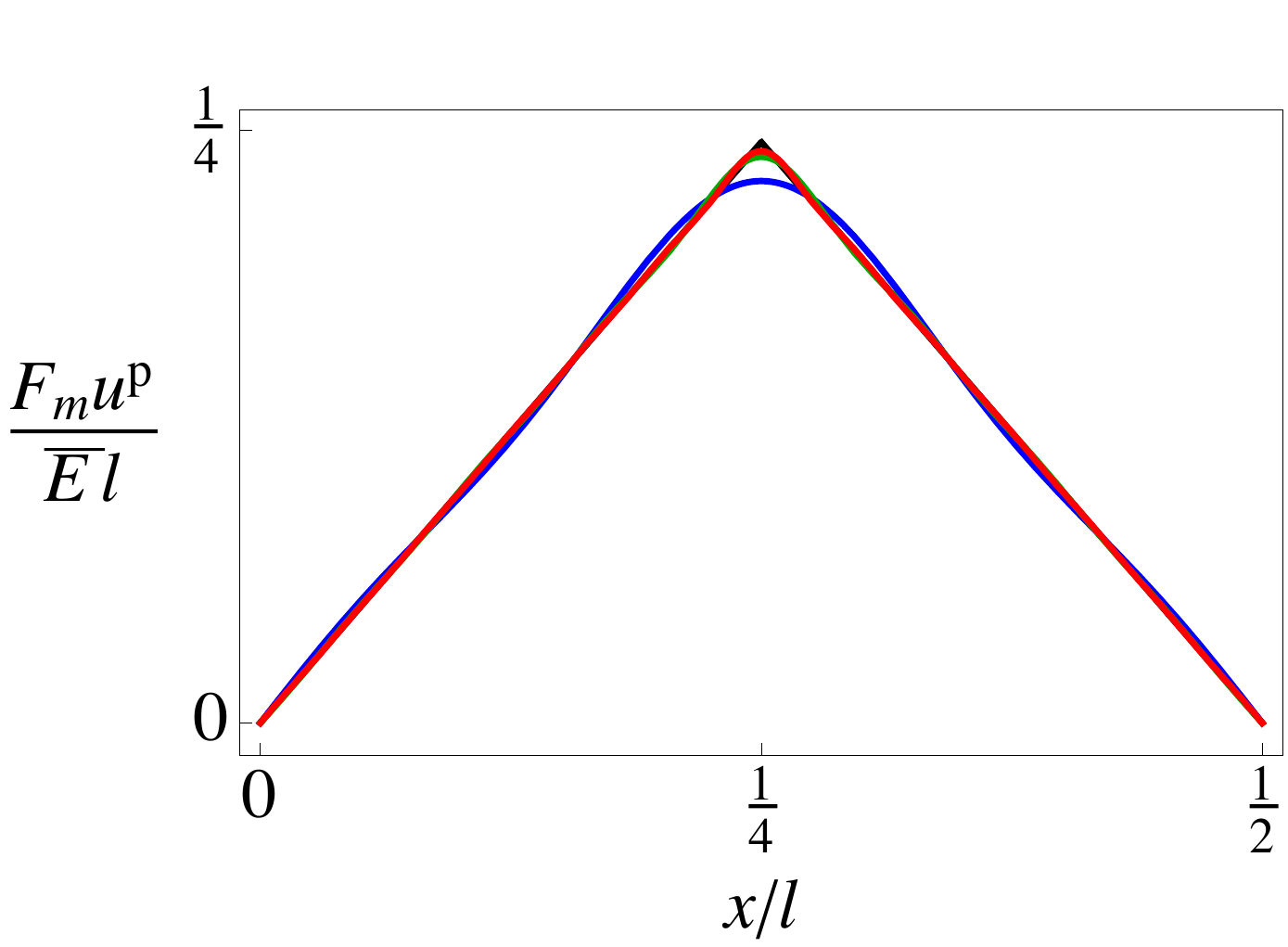}
\caption{Normalized Fourier series approximation 
\(\mathcal{F}_{\!\!m\,}\tilde{E}(x)/\bar{E}\) (left) and 
\(\mathcal{F}_{\!\!m\,}u^{\smash{\mathrm{p}}}(x)/\bar{E}l\) (right) 
of \(\tilde{E}(x)\) and \(u^{\smash{\mathrm{p}}}(x)\), respectively, 
on the left half unit cell \(\lbrack 0,l/2\rbrack\) 
for discontinuous (\(\epsilon=0\)) compliance and different degrees of 
truncation \(m<\infty\). In particular, \(m=5\) (blue), \(m=15\) (green), and 
\(m=25\) (red). The corresponding normalized analytic relations 
\(\tilde{E}(x)/\bar{E}\) and \(u^{\smash{\mathrm{p}}}(x)/\bar{E}l\) 
from \eqref{equ:FieStnDisIncMat1D} are shown (black) for comparison.} 
\label{fig:SerFouStnDisFluDsc}
\end{figure}

As a first step toward Fourier-series-based numerical solution of the above BVP, 
consider next the truncated Fourier series approximation of \(\tilde{E}\) 
and \(u^{\smash{\mathrm{p}}}\) from \eqref{equ:FieStnDisIncMat1D}. 
The truncated Fourier series of any \(f\in L^{2}(0,l)\) is given by 
\begin{equation}
\mathcal{F}_{\!\!m\,}f(x)
:=\sum\nolimits_{\kappa=-m}^{m}
e^{\imath k_{\kappa}x}
\hat{f}_{\kappa}
\,,\quad
\hat{f}_{\kappa}
:=\frac{1}{l}
\int_{0}^{\,l}
e^{-\imath k_{\kappa}x}
f(x)
\ dx
\,,
\label{equ:RepFieSerFor1D}
\end{equation} 
\cite[e.g.,][]{Trefethen2000,Kopriva2009,Liu2011} with \(\imath:=\sqrt{-1}\) 
and \(k_{\kappa}:=2\pi\kappa/l\). As is well-known, 
\(\mathcal{F}_{\!\!\infty\,}f(x)\neq f(x)\) in general at one or more 
\(x\in\lbrack 0,l\rbrack\). In particular, for \(f\) discontinuous on 
\(\lbrack 0,l\rbrack\), then, \(\mathcal{F}_{\!\!m\,}f\) deviates from \(f\) 
due to (i) \(\mathcal{F}_{\!\!\infty\,}f(x)\neq f(x)\) at points of discontinuity 
\(x\in\lbrack 0,l\rbrack\), and (ii) truncation of \(\mathcal{F}_{\!\!\infty\,}f\). 
Since (i) is related to the Gibbs phenomenon, it will be referred to as Gibbs 
error in what follows. In the sharp (\(\epsilon=0\)) MI interface case, then, 
\(\mathcal{F}_{\!\!m\,}\tilde{E}\) is affected by (i) and (ii), whereas 
\(\mathcal{F}_{\!\!m\,}u^{\smash{\mathrm{p}}}\) is affected only by (ii). 
These are displayed in Figure \ref{fig:SerFouStnDisFluDsc}. 
As expected, 
\(
\lim_{m\to\infty}\mathcal{F}_{\!\!m\,}\tilde{E}(x)\neq\tilde{E}(x)
\) 
at \(x=l/4\); rather, 
\(
\lim_{m\to\infty}\mathcal{F}_{\!\!m\,}\tilde{E}(l/4)
=\frac{1}{2}\lim_{x\,\uparrow\,l/4}\tilde{E}(x)
+\frac{1}{2}\lim_{x\,\downarrow\,l/4}\tilde{E}(x)
\). 
Being due to approximation of discontinuous \(\tilde{E}\) by 
\(\mathcal{F}_{\!\!\infty\,}\tilde{E}\), note that Gibbs error is 
unaffected by truncation of \(\mathcal{F}_{\!\!\infty\,}\tilde{E}\) 
to \(\mathcal{F}_{\!\!m\,}\tilde{E}\), i.e., independent of \(m\). 
In addition, it is independent of the magnitude of \(S_{\!\mathrm{IM}}\). 

\section{Numerical solution algorithms in 1D} 
\label{sec:AlgStr1D} 

\subsection{Algorithms based on trapezoidal discretization} 
\label{sec:AlgAppTra}

As discussed in the introduction, algorithms 
for the numerical solution of the current BVP are based on 
(i) discretization of \eqref{equ:RepFieSerFor1D} in 1D, and 
(ii) tensor-product generalization of these to 3D. 
More specifically, two discretizations of \eqref{equ:RepFieSerFor1D} 
are considered and compared in this work. 
The most common of these is trapezoidal discretization of 
\eqref{equ:RepFieSerFor1D}${}_{2}$ based on that 
\(
\lbrack 0,l\rbrack
=\bigcup_{i=1}^{n}
\lbrack 
x_{i},x_{i+1}
\rbrack
\) 
of \(\lbrack 0,l\rbrack\) with nodes at \(x_{i}=(i-1)h\) and nodal spacing \(h=l/n\). 
In this case, \eqref{equ:RepFieSerFor1D}${}_{2}$ and 
\(\hat{f}_{\smash{\kappa}}\) discretizes to 
\begin{equation}
\hat{f}_{\smash{\kappa}}^{\mathrm{t}}
:=n^{-1/2}\check{f}_{\kappa}^{\mathrm{t}}
\,,\quad
\check{f}_{\kappa}^{\mathrm{t}}
:=n^{-1/2}
\sum\nolimits_{i=1}^{n}
e^{-\imath k_{\kappa}x_{i}}
f_{i}^{\mathrm{t}}
\,,
\label{equ:RepFieSerForTra1D}
\end{equation} 
("t" stands for "trapezoidal"). 
As detailed in Appendix \ref{app:IntForAppTra1D}, \eqref{equ:RepFieSerForTra1D} 
induces the trapezoidal discretization 
(TD)\footnote{In \eqref{equ:FieFouAppTra1D} and 
all relations to follows, repeated indices, e.g., 
\(i\) or \(\mu\), imply a sum over these from \(1\) to \(n\) 
\textit{when this makes sense}.} 
\begin{equation}
f_{\mathrm{T}}(x)
:=\varphi_{\mu}^{\mathrm{t}}(x)
\,\check{f}_{\mu}^{\mathrm{t}}
\,,\quad 
\varphi_{\mu}^{\mathrm{t}}(x)
:=n^{-1/2}
e^{q_{\mu}x}
\,,\quad 
f_{i}^{\mathrm{t}}
=F_{\!\smash{\mu i}\,}^{\mathrm{t}+}
\check{f}_{\mu}^{\mathrm{t}}
\,,\quad
\check{f}_{\mu}^{\mathrm{t}}
=F_{\!\smash{\mu i}\,}^{\mathrm{t}-}
f_{i}^{\mathrm{t}}
\,,
\label{equ:FieFouAppTra1D} 
\end{equation} 
of \eqref{equ:RepFieSerFor1D} via \eqref{equ:IntFouApp1D} for \(i,\mu=1,\ldots,n\). 
Here, 
\(
\check{f}_{\smash{\mu}}^{\mathrm{t}}
\equiv
\check{f}_{\smash{\kappa=\mu-1-m}}^{\mathrm{t}}
\), 
\(q_{\mu}:=\imath k_{\mu-1-m}\), 
\(f_{\smash{i}}^{\mathrm{t}}\equiv f(x_{i})\), and 
\(F_{\!\smash{\mu i}}^{\mathrm{t}\pm}:=n^{-1/2}e^{\pm q_{\mu}x_{i}}\) 
from \eqref{equ:TraFouDisMat}. If \(\bar{f}\) is prescribed, note that 
\(
\check{f}_{\smash{\mu=m+1}}^{\mathrm{t}}
=\check{f}_{\smash{\kappa=0}}^{\mathrm{t}}
=n^{-1/2}\sum_{i=1}^{n}f_{\smash{i}}^{\mathrm{t}}
=n^{1/2}\bar{f}
\) 
is determined for a given discretization. For notational simplicity, 
the "\(\mathrm{t}\)" superscript for "trapezoidal" is dropped in what follows.

As discussed in the introduction, following previous work 
\cite[e.g.,][]{Dre00,Mou14,Willot2015}, finite difference discretization  
(FDD) of differential operators in the context of \eqref{equ:FieFouAppTra1D} 
is employed here for improved robustness in convergence. For example, 
\begin{equation}
u_{\smash{\mathrm{T}}}^{\mathrm{p}}(x)
=\varphi_{\mu}(x)\,\check{u}_{\mu}^{\smash{\mathrm{p}}}
\,,\quad
\nabla^{a}u_{\mathrm{T}}^{\smash{\mathrm{p}}}(x)
=\varphi_{\mu}(x)\,\check{u}_{\mu}^{\smash{\mathrm{p}}}
\otimes 
q_{\mu}^{a}
\,,
\label{equ:DisFouAppTra1D}
\end{equation} 
via FDD of \(\nabla\) in terms of the effective wavenumber 
\(q_{\smash{\mu}}^{a}\). In particular, 
\begin{equation}
\begin{array}{rclclcl}
\nabla^{\mathrm{FD}}u_{\mathrm{T}}^{\smash{\mathrm{p}}}(x)
&:=&
h^{-1}
\lbrack 
u_{\mathrm{T}}^{\smash{\mathrm{p}}}(x+h)
-u_{\mathrm{T}}^{\smash{\mathrm{p}}}(x)
\rbrack
&=&
\varphi_{\mu}(x)
\,\skew2\check{u}_{\mu}^{\smash{\mathrm{p}}}
\otimes 
q_{\mu}^{\mathrm{FD}}
&=:&
\tilde{E}_{\mathrm{FD}}(x)
\,,\\
\nabla^{\mathrm{BD}}u_{\mathrm{T}}^{\smash{\mathrm{p}}}(x)
&:=&
h^{-1}
\lbrack 
u_{\mathrm{T}}^{\smash{\mathrm{p}}}(x)
-u_{\mathrm{T}}^{\smash{\mathrm{p}}}(x-h)
\rbrack
&=&
\varphi_{\mu}(x)
\,\skew2\check{u}_{\mu}^{\smash{\mathrm{p}}}
\otimes 
q_{\mu}^{\mathrm{BD}}
&=:&
\tilde{E}_{\mathrm{BD}}(x)
\,,\\
\nabla^{\mathrm{CD}}
u_{\mathrm{T}}^{\smash{\mathrm{p}}}(x)
&:=&
\frac{1}{2}h^{-1}
\lbrack 
u_{\mathrm{T}}^{\smash{\mathrm{p}}}(x+h)
-u_{\mathrm{T}}^{\smash{\mathrm{p}}}(x-h)
\rbrack
&=&
\varphi_{\mu}(x)
\,\skew2\check{u}_{\mu}^{\smash{\mathrm{p}}}
\otimes 
q_{\mu}^{\mathrm{CD}}
&=:&
\tilde{E}_{\mathrm{CD}}(x)
\,,\\
\nabla^{\mathrm{hC}}u_{\mathrm{T}}^{\smash{\mathrm{p}}}(x)
&:=&
h^{-1}
\lbrack 
u_{\mathrm{T}}^{\smash{\mathrm{p}}}(x+\tfrac{1}{2}h)
-u_{\mathrm{T}}^{\smash{\mathrm{p}}}(x-\tfrac{1}{2}h))
\rbrack
&=&
\varphi_{\mu}(x)
\,\skew2\check{u}_{\mu}^{\smash{\mathrm{p}}}
\otimes 
q_{\mu}^{\mathrm{hC}}
&=:&
\tilde{E}_{\mathrm{hC}}(x)
\,,
\end{array}
\label{equ:AppDifFinDFStn1D}
\end{equation} 
for forward (\(a=\mathrm{FD}\)), 
backward (\(a=\mathrm{BD}\)), 
central (\(a=\mathrm{CD}\)), 
and half central (\(a=\mathrm{hC}\)), 
difference discretization, respectively, with 
\begin{equation}
q_{\mu}^{\mathrm{FD}}
:=h^{-1}(e^{q_{\mu}h}-1)
\,,\quad 
q_{\mu}^{\mathrm{BD}}
:=h^{-1}(1-e^{-q_{\mu}h})
\,,
\label{equ:FreDifFinFBDF1D}
\end{equation}
and
\begin{equation}
q_{\mu}^{\mathrm{CD}}
:=h^{-1}\sinh(q_{\mu}h)
\,,\quad 
q_{\mu}^{\mathrm{hC}}
:=2h^{-1}\sinh(\tfrac{1}{2}q_{\mu}h)
\,.
\label{equ:FreDifFinCDDF1D}
\end{equation} 
Analogous to \eqref{equ:DisFouAppTra1D}, TD of \(T(x)\) yields 
\begin{equation}
T_{\!\mathrm{T}\,}(x)
=\varphi_{\mu}(x)
\ \check{T}_{\!\mu\,}
\,,\quad
\mathrm{div}^{b}T_{\!\mathrm{T}\,}(x)
=\varphi_{\mu}(x)\ \check{T}_{\!\mu\,}q_{\mu}^{b}
\,,
\label{equ:StrFouAppTra1D}
\end{equation} 
via FDD of the divergence operator.  Together, 
\eqref{equ:DisFouAppTra1D}${}_{2}$ and \eqref{equ:StrFouAppTra1D}${}_{2}$ 
induce the preconditioned form 
\begin{equation}
-\mathrm{div}^{b}
C_{\!\mathrm{H}}
\lbrack
\nabla^{a}u_{\mathrm{T}}^{\smash{\mathrm{p}}}(x)
\rbrack
=\varphi_{\mu}(x)
\ \check{G}_{\mathrm{H}}^{-1}(q_{\mu}^{a},q_{\mu}^{b})
\,\skew2\check{u}_{\mu}^{\smash{\mathrm{p}}}
\label{equ:OpeFunGreAlgMod1D}
\end{equation} 
of the homogenous elliptic operator with 
\begin{equation}
\check{G}_{\mathrm{H}}^{-1}(q_{\mu}^{a},q_{\mu}^{b})
\,\skew2\check{u}_{\mu}^{\smash{\mathrm{p}}}
:=-C_{\mathrm{H}}
\lbrack 
\skew2\check{u}_{\mu}^{\smash{\mathrm{p}}}\otimes q_{\mu}^{a}
\rbrack 
\,q_{\mu}^{b}
\,,\quad
\check{G}_{\mathrm{H}}(q_{\mu}^{a},q_{\mu}^{b})
=-1/(C_{\mathrm{H}}\,q_{\mu}^{a}\,q_{\mu}^{b})
\,,
\label{equ:FunGreModFD1D}
\end{equation} 
the corresponding Green function (\(q_{\mu}^{a}\neq 0, q_{\mu}^{b}\neq 0\)). 

In what follows, \(q_{\mu}^{b}\) is determined from a given choice of 
\(a\) in two different ways. The first is based on conjugacy, i.e., 
\begin{equation}
q_{\mu}^{b}=q_{\mu}^{a\ast}
\label{equ:EffNumWavConWil1D}
\end{equation}
\cite[e.g.,][Equations (19) and (24)]{Willot2015}. 
For example, \(q_{\mu}^{b}=-q_{\smash{\mu}}^{\mathrm{BD}}\) 
for \(a=\mathrm{FD}\) from \eqref{equ:FreDifFinFBDF1D}, or 
\(q_{\mu}^{b}=-q_{\smash{\mu}}^{a}\) for 
\(a=\mathrm{F},\mathrm{CD},\mathrm{hC}\) 
via \eqref{equ:FreDifFinCDDF1D}. The second way is based on the 
choice \(a=\mathrm{FD}\) and the observation that 
\(E_{\mathrm{FD}}(x)=\nabla^{\mathrm{FD}}u(x)\) 
determines \(T(x+\tfrac{1}{2}h)\) via the mean-value theorem. 
The choice \(\mathrm{div}^{\mathrm{BD}}T(x+\tfrac{1}{2}h)\) and identity
\begin{equation}
\nabla^{\mathrm{BD}}\!f(x+\tfrac{1}{2}h)
=\nabla^{\mathrm{hC}}\!f(x)
\label{equ:BDhCIde1D}
\end{equation}
then result in the combination \((a,b)=(\mathrm{FD},\mathrm{hC})\). 
For this choice, then, both the stress divergence and displacement are 
calculated at \(x\). Note that this choice does not satisfy 
\eqref{equ:EffNumWavConWil1D}. 

The above results are incorporated in Algorithm \ref{alg:AlgDisIntFouDifFinFunGre1D}. 
\begin{algorithmT}[H]
\begin{enumerate}
\setlength{\itemsep}{-0.1mm}
\item given:~\(n\), \(C(x)\), \(\bar{E}\), \(C_{\mathrm{H}}\), 
\(
\check{G}_{\mathrm{H}\,\mu}^{ab}
:=\check{G}_{\mathrm{H}}(q_{\smash{\mu}}^{a},q_{\smash{\mu}}^{b})
\)
\item initialization \(\iota=0\)
\vspace{-3mm}
\begin{itemize}
\setlength{\itemsep}{-0.5mm}
\item for \(i=1,\ldots,n\)
\newline
\(
{}\ 
C_{i}=C(x_{i})
\); 
\(
{}\ 
T_{\!\smash{i}}^{(\iota)}=C_{i}\bar{E}
\);
\item for \(\mu=1,\ldots,n\)
\newline
\(
{}\ 
\skew2\check{u}_{\smash{\mu}}^{\mathrm{p}(\iota)}
=0
\);
\(
{}\ 
\check{T}_{\!\smash{\mu}}^{(\iota)}
=F_{\!\smash{\mu i}\,}^{-}T_{\!\smash{i}}^{(\iota)}
\);
\(
{}\ 
\Delta\skew2\check{u}_{\smash{\mu}}^{\mathrm{p}(\iota)}
=\check{G}_{\smash{\mathrm{H}\,\mu}}^{ab}
\check{T}_{\!\smash{\mu}}^{(\iota)}q_{\smash{\mu}}^{b}
\);
\item \(\iota\!+\!\!+\);
\end{itemize}
\item while 
\(
\sum_{\mu=1}^{n}
|\Delta\skew2\check{u}_{\smash{\mu}}^{\mathrm{p}(\iota)}
-\Delta\skew2\check{u}_{\smash{\mu}}^{\mathrm{p}(\iota-1)}|
\geqslant\mathrm{tol}
\) 
\& \(\iota\leqslant\mathrm{maxit}\)
\vspace{-2mm}
\begin{itemize}
\setlength{\itemsep}{-0.5mm}
\item for \(i=1,\ldots,n\)
\newline 
\(
{}\ 
\tilde{E}_{\smash{i}}^{(\iota)}
=F_{\!\smash{\mu i}\,}^{+}
\skew2\check{u}_{\smash{\mu}}^{\mathrm{p}(\iota)}
\otimes 
q_{\mu}^{a}
\);
\(
{}\ 
T_{\!\smash{i}}^{(\iota)}
=T_{\!\smash{i}}^{(0)}+C_{i}\tilde{E}_{\smash{i}}^{(\iota)};
\)
\item for \(\mu=1,\ldots,n\)
\newline
\(
{}\ 
\check{T}_{\!\smash{\mu}}^{(\iota)}
=F_{\!\smash{\mu i}\,}^{-}T_{\!\smash{i}}^{(\iota)}
\);
\(
{}\ 
\Delta\skew2\check{u}_{\smash{\mu}}^{\mathrm{p}(\iota)}
=\check{G}_{\smash{\mathrm{H}\,\mu}}^{ab}
\check{T}_{\!\smash{\mu}}^{(\iota)}q_{\smash{\mu}}^{b}
\);
\(
{}\ 
\skew2\check{u}_{\mu}^{\smash{\mathrm{p}}(\iota)}
+\!=
\Delta\skew2\check{u}_{\mu}^{\smash{\mathrm{p}}(\iota)}
\);
\item \(\iota\!+\!\!+\); 
\end{itemize}
\end{enumerate}
\caption{}
\label{alg:AlgDisIntFouDifFinFunGre1D}
\end{algorithmT} 
\vspace{-3mm}
The direct Fourier case is included here by defining 
\(q_{\smash{\mu}}^{\mathrm{F}}:=q_{\mu}\) ("F" for "Fourier"). 
Rather than being strain-based like the "basic scheme" of 
\cite{Suq97} and \cite{Mich00,Mich01}, note that Algorithm 
\ref{alg:AlgDisIntFouDifFinFunGre1D} is displacement-based. 
As evident from \eqref{equ:FunGreModFD1D}, 
\(\check{G}_{\mathrm{H}}^{-1}
(q_{\smash{\mu}}^{a},q_{\smash{\mu}}^{b})\) 
is not invertible when either \(q_{\smash{\mu}}^{a}\) 
or \(q_{\smash{\mu}}^{b}\) vanish. 
For \(n\) even, \eqref{equ:FreDifFinFBDF1D} and \eqref{equ:FreDifFinCDDF1D} 
imply \(q_{\smash{\mu}}^{a,b}=0\) at \(\mu=m+1\) for 
\(a,b=\mathrm{F},\mathrm{FD},\mathrm{BD},\mathrm{hC}\), 
and at \(\mu=1,m+1\) for \(a,b=\mathrm{CD}\). For this latter case, 
we follow \citet[][Equation (29)]{Willot2015} and formally set  
\(\check{G}_{\mathrm{H}}
(q_{\smash{\mu}}^{a},q_{\smash{\mu}}^{b})\equiv 0\) 
for \(\mu=1\). For \(n\) odd, \(q_{\smash{\mu}}^{a,b}=0\) only at \(\mu=m+1\) 
for all cases. 

\subsection{Algorithms based on piecewise-constant discretization}
\label{sec:ConPieFieApp1D}

Assume now that \(f(x)\) in \eqref{equ:RepFieSerFor1D}${}_{2}$ is 
constant on each subinterval 
\(
\lbrack 
x_{i},x_{i+1}
\rbrack
\) 
of 
\(
\lbrack 0,l\rbrack
\) 
with value 
\(
f_{\smash{i}}^{\mathrm{c}}
\equiv
f(x_{\smash{i}}^{\mathrm{c}})
\) 
at the mid-point or "center" 
\(
x_{i}^{\mathrm{c}}=(i-\frac{1}{2})\,h
\) 
of \(\lbrack x_{i},x_{i+1}\rbrack\). In this case, \(\hat{f}_{\kappa}\) in 
\eqref{equ:RepFieSerFor1D}${}_{2}$ reduces to 
\begin{equation}
\hat{f}_{\kappa}^{\mathrm{c}}
:=n^{-1/2}s_{\kappa}\,\check{f}_{\kappa}^{\mathrm{c}}
\,,\quad 
s_{\kappa}
:=\mathop{\mathrm{sinc}}(\pi\kappa/n)
\,,\quad 
\check{f}_{\kappa}^{\mathrm{c}}
:=n^{-1/2}
\sum\nolimits_{i=1}^{n}
e^{-\imath k_{\kappa}x_{i}^{\mathrm{c}}}
f_{i}^{\mathrm{c}}
\,,
\label{equ:RepFieSerForPieCon1D}
\end{equation} 
with 
\(
\mathop{\mathrm{sinc}}(x):=\sin(x)/x
\) 
the (unnormalized) sinc function. As shown by comparison of 
\(\hat{f}_{\kappa}^{\mathrm{c}}\) with 
\(\hat{f}_{\kappa}^{\mathrm{t}}\) from 
\eqref{equ:RepFieSerForTra1D}, differences between these include 
(i) the dependence of \(\hat{f}_{\kappa}^{\mathrm{c}}\) on 
the sinc "filter" \(s_{\kappa}\), and 
(ii) different discretizations of \(\lbrack 0,l\rbrack\). 
Leaving the details to Appendix \ref{app:DisCenIntSub1DApp}, 
\eqref{equ:RepFieSerForPieCon1D} results in the piecewise-constant 
discretization (PCD) 
\begin{equation}
f_{\mathrm{PC}}(x)
:=\varphi_{\mu}^{\mathrm{c}}(x)
\,\check{f}_{\mu}^{\mathrm{c}}
\,,\quad 
\varphi_{\mu}^{\mathrm{c}}(x)
:=s_{\mu}^{\mathrm{c}}
\,\varphi_{\mu}^{\mathrm{t}}(x)
\,,\quad
f_{i}^{\mathrm{c}}
=F_{\!\smash{\mu i}\,}^{\mathrm{c}+}
\check{f}_{\mu}^{\mathrm{c}}
\,,\quad 
\check{f}_{\mu}^{\mathrm{c}}
=F_{\!\smash{\mu i}\,}^{\mathrm{c}-}f_{i}^{\mathrm{c}}
\,,
\label{equ:DisFouPC1D} 
\end{equation} 
of \eqref{equ:RepFieSerFor1D} via \eqref{equ:SerForPCTra1D} for 
\(i,\mu=1,\ldots,n\), analogous to \eqref{equ:FieFouAppTra1D} in the TD 
case, with 
\(
s_{\smash{\mu}}^{\mathrm{c}}
:=s_{\mu-1-m}
\) 
and 
\(
F_{\!\smash{\mu i}\,}^{\mathrm{c}\pm}
:=n^{-1/2}e^{\pm q_{\mu}x_{i}^{\mathrm{c}}}
\). 
In particular, 
\(
s_{\smash{m+1}}^{\mathrm{c}}=s_{0}^{}=1
\) 
and 
\(
\check{f}_{\smash{\mu=m+1}}^{\mathrm{c}}
=\check{f}_{\smash{\kappa=0}}^{\mathrm{c}}
=n^{-1/2}\sum_{i=1}^{n}f_{i}^{\mathrm{c}}
=n^{1/2}\bar{f}
\). 
Analogous to \eqref{equ:DisFouAppTra1D} and \eqref{equ:StrFouAppTra1D}, 
then, we have 
\begin{equation}
u_{\mathrm{PC}}^{\smash{\mathrm{p}}}(x)
=\varphi_{\smash{\mu}}^{\mathrm{c}}(x)
\ \check{u}_{\smash{\mu}}^{\mathrm{pc}}
\,,\quad 
T_{\!\mathrm{PC}}(x)
=\varphi_{\smash{\mu}}^{\mathrm{c}}(x)
\ \check{T}_{\!\smash{\mu}}^{\mathrm{c}}
\,,
\label{equ:DisFouConPie1D}
\end{equation}
and 
\begin{equation}
\nabla^{a}\!u_{\mathrm{PC}}^{\mathrm{p}}(x)
=\varphi_{\mu}^{\mathrm{c}}(x)
\ \check{u}_{\mu}^{\mathrm{pc}}\otimes q_{\mu}^{a}
\,,\ \ 
\mathrm{div}^{b}
T_{\mathrm{PC}}^{\mathrm{c}}(x)
=\varphi_{\mu}^{\mathrm{c}}(x)
\,\check{T}_{\!\mu}^{\mathrm{c}}\,q_{\mu}^{b}
\,.
\end{equation}
Further, 
\begin{equation}
-\mathrm{div}^{b}
C_{\!\mathrm{H}}
\lbrack
\nabla^{a}\!u_{\mathrm{PC}}^{\smash{\mathrm{p}}}(x)
\rbrack
=\varphi_{\mu}^{\mathrm{c}}(x)
\ \check{G}_{\mathrm{H}}^{-1}(q_{\mu}^{a},q_{\mu}^{b})
\,\check{u}_{\mu}^{\mathrm{pc}}
\,,
\end{equation}
analogous to \eqref{equ:OpeFunGreAlgMod1D}. Completely analogous to 
the case of Algorithm \ref{alg:AlgDisIntFouDifFinFunGre1D}, then, these 
relations and the fact that \(s_{\smash{\mu}}^{\mathrm{c}}\neq 0\) 
can be employed to obtain Algorithm \ref{alg:AlgDisIntFouDifFinFunGrePC1D}. 
\begin{algorithmP}[H]
\begin{enumerate}
\setlength{\itemsep}{-0.1mm}
\item given:~\(n\), \(C(x)\), \(\bar{E}\), \(C_{\mathrm{H}}\), 
\(
\check{G}_{\mathrm{H}\,\mu}^{ab}
:=\check{G}_{\mathrm{H}}(q_{\smash{\mu}}^{a},q_{\smash{\mu}}^{b})
\)
\item initialization \(\iota=0\) 
\vspace{-3mm}
\begin{itemize}
\setlength{\itemsep}{-0.5mm}
\item for \(i=1,\ldots,n\)
\newline
\(
{}\ 
C_{\smash{i}}^{\mathrm{c}}=C(x_{\smash{i}}^{\mathrm{c}})
\); 
\(
{}\ 
T_{\!\smash{i}}^{\mathrm{c}(\iota)}
=C_{\smash{i}}^{\mathrm{c}}\bar{E}
\);
\item for \(\mu=1,\ldots,n\)
\newline 
\(
{}\ \ 
\skew2\check{u}_{\mu}^{\smash{\mathrm{pc}(\iota)}}
=0
\);
\(
{}\ 
\check{T}_{\!\smash{\mu}}^{\mathrm{c}(\iota)}
=F_{\!\smash{\mu i}\,}^{-}T_{\!\smash{i}}^{\mathrm{c}(\iota)}
\);
\(
{}\ 
\Delta\skew2\check{u}_{\smash{\mu}}^{\mathrm{pc}(\iota)}
=\check{G}_{\smash{\mathrm{H}\,\mu}}^{ab}
\check{T}_{\!\smash{\mu}}^{\mathrm{c}(\iota)}q_{\smash{\mu}}^{b}
\);
\item \(\iota\!+\!\!+\);
\end{itemize}
\item while 
\(
\sum_{\mu=1}^{n}
|\Delta\skew2\check{u}_{\smash{\mu}}^{\mathrm{pc}(\iota)}
-\Delta\skew2\check{u}_{\smash{\mu}}^{\mathrm{pc}(\iota-1)}|
\geqslant
\mathrm{tol}
\) 
\& \(\iota\leqslant\mathrm{maxit}\)
\vspace{-2mm}
\begin{itemize}
\setlength{\itemsep}{-0.5mm}
\item for \(i=1,\ldots,n\)
\newline 
\(
{}\ 
\tilde{E}_{\smash{i}}^{\mathrm{c}(\iota)}
=F_{\!\smash{\mu i}\,}^{\mathrm{c}+}
\skew2\check{u}_{\smash{\mu}}^{\smash{\mathrm{pc}(\iota)}}
\otimes 
q_{\smash{\mu}}^{a}
\); 
\(
{}\ 
T_{\!\smash{i}}^{\mathrm{c}(\iota)}
=T_{\!\smash{i}}^{\mathrm{c}(0)}
+C_{\smash{i}}^{\mathrm{c}}\tilde{E}_{\smash{i}}^{\mathrm{c}(\iota)};
\)
\item for \(\mu=1,\ldots,n\)
\newline
\(
{}\ 
\check{T}_{\!\smash{\mu}\,}^{\mathrm{c}(\iota)}
=F_{\!\smash{\mu i}\,}^{\mathrm{c}-}T_{\!\smash{i}}^{\mathrm{c}(\iota)}
\);
\(
{}\ 
\Delta\skew2\check{u}_{\smash{\mu}}^{\smash{\mathrm{pc}}(\iota)}
=\skew4\check{G}_{\smash{\mathrm{H}\,\mu}}^{ab}
\check{T}_{\!\smash{\mu}}^{\mathrm{c}(\iota)}
q_{\smash{\mu}}^{b}
\);
\(
{}\ 
\skew2\check{u}_{\smash{\mu}}^{\smash{\mathrm{pc}}(\iota)}
+\!=
\Delta\skew2\check{u}_{\smash{\mu}}^{\smash{\mathrm{pc}}(\iota)}
\);
\item \(\iota\!+\!\!+\); 
\end{itemize}
\end{enumerate}
\caption{}
\label{alg:AlgDisIntFouDifFinFunGrePC1D}
\end{algorithmP} 
As evident, this algorithm is independent of 
\(
s_{\smash{\mu}}^{\mathrm{c}}
=s_{\mu-1-m}
=\mathop{\mathrm{sinc}}(\pi(\mu-1-m)/n)
\) 
from \eqref{equ:RepFieSerForPieCon1D}${}_{2}$. 
This is in contrast to the PCD-based algorithm of \cite{Eloh2019}, to which 
we now turn. 

\subsection{Algorithm of \cite{Eloh2019}}
\label{sec:ConPieFieAppElo1D}

Following \cite{Brisard2010}, \cite{Eloh2019} recently developed an 
algorithm also based on PCD and \eqref{equ:RepFieSerForPieCon1D} 
quite different than Algorithm \ref{alg:AlgDisIntFouDifFinFunGrePC1D}. 
To this end, they formulate their algorithm based on 
\(
\mathcal{F}_{\!\!\infty\,}^{\mathrm{c}}f(x)
:=\sum\nolimits_{\kappa=-\infty}^{\infty}
e^{\imath k_{\kappa}x}
\hat{f}_{\smash{\kappa}}^{\mathrm{c}}
\) 
from \eqref{equ:RepFieSerFor1D} for \(m=\infty\) and 
\eqref{equ:RepFieSerForPieCon1D}, and truncate the result. 
Equivalently, as done here, one can work with the truncated 
form 
\begin{equation}
\mathcal{F}_{\!\!p\,}^{\mathrm{c}}f(x)
:=\sum\nolimits_{\kappa=-p}^{p-1}
e^{\imath k_{\kappa}x}
\hat{f}_{\kappa}^{\mathrm{c}}
=n^{-1/2}
\sum\nolimits_{\omega=0}^{n-1}
e^{\imath k_{\omega}x}
\sum\nolimits_{\nu=-m}^{m-1}
e^{\imath k_{\nu n}x}
s_{\nu n+\omega}
\,\check{f}_{\nu n+\omega}^{\mathrm{c}}
\label{equ:ConPieSerForTru}
\end{equation} 
of \(\mathcal{F}_{\!\!\infty\,}^{\mathrm{c}}f(x)\) from the 
start (for \(p=nm\) even). Combining this with (the second of) 
\begin{equation}
\check{f}_{\nu n+\omega}^{\mathrm{c}}
=(-1)^{\nu}
\check{f}_{\omega}^{\mathrm{c}}
\,,\quad
n^{-1/2}
\sum\nolimits_{\omega=0}^{n-1}
e^{\imath k_{\omega}x_{i}^{\mathrm{c}}}
\check{f}_{\omega}^{\mathrm{c}}
=f(x_{i}^{\mathrm{c}})
=f_{i}^{\mathrm{c}}
\,,
\label{equ:CenTraForDisConBac}
\end{equation} 
from \eqref{equ:RepFieSerForPieCon1D}${}_{3}$ results in the PCD 
\begin{equation} 
f_{\mathrm{El}}(x)
:=\varphi_{\omega}^{\mathrm{El}}(x)
\,\check{f}_{\omega}^{\mathrm{c}}
\,,\quad
\varphi_{\omega}^{\mathrm{El}}(x)
:=n^{-1/2}
\sum\nolimits_{\nu=-m}^{m-1}
e^{\imath k_{\nu n+\omega}x}
\,(-1)^{\nu}
s_{\nu n+\omega}
\,,
\label{equ:DisPieCon1D}
\end{equation} 
(sum over repeated \(\omega\)) of \eqref{equ:RepFieSerFor1D} 
for \(\omega=0,\ldots,n-1\), 
analogous to \eqref{equ:DisFouPC1D} ("El" stands for "Eloh"). 
As in the basic scheme \cite[e.g.,][]{Suq97}, they work with (i) strain 
\(
\tilde{E}_{\mathrm{El}}(x)
=\varphi_{\omega}^{\mathrm{El}}(x)
\,\check{\tilde{E}}_{\omega}^{\mathrm{c}}
\) 
as the primary discretant, and 
(ii) the Fourier transform\footnote{To be more precise, again like 
\cite{Suq97}, \cite{Eloh2019} work with the 
polarization stress \(T-C_{\mathrm{H}}E\) instead of 
\(T\) directly as done in Algorithm \ref{alg:AlgDisConPie1D}. This is also true for 
the 3D case and Algorithm \ref{alg:AlgDisConPie3D} below.}  
\begin{equation}
\skew4\hat{\Gamma}_{\!\!\mathrm{H}}(k)
\,\hat{T}(k)
=-\skew4\hat{G}_{\mathrm{H}}(k)
\,\hat{T}(k)
\,k\otimes k
=0
\,,\quad
k\neq 0
\,,
\label{equ:SchLipEquMecFouTra1D}
\end{equation}
of mechanical equilbrium
\(
\Gamma_{\!\!\mathrm{H}}\ast T
:=\nabla G_{\mathrm{H}}\ast\mathop{\mathrm{div}}T
=0
\) 
in pre-conditioned Lippmann-Schwinger form. 
Again leaving the details to Appendix \ref{app:DisCenIntSub1DApp}, this 
results in Algorithm \ref{alg:AlgDisConPie1D}. 
\begin{algorithmD}[H]
\begin{enumerate}
\setlength{\itemsep}{-0.1mm}
\item given:~\(n\), \(m\), \(C(x)\), \(\bar{E}\), \(C_{\mathrm{H}}\), 
\(
\skew4\check{\Gamma}_{\!\!\smash{\mathrm{H}\,\omega=0}}^{\mathrm{c}}
=0
\),
\(
\skew4\check{\Gamma}_{\!\!\smash{\mathrm{H}\,\omega\neq 0}}^{\mathrm{c}}
=\sum\nolimits_{\nu=-m}^{m-1}
s_{\nu n+\omega}
\ \skew4\hat{\Gamma}_{\!\!\mathrm{H}}(k_{\nu n+\omega})
\)
\item initialization \(\iota=0\)
\vspace{-3mm}
\begin{itemize}
\setlength{\itemsep}{-0.5mm}
\item for \(i=1,\ldots,n\)
\newline
\(
{}\ 
C_{\smash{i}}^{\mathrm{c}}
=C(x_{\smash{i}}^{\mathrm{c}})
\); 
\(
{}\ 
T_{\!\smash{i}}^{\mathrm{c}(\iota)}
=C_{\smash{i}}^{\mathrm{c}}\bar{E}
\);
\item for \(\omega=0,\ldots,n-1\)
\newline
\(
{}\ 
\check{E}_{\smash{\omega}}^{\mathrm{c}(\iota)}
=0
\); 
\(
{}\ 
\check{T}_{\!\smash{\omega}}^{\mathrm{c}(\iota)}
=F_{\!\smash{\omega i}\,}^{-}T_{\!\smash{i}}^{\mathrm{c}(\iota)}
\);
\(
{}\ 
\Delta\check{E}_{\smash{\omega}}^{\mathrm{c}(\iota)}
=\skew4\check{\Gamma}_{\!\!\smash{\mathrm{H}\,\omega}\,}^{\mathrm{c}}
\skew4\check{T}_{\!\smash{\omega}}^{\mathrm{c}(\iota)}
\);
\item \(\iota\!+\!\!+\);
\end{itemize}
\item while
\(
\sum_{\omega=1}^{n-1}
|\Delta\check{E}_{\omega}^{\mathrm{c}(\iota)}
-\Delta\check{E}_{\omega}^{\mathrm{c}(\iota-1)}|
\geqslant
\mathrm{tol}
\) 
\& \(\iota\leqslant\mathrm{maxit}\) 
\vspace{-3mm}
\begin{itemize}
\setlength{\itemsep}{-1mm}
\item for \(i=1,\ldots,n\) 
\newline
\(
{}\ 
\tilde{E}_{\smash{i}}^{\mathrm{c}(\iota)}
=\check{F}_{\smash{\omega i}}^{\mathrm{c}+}
\check{E}_{\smash{\omega}}^{\mathrm{c}(\iota)}
\);
\(
{}\ 
T_{\!\smash{i}}^{\mathrm{c}(\iota)}
=T_{\!\smash{i}}^{\mathrm{c}(0)}
+C_{\smash{i}}^{\mathrm{c}}\tilde{E}_{\smash{i}}^{\mathrm{c}(\iota)}
\);
\item for \(\omega=0,\ldots,n-1\)
\newline
\(
{}\ 
\skew4\check{T}_{\!\smash{\omega}}^{\mathrm{c}(\iota)}
=\check{F}_{\!\smash{\omega i}\,}^{\mathrm{c}-}
T_{\!\smash{i}}^{\mathrm{c}(\iota)}
\);
\(
{}\ 
\Delta\check{E}_{\smash{\omega}}^{\mathrm{c}(\iota)}
=\skew4\check{\Gamma}_{\!\!\smash{\mathrm{H}\omega}\,}^{\mathrm{c}}
\skew4\check{T}_{\!\smash{\omega}}^{\mathrm{c}(\iota)}
\);
\(
{}\ 
\check{E}_{\smash{\omega}}^{\mathrm{c}(\iota)}
+\!=
\Delta\check{E}_{\smash{\omega}}^{\mathrm{c}(\iota)}
\);
\item \(\iota\!+\!\!+\);
\end{itemize}
\end{enumerate}
\caption{}
\label{alg:AlgDisConPie1D}
\end{algorithmD} 
Here,  
\(
\check{F}_{\!\smash{\omega i}}^{\mathrm{c}\pm}
:=n^{-1/2}e^{\pm\imath k_{\omega}x_{\smash{i}}^{\mathrm{c}}}
\) 
and 
\(
k_{\nu n+\omega}
=2\pi(\nu n+\omega)/l
=2\pi(\nu+\omega/n)/h
\). 
Again, a sum on repeated \(i\), \(\omega\) is employed. Note the dependence 
of the discretized Lippmann-Schwinger operator 
\(
\skew4\check{\Gamma}_{\!\!\smash{\mathrm{H}\,\omega}}^{\mathrm{c}}
\) 
on \(s_{-m n+\omega}\), \(\ldots\), \(s_{(m-1) n+\omega}\). As already 
noted above, this is one difference between the two PCD-based algorithms 
\ref{alg:AlgDisIntFouDifFinFunGrePC1D} and \ref{alg:AlgDisConPie1D}. 

\section{Computational comparisons in 1D} 
\label{sec:Res1D}

Employing the algorithms just discussed, the 1D BVP for the periodic 
MI case is now solved numerically and compared with the analytical solution. 
To this end, the unit cell from the analytic case in Section \ref{sec:ConAna1D} 
as based on \(C_{\mathrm{I}}/C_{\mathrm{M}}=100\) is employed, 
and the homogeneous stiffness \(C_{\mathrm{H}}\) is determined by 
\(C_{\mathrm{H}}=\frac{1}{2}(C_{\mathrm{M}}+C_{\mathrm{I}})\). 

Since \(C(x)\) is not defined on the MI interface in the discontinuous case, 
note that this interface always lies between two nodes, one in the matrix, 
and the other in the inclusion. This is in contrast to the smooth case, for which 
\(C(x)\) is defined everywhere. The stiffness profile \(C(x)=1/S\!(x)\) for the 
smooth interface is based on \(\nu(x)\) with \(\epsilon/l=1/100\). 
To be comparible with the analytic results in Figure \ref{fig:StnDisIncMat}, 
the following numerical results are based on deformation control and 
\(\bar{E}=1\). 
In the context of \eqref{equ:EffNumWavConWil1D}, results are obtained 
and compared for \(a=\mathrm{F},\mathrm{CD},\mathrm{FD}\). 
These are obtained as well for \((a,b)=(\mathrm{FD},\mathrm{hC})\). 
In all cases, the tolerance \(\mathrm{tol}\) is machine precision. 

\begin{figure}[H]
\centering
\includegraphics[width=0.48\textwidth]{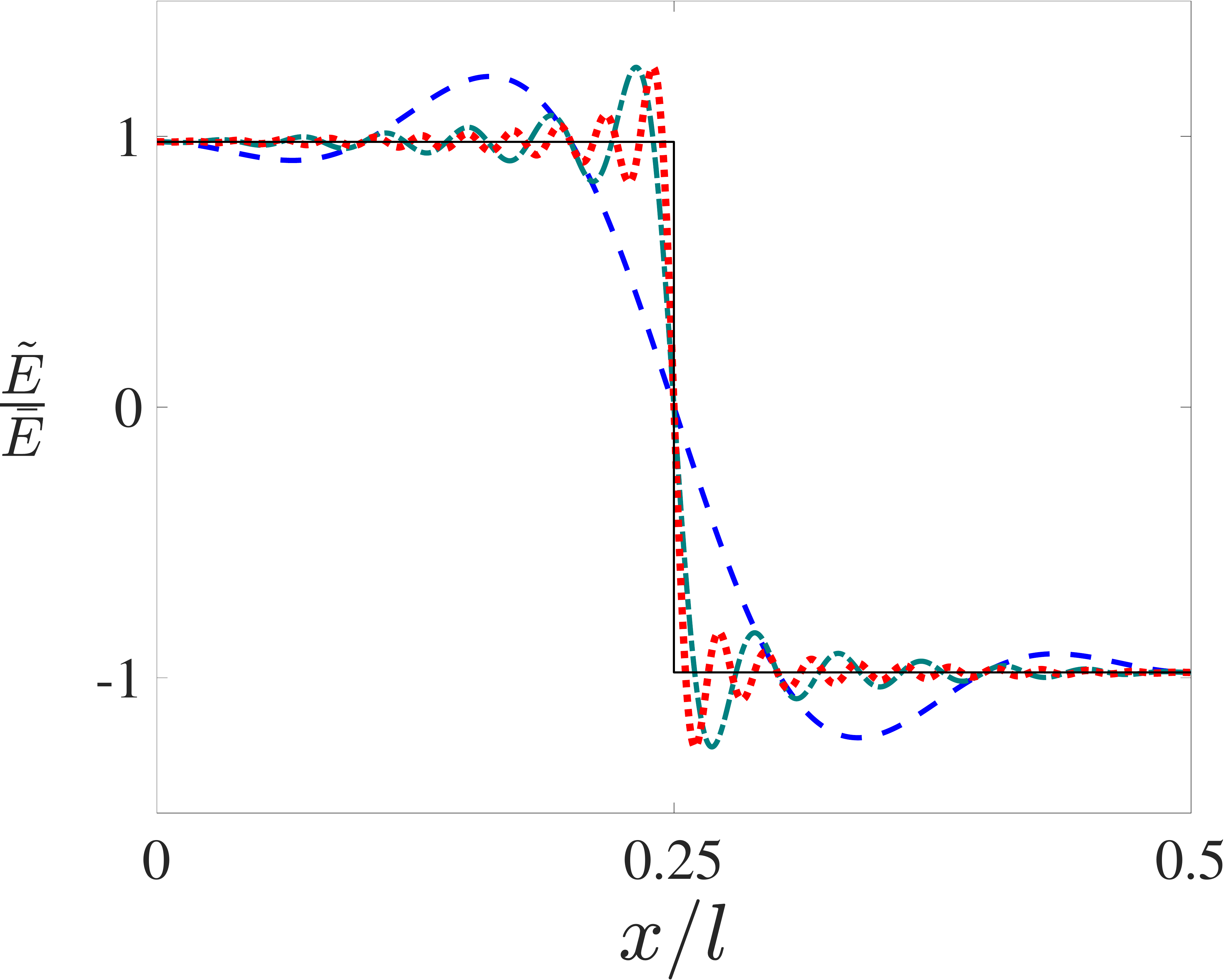}
\hspace{1mm}
\includegraphics[width=0.48\textwidth]{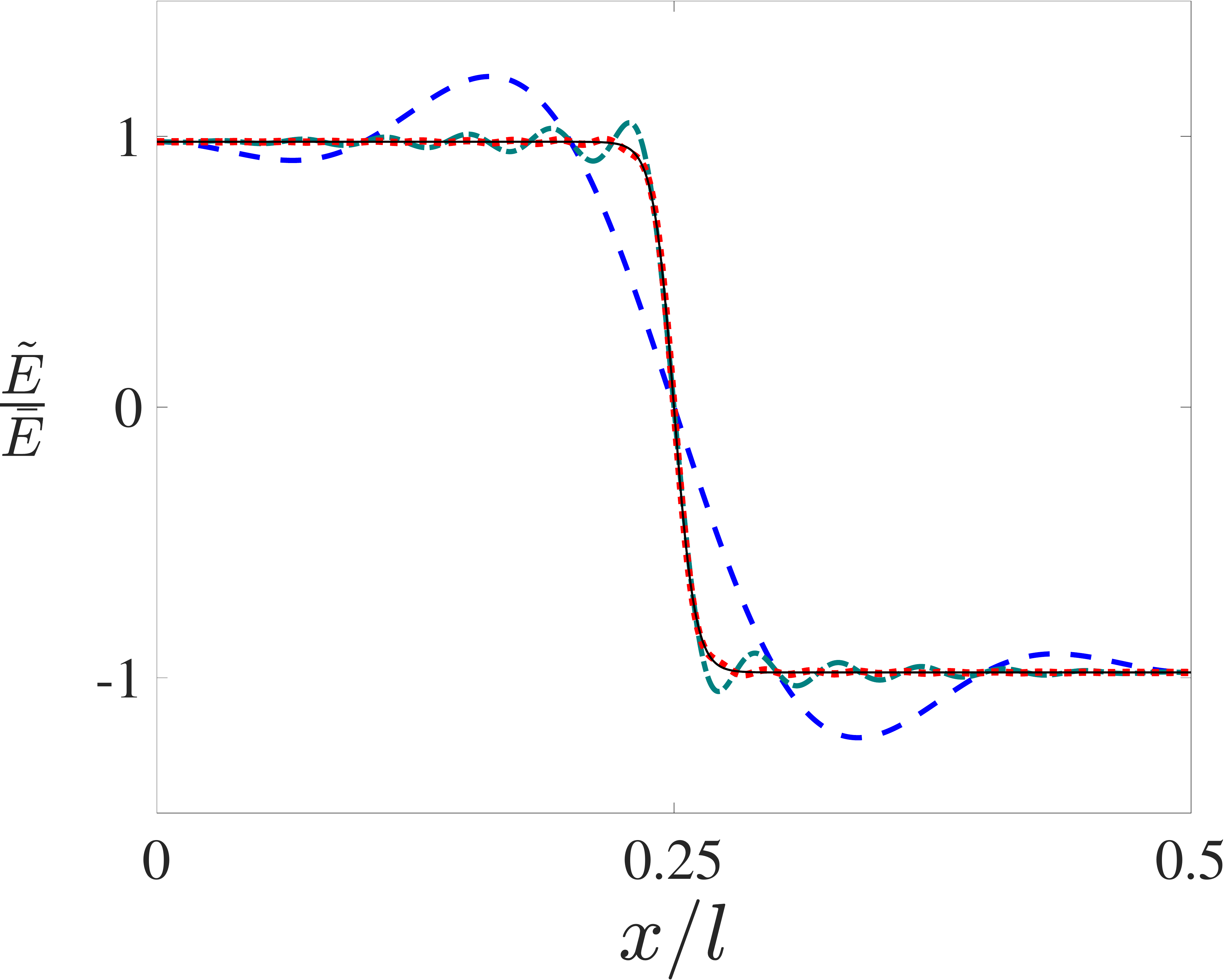}
\caption{Comparison of results for 
\(
\tilde{E}_{a}(x)
=\varphi_{\mu}(x)
\,\check{u}_{\mu}^{\smash{\mathrm{p}}}
\otimes 
q_{\mu}^{a}
\) 
obtained from Algorithm \ref{alg:AlgDisIntFouDifFinFunGre1D}  
with the analytic result \(\tilde{E}(x)\) (black curve) from 
\eqref{equ:FieStnDisIncMat1D}${}_{1}$ across a sharp (left) and 
a smooth (right) MI interface at \(x/l=1/4\) for the discretizations
(i) \(n=10\) (\(m=5\); blue curve), 
(ii) \(n=50\) (\(m=25\); green curve), 
(iii) \(n=90\) (\(m=45\); red curve). As done in Figures 
\ref{fig:StnDisIncMat} and \ref{fig:SerFouStnDisFluDsc} above, 
results are shown here and in what follows for the left half of the unit cell. 
} 
\label{fig:StnShaSmoStr1D}
\end{figure}

To begin, consider the results for \(\tilde{E}(x)\) in Figure \ref{fig:StnShaSmoStr1D}. 
\vspace{-3mm}
As implied by these results, all choices considered for 
\(q_{\mu}^{a}\) and \(q_{\mu}^{b}\) in the context of Algorithm 
\ref{alg:AlgDisIntFouDifFinFunGre1D} yield the same 
solution for \(\tilde{E}_{a}(x)\) at both sharp 
(Figure \ref{fig:StnShaSmoStr1D}, left) and smooth 
(Figure \ref{fig:StnShaSmoStr1D}, right) MI interfaces. 
Moreover, as expected from Fourier series approximation  
of the analytic solution and series truncation 
(Figure \ref{fig:SerFouStnDisFluDsc}, left), 
\(\tilde{E}_{a}(x)\) at the sharp MI interface (left) exhibits 
Gibbs error whose magnitude is independent of numerical resolution. 
This is in contrast to \(\tilde{E}_{a}(x)\) at the smooth MI interface 
(Figure \ref{fig:StnShaSmoStr1D}, right), which does converge in this 
fashion, again as expected. 

\begin{figure}[H]
\centering
\includegraphics[width=0.95\textwidth]{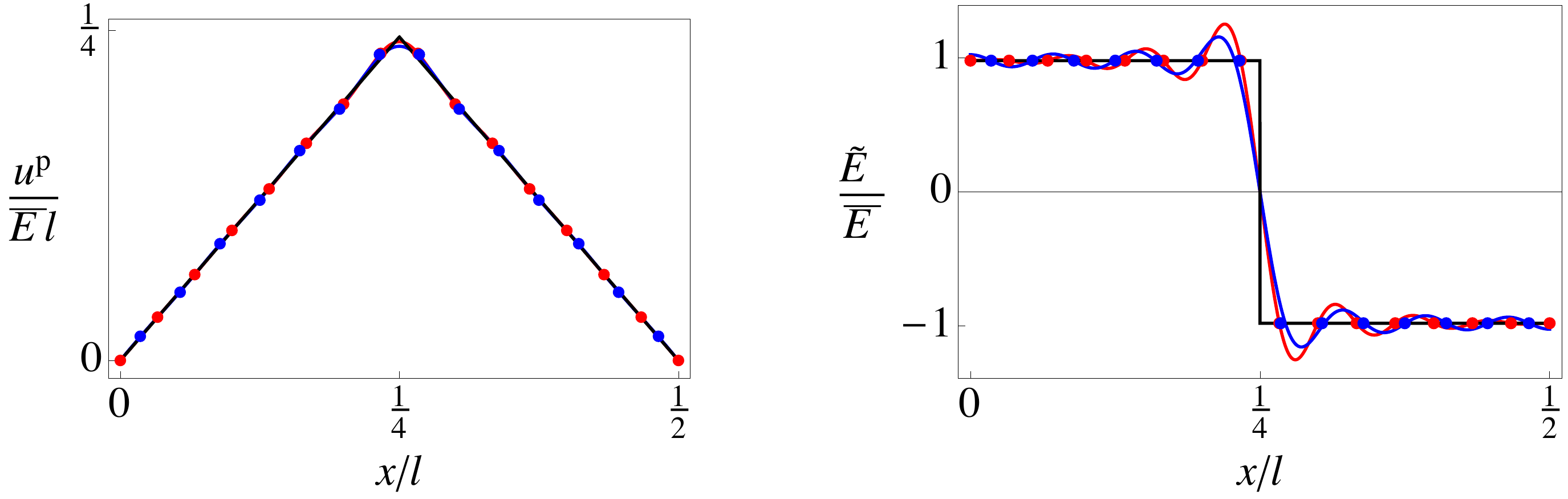}
\caption{Left:~comparison of results for 
\(
u_{\smash{\mathrm{T}}}^{\mathrm{p}}(x)
=\varphi_{\mu}(x)\,\skew2\check{u}_{\smash{\mu}}^{\mathrm{p}}
\) 
from Algorithm \ref{alg:AlgDisIntFouDifFinFunGre1D} (red points, curve) and for 
\(
u_{\smash{\mathrm{PC}}}^{\mathrm{p}}(x)
=\varphi_{\smash{\mu}}^{\mathrm{c}}(x)
\,\check{u}_{\smash{\mu}}^{\mathrm{pc}}
\) 
from Algorithm \ref{alg:AlgDisIntFouDifFinFunGrePC1D} (blue points, curve) 
with analytic solution (black curve) across a sharp MI interface at \(x/l=0.25\). 
Right:~comparison of results for 
\(
\tilde{E}_{\mathrm{T}}(x)
:=\nabla^{\smash{\mathrm{F}}}\!u_{\smash{\mathrm{T}}}^{\mathrm{p}}(x)
=\varphi_{\mu}(x)
\,\check{u}_{\smash{\mu}}^{\smash{\mathrm{p}}}
\otimes 
q_{\smash{\mu}}^{\mathrm{F}}
\) 
from Algorithm \ref{alg:AlgDisIntFouDifFinFunGre1D} (red points, curve) and for 
\(
\tilde{E}_{\mathrm{PC}}(x)
:=\nabla^{\mathrm{F}}\!u_{\smash{\mathrm{PC}}}^{\mathrm{p}}(x)
=\varphi_{\smash{\mu}}^{\mathrm{c}}(x)
\,\check{u}_{\smash{\mu}}^{\mathrm{pc}}
\otimes 
q_{\mu}^{\mathrm{F}}
\) 
from Algorithm \ref{alg:AlgDisIntFouDifFinFunGrePC1D} (blue points, curve) 
with analytic solution (black curve). Results are based on discretizations of 
\(n=30\) (\(m=15\); red points) and \(n=32\) (\(m=16\); blue points). 
} 
\label{fig:DisParStnFluDsc1D}
\end{figure}

\begin{figure}[H]
\centering
\includegraphics[width=0.42\textwidth]{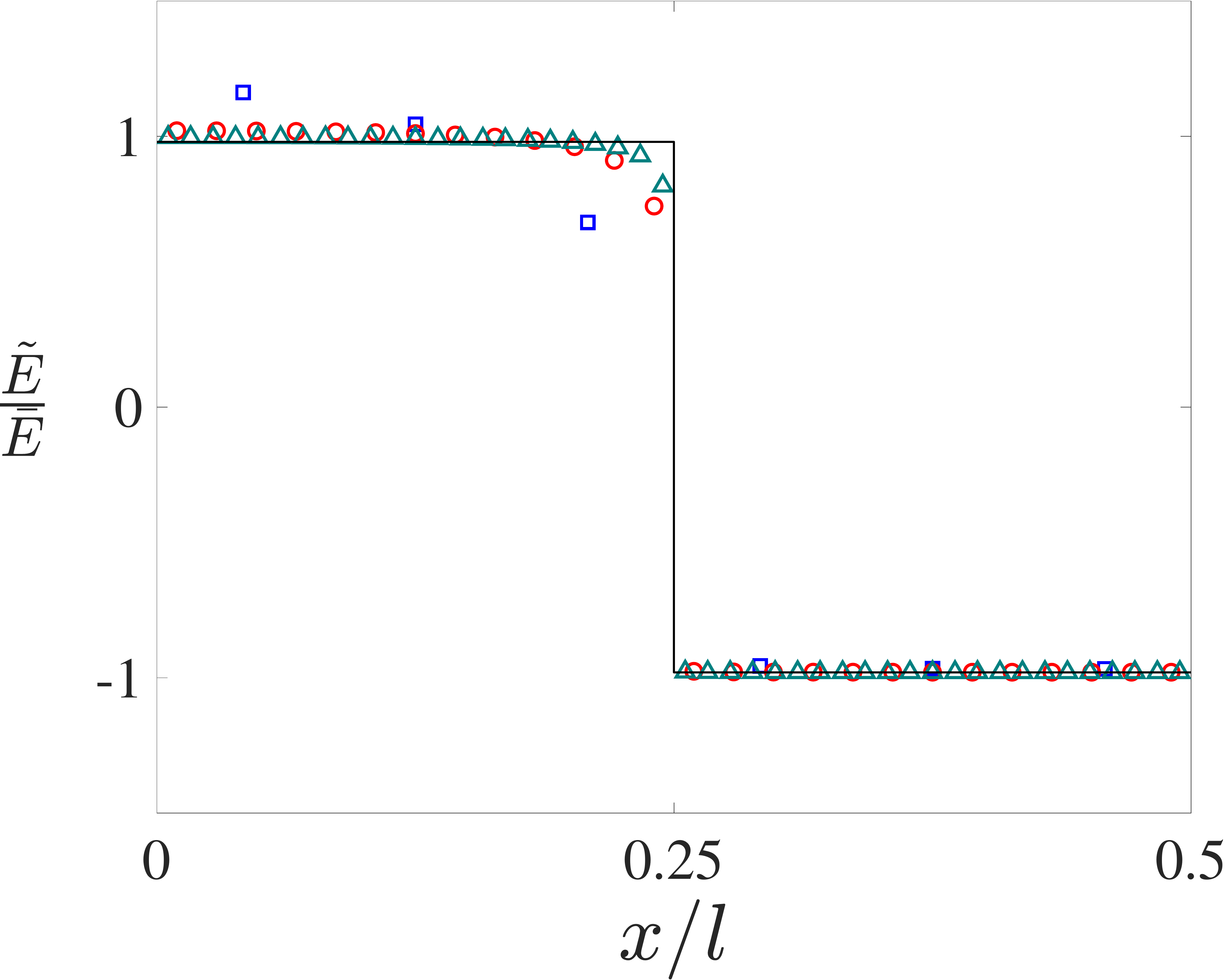}
\hspace{5mm}
\includegraphics[width=0.42\textwidth]{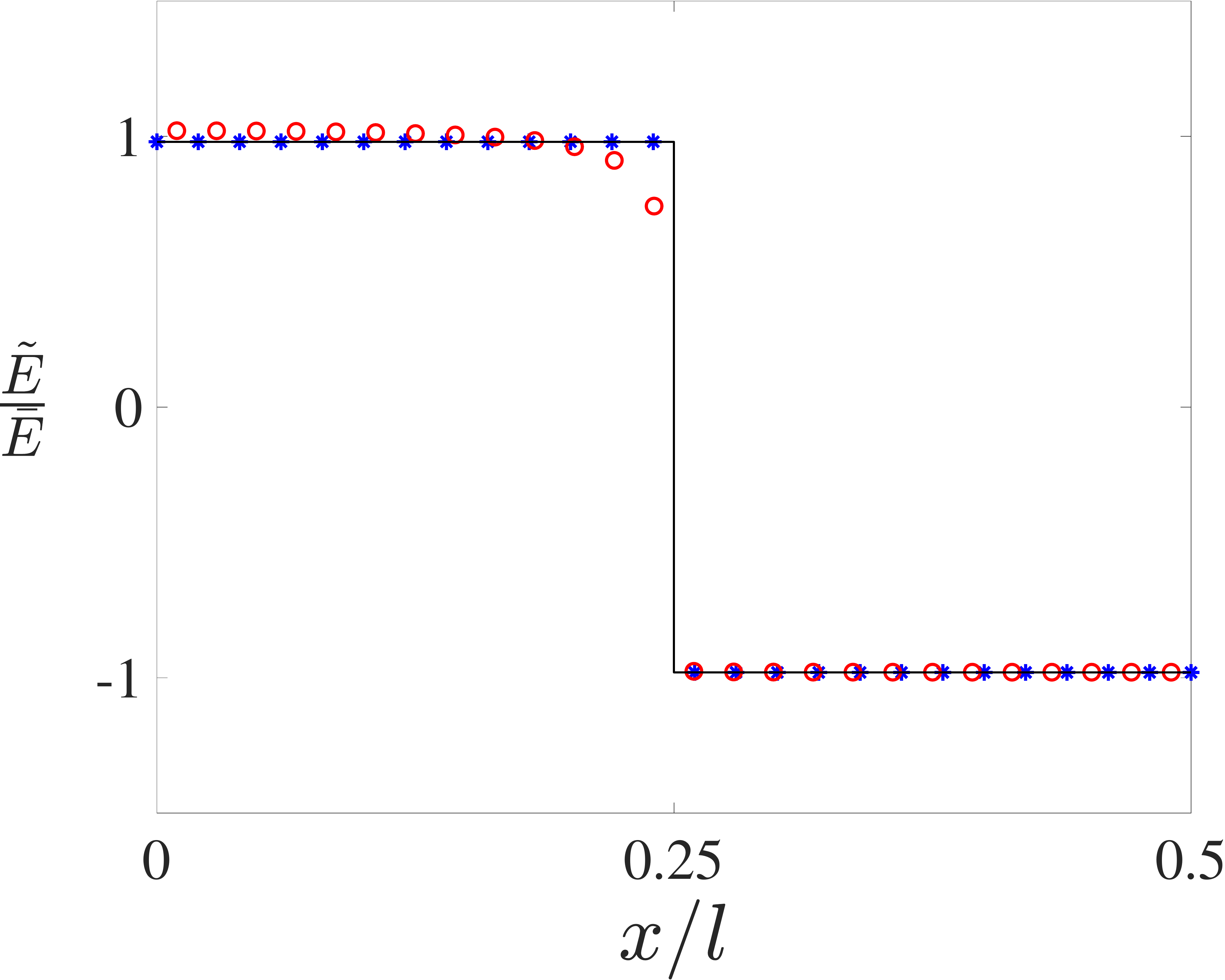}
\caption{Left:~Results for nodal \(\tilde{E}_{i}^{\mathrm{c}}\) from 
Algorithm \ref{alg:AlgDisConPie1D} across a sharp MI interface at \(x/l=1/4\) 
for the discretizations
(i) \(n=10\) (\(m=5\); blue squares), 
(ii) \(n=50\) (\(m=25\); red circles), 
(iii) \(n=90\) (\(m=45\); green triangles), compared with the analytic 
result \(\tilde{E}(x)\) (black curve) from \eqref{equ:FieStnDisIncMat1D}${}_{1}$. 
Righ:~Comparison of results for nodal \(\tilde{E}_{i}\) from 
Algorithm \ref{alg:AlgDisIntFouDifFinFunGre1D} (blue asterisks) 
and those for \(\tilde{E}_{i}^{\mathrm{c}}\) from Algorithm 
\ref{alg:AlgDisConPie1D} (red circles) for the discretization \(n=50\) (\(m=25\)). 
} 
\label{fig:StnFBDGO1D}
\end{figure}

Consider next the comparison of results from Algorithms 
\ref{alg:AlgDisIntFouDifFinFunGre1D} and \ref{alg:AlgDisIntFouDifFinFunGrePC1D} 
for the case of a sharp MI interface shown in Figure \ref{fig:DisParStnFluDsc1D} 
\vspace{-3mm}
As seen in particular in the strain results on the right, the 
dependence of \(\varphi_{\mu}^{\mathrm{c}}(x)\) in \eqref{equ:DisFouPC1D} 
on \(s_{\mu}^{\mathrm{c}}\) dampens Gibbs oscillations / error in 
\(\tilde{E}_{\mathrm{PC}}(x)\) in comparison to \(\tilde{E}_{\mathrm{T}}(x)\). 
As expected, this has no effect on the numerical (i.e., collocation) solution at 
the nodes, in contrast to the analogous results from 
Algorithm \ref{alg:AlgDisConPie1D} in Figure \ref{fig:StnFBDGO1D}. 
\vspace{-3mm}
As shown on the left, the nodal results from Algorithm \ref{alg:AlgDisConPie1D} 
deviate slightly from those of the anayltic solution in the (relatively soft) matrix, 
with maximum deviation in the matrix next to the MI interface. Note that this 
deviation decreases with increasing numerical resolution. This is in contrast 
to the analogous solutions from Algorithm \ref{alg:AlgDisIntFouDifFinFunGre1D} 
(and so Algorithm \ref{alg:AlgDisIntFouDifFinFunGrePC1D}) as shown by the 
comparison on the right. One difference between the PCD-based Algorithms 
\ref{alg:AlgDisIntFouDifFinFunGrePC1D} and \ref{alg:AlgDisConPie1D} that 
is likely playing a role here is the dependence of the discretized 
Lippmann-Schwinger operator 
\(
\skew4\check{\Gamma}_{\!\!\smash{\mathrm{H}\,\omega}}^{\mathrm{c}}
\) 
on \(s_{\nu n+\omega}\) in Algorithm \ref{alg:AlgDisConPie1D}. These and other 
differences in 1D become even more evident in the context of 
tensor-product-based generalization of the above 1D algorithms to 3D, to 
which we now turn. 

\section{Algorithms for strong mechanical equilibrium in 3D}
\label{sec:AlgStr3D}

\subsection{Basics} 
\label{sec:Bas3D}

Let \(\bm{u}\), \(\bm{H}=\nabla\bm{u}\), and 
\(\bm{E}=\mathop{\mathrm{sym}}\bm{H}\) 
represent the 3D displacement, distortion and strain, 
respectively. In this case, let \(f=\bar{f}+\tilde{f}\) now 
represent the 3D generalization of \eqref{equ:FieMeaCelUni1D} 
with respect to the unit cell \(U\). Likewise, 
\begin{equation}
\bm{u}
=\bm{u}^{\smash{\mathrm{h}}}
+\bm{u}^{\smash{\mathrm{p}}}
\,,\quad
\bm{u}^{\smash{\mathrm{h}}}(\bm{x})
=\bm{c}+\skew3\bar{\bm{H}}\bm{x}
\,,\quad
\bm{u}^{\smash{\mathrm{p}}}(\bm{x})
=\int\skew3\tilde{\bm{H}}(\bm{x})\,d\bm{x}
\,,\quad
\nabla\bm{u}^{\smash{\mathrm{p}}}
=\skew3\tilde{\bm{H}}
\,,
\label{equ:DisFluParHom3D}
\end{equation} 
is the generalization of \eqref{equ:DisFluParHom1D} to 3D 
with \(\bar{\bm{u}}^{\smash{\mathrm{p}}}=\bm{0}\), and so 
\(
\bar{\bm{u}}^{\smash{\mathrm{h}}}
=\bm{c}+\skew3\bar{\bm{H}}\bar{\bm{x}}
=\bar{\bm{u}}
\). 
As in the 1D case, \(\bm{c}\), \(\skew3\bar{\bm{H}}\), and so 
\(\bm{u}^{\smash{\mathrm{h}}}\), are known, and 
attention is focused on \(\bm{u}^{\smash{\mathrm{p}}}\). On 
this basis, quasi-static mechanical equilibrium and isotropic, linear 
elastic material behavior continue to apply. Let \(\bm{T}=\ssans{C}\bm{E}\) 
be the linear elastic stress, and 
\(
\ssans{C}(\bm{x})
=\lambda(\bm{x})\,\bm{I}\otimes\bm{I}
+\mu(\bm{x})\,(\bm{I}\,\square\,\bm{I}+\bm{I}\,\triangle\,\bm{I})
\) 
the isotropic stiffness. For the case of discontinuous \(\ssans{C}(\bm{x})\), 
the following hold at the MI interface (with unit normal \(\bm{n}\)):~
(i) no cracking 
\(
\bm{u}_{\mathrm{I}}
=\bm{u}_{\mathrm{M}}
\), 
(ii) kinematic compatibility 
\(
\bm{H}_{\mathrm{I}}
=\bm{H}_{\mathrm{M}}+\bm{h}\otimes\bm{n}
\), 
and 
(iii) mechanical equilibrium 
\(
\bm{T}_{\!\mathrm{I}}\bm{n}
=\bm{T}_{\!\mathrm{M}}\bm{n}
\). 
Both \(\lambda(\bm{x})\) and \(\mu(\bm{x})\) are determined by 
direct tensor-product generalization of analogous 1D relations like 
\eqref{equ:FieComIncMat1D} based on \(\theta\) in the discontinuous, 
and on \(\nu\) in the continuous, case. 

\subsection{Algorithm based on trapezoidal discretization} 
\label{sec:AppTra3D}

Let \(r=1,2,3\). As in the 1D case, consider a uniform discretization of 
\(\lbrack 0,l_{r}\rbrack\) based on \(n_{r}+1\) nodes 
\(x_{\smash{i_{r}}}=i_{r}h_{r}\) with  
\(i_{r}=0,\ldots,n_{r}\) and spacing 
\(h_{r}\) such that \(l_{r}=n_{r}h_{r}\) 
(\(n_{r}\) subintervals). 
Direct tensor-product-based generalization of \eqref{equ:FieFouAppTra1D}${}_{1}$
to 3D yields  
\begin{equation}
f_{\smash{\mathrm{T}}}(\mathbf{x})
=\varphi_{\smash{\mu_{1}}}(x_{1})
\,\varphi_{\smash{\mu_{2}}}(x_{2})
\,\varphi_{\smash{\mu_{3}}}(x_{3})
\,\check{f}_{\smash{\mu_{1}\mu_{2}\mu_{3}}}
=:\varphi_{\bm{\mu}}(\mathbf{x})
\star
\check{f}_{\bm{\mu}}
\label{equ:DisFouSca3D}
\end{equation} 
(sum on repeated indices) 
in terms of Rayleigh product notation with 
\(
\mathbf{x}
:=(x_{1},x_{2},x_{3})
\) 
and 
\(
\mu_{r}=1,\ldots,n_{r}\), 
\(r=1,2,3\). 
Here, 
\begin{equation}
\begin{array}{rclclcl}
f_{\mathbf{i}}
&:=&
f_{\smash{i_{1}i_{2}i_{3}}}
&=&
F_{\!\smash{\mu_{1}i_{1}}\,}^{+}
F_{\!\smash{\mu_{2}i_{2}}\,}^{+}
F_{\!\smash{\mu_{3}i_{3}}\,}^{+}
\,\check{f}_{\smash{\mu_{1}\mu_{2}\mu_{3}}}
&=:&
F_{\!\bm{\mu}\mathbf{i}\,}^{+}
\star
\check{f}_{\bm{\mu}}
\,,\\
\check{f}_{\bm{\mu}}
&:=&
\check{f}_{\smash{\mu_{1}\mu_{2}\mu_{3}}}
&=&
F_{\!\smash{\mu_{1}i_{1}}\,}^{-}
F_{\!\smash{\mu_{2}i_{2}}\,}^{-}
F_{\!\smash{\mu_{3}i_{3}}\,}^{-}
\,f_{\smash{i_{1}i_{2}i_{3}}}
&=:&
F_{\!\bm{\mu}\mathbf{i}\,}^{-}
\star 
f_{\mathbf{i}}
\,,
\end{array}
\label{equ:DisFouMat3D}
\end{equation} 
(sum on repeated indices) via \eqref{equ:FieFouAppTra1D}${}_{3,4}$. Given 
these, generalization of Algorithm \ref{alg:AlgDisIntFouDifFinFunGre1D} to 3D 
is based in particular on those 
\begin{equation}
\nabla^{\mathbf{a}}
\mathbf{u}_{\smash{\mathrm{T}}}^{\mathrm{p}}(\mathbf{x})
=\varphi_{\bm{\mu}}(\mathbf{x})
\star
\check{\mathbf{u}}_{\bm{\mu}}^{\smash{\mathrm{p}}}
\otimes
\mathbf{q}_{\bm{\mu}}^{\mathbf{a}}
\,,\quad
\mathrm{div}^{\mathbf{b}}
\mathbf{T}_{\mathrm{T}}(\mathbf{x})
=\varphi_{\bm{\mu}}(\mathbf{x})
\star
\,\check{\mathbf{T}}_{\!\bm{\mu}\,}
\mathbf{q}_{\bm{\mu}}^{\mathbf{b}}
\,,
\label{equ:DivGraFouTraFDD3D}
\end{equation} 
of \eqref{equ:DisFouAppTra1D}${}_{1}$ and \eqref{equ:StrFouAppTra1D}${}_{2}$, 
respectively. Here,  
\(
\check{\mathbf{u}}_{\smash{\bm{\mu}}}^{\mathrm{p}}
:=(\check{u}_{\smash{1\bm{\mu}}}^{\mathrm{p}},
\check{u}_{\smash{2\bm{\mu}}}^{\mathrm{p}},
\check{u}_{\smash{3\bm{\mu}}}^{\mathrm{p}})
\), 
\(
\check{\mathbf{T}}_{\!\bm{\mu}\,}
\) 
is the \(3\times 3\) symmetric matrix of stress components, and 
\(
\mathbf{q}_{\smash{\bm{\mu}}}^{\mathbf{a}}
:=(q_{\smash{\mu_{1}}}^{a_{1}},
q_{\smash{\mu_{2}}}^{a_{2}},
q_{\smash{\mu_{3}}}^{a_{3}})
\) 
with 
\(
q_{\smash{\mu_{r}}}
:=\imath k_{\smash{\mu_{r}-1-m_{r}}}
\). 
As in the 1D case 
\eqref{equ:FunGreModFD1D}, 3D quasi-static momentum balance is 
formulated algorithmically in preconditioned form based on the Green 
function 
\(
\check{\mathbf{G}}_{\mathrm{H}}
(\mathbf{q}_{\smash{\bm{\mu}}}^{\mathbf{a}},
\mathbf{q}_{\smash{\bm{\mu}}}^{\mathbf{b}})
\) 
of the corresponding operator 
\begin{equation}
-\mathrm{div}^{\mathbf{b}}
\,\textsf{C}_{\mathrm{H}}
\nabla_{\!\mathbf{h}}^{\mathbf{a}}\mathbf{u}(\mathbf{x})
=-\varphi_{\bm{\mu}}(\mathbf{x})
\star
\textsf{C}_{\mathrm{H}}
\lbrack
\check{\mathbf{u}}_{\bm{\mu}}
\otimes
\mathbf{q}_{\bm{\mu}}^{\mathbf{a}}
\rbrack
\mathbf{q}_{\bm{\mu}}^{\mathbf{b}}
=\varphi_{\bm{\mu}}(\mathbf{x})
\star
\check{\mathbf{G}}_{\mathrm{H}}^{-1}
(\mathbf{q}_{\bm{\mu}}^{\mathbf{a}},
\mathbf{q}_{\bm{\mu}}^{\mathbf{b}})
\,\check{\mathbf{u}}_{\bm{\mu}}
\,.
\label{equ:OpeFunGreAlgMod3D}
\end{equation} 
Given these relations, direct componentwise generalization of Algorithm 
\ref{alg:AlgDisIntFouDifFinFunGre1D} yields 
Algorithm \ref{alg:AlgDisIntFouDifFinFunGre3D}.  
\begin{algorithmT}[H]
\begin{enumerate}
\setlength{\itemsep}{-0.1mm}
\item given:~\(n_{1},\ldots,n_{3}\), 
\(\ssans{C}(\bm{x})\), 
\(\bar{\bm{E}}\),
\(\ssans{C}_{\mathrm{H}}\), 
\(
\check{\mathbf{G}}_{\smash{\mathrm{H}\bm{\mu}}}^{\mathbf{ab}}
=\hat{\mathbf{G}}_{\mathrm{H}}
(\mathbf{q}_{\smash{\bm{\mu}}}^{\mathbf{a}},
\mathbf{q}_{\smash{\bm{\mu}}}^{\mathbf{b}})
\)
\item initialization \(\iota=0\)
\vspace{-3mm}
\begin{itemize}
\setlength{\itemsep}{-0.1mm}
\item for \(i_{1}=1,\ldots,n_{1}\), \(\ldots\), \(i_{3}=1,\ldots,n_{3}\) 
\newline 
\(
{}\ 
\textsf{C}_{\mathbf{i}}=\textsf{C}(\mathbf{x}_{\mathbf{i}})
\); 
\(
{}\ 
\mathbf{T}_{\!\smash{\mathbf{i}}}^{(\iota)}
=\textsf{C}_{\mathbf{i}\,}\bar{\mathbf{E}}
\); 
\item for \(\mu_{1}=1,\ldots,n_{1}\), \(\ldots\), \(\mu_{3}=1,\ldots,n_{3}\)
\newline
\(
{}\ 
\check{\mathbf{u}}_{\smash{\bm{\mu}}}^{\mathrm{p}(\iota)}
=\bm{0}
\); 
\(
{}\ 
\check{\mathbf{T}}_{\!\smash{\bm{\mu}}}^{(\iota)}
=F_{\!\smash{\bm{\mu}\mathbf{i}}}^{-}
\star
\mathbf{T}_{\!\smash{\mathbf{i}}}^{(\iota)}
\); 
\(
{}\ 
\Delta\check{\mathbf{u}}_{\bm{\mu}}^{\smash{\mathrm{p}}(\iota)}
=\check{\mathbf{G}}_{\smash{\mathrm{H}\bm{\mu}}}^{\mathbf{ab}}
\check{\mathbf{T}}_{\!\smash{\bm{\mu}}}^{(\iota)}
\mathbf{q}_{\smash{\bm{\mu}}}^{\mathbf{b}}
\); 
\item \(\iota\!+\!\!+\);
\end{itemize}
\item while
\(
\sum_{\mu_{1}=1}^{n_{1}}
\cdots
\sum_{\mu_{3}=1}^{n_{3}}
|\Delta\check{\mathbf{u}}_{\smash{\bm{\mu}}}^{\mathrm{p}(\iota)}
-\Delta\check{\mathbf{u}}_{\smash{\bm{\mu}}}^{\mathrm{p}(\iota-1)}|
\geqslant\mathrm{tol}
\) 
\& \(\iota\leqslant\mathrm{maxit}\) 
\vspace{-2mm}
\begin{itemize}
\setlength{\itemsep}{-0.1mm}
\item for \(i_{1}=1,\ldots,n_{1}\), \(\ldots\), \(i_{3}=1,\ldots,n_{3}\) 
\newline 
\(
{}\ 
\tilde{\mathbf{E}}_{\smash{\mathbf{i}}}^{(\iota)}
=F_{\!\smash{\bm{\mu}\mathbf{i}}}^{+}
\star
\mathop{\mathrm{sym}}
(\check{\mathbf{u}}_{\smash{\bm{\mu}}}^{\mathrm{p}(\iota)}
\otimes
\mathbf{q}_{\smash{\bm{\mu}}}^{\mathbf{a}(\iota)})
\);
\(
{}\ 
\mathbf{T}_{\!\smash{\mathbf{i}}}^{(\iota)}
=\mathbf{T}_{\!\smash{\mathbf{i}}}^{(0)}
+\textsf{C}_{\mathbf{i}}
\tilde{\mathbf{E}}_{\smash{\mathbf{i}}}^{\mathbf{a}(\iota)}
\);
\item for \(\mu_{1}=1,\ldots,n_{1}\), \(\ldots\), \(\mu_{3}=1,\ldots,n_{3}\)
\newline
\(
{}\ 
\check{\mathbf{T}}_{\!\smash{\bm{\mu}}}^{(\iota)}
=F_{\!\smash{\bm{\mu}\mathbf{i}}}^{-}
\star
\mathbf{T}_{\!\smash{\mathbf{i}}}^{(\iota)}
\); 
\(
{}\ 
\Delta\check{\mathbf{u}}_{\bm{\mu}}^{\smash{\mathrm{p}}(\iota)}
=\check{\mathbf{G}}_{\smash{\mathrm{H}\bm{\mu}}}^{\mathbf{ab}}
\check{\mathbf{T}}_{\!\smash{\bm{\mu}}}^{(\iota)}
\mathbf{q}_{\smash{\bm{\mu}}}^{\mathbf{b}}
\); 
\(
{}\ 
\check{\mathbf{u}}_{\smash{\bm{\mu}}}^{\mathrm{p}(\iota)}
+\!=
\Delta\check{\mathbf{u}}_{\smash{\bm{\mu}}}^{\mathrm{p}(\iota)}
\);
\item \(\iota\!+\!\!+\);
\end{itemize}
\end{enumerate}
\caption{}
\label{alg:AlgDisIntFouDifFinFunGre3D}
\end{algorithmT}
For the current isotropic case, 
\(
\ssans{C}_{\mathrm{H}}
=\lambda_{\mathrm{H}}\,\bm{I}\otimes\bm{I}
+\mu_{\mathrm{H}}
\,(\bm{I}\,\square\,\bm{I}+\bm{I}\,\triangle\,\bm{I})
\), 
and so 
\begin{equation}
\check{\mathbf{G}}_{\mathrm{H}}^{-1}
(\mathbf{q}_{\bm{\mu}}^{\mathbf{a}},
\mathbf{q}_{\bm{\mu}}^{\mathbf{b}})
:=-\mu_{\mathrm{H}}
\,(\mathbf{q}_{\bm{\mu}}^{\mathbf{a}}
\cdot
\mathbf{q}_{\bm{\mu}}^{\mathbf{b}})
\,\mathbf{I}
-\mu_{\mathrm{H}}
\,\mathbf{q}_{\bm{\mu}}^{\mathbf{a}}
\otimes
\mathbf{q}_{\bm{\mu}}^{\mathbf{b}}
-\lambda_{\mathrm{H}}
\,\mathbf{q}_{\bm{\mu}}^{\mathbf{b}}
\otimes
\mathbf{q}_{\bm{\mu}}^{\mathbf{a}}
\,.
\label{equ:EquMecFunGre3D}
\end{equation} 
Since  
\(
\det\check{\mathbf{G}}_{\smash{\mathrm{H}}}^{-1}
(\mathbf{q}_{\smash{\bm{\mu}}}^{\mathbf{a}},
\mathbf{q}_{\smash{\bm{\mu}}}^{\mathbf{b}})
=-(\lambda_{\mathrm{H}}+\mu_{\mathrm{H}}+3)
\,(\mathbf{q}_{\smash{\bm{\mu}}}^{\mathbf{a}}
\cdot
\mathbf{q}_{\smash{\bm{\mu}}}^{\mathbf{b}})
\), 
note that 
\(
\hat{\mathbf{G}}_{\smash{\mathrm{H}}}^{-1}
\) 
is invertible for 
\(
\mathbf{q}_{\bm{\mu}}^{\mathbf{a}}
\cdot
\mathbf{q}_{\bm{\mu}}^{\mathbf{b}}
\neq 
0
\). 
In the context of conjugacy (e.g., \eqref{equ:EffNumWavConWil3D} below in 3D), 
note that 
\(
\mathbf{q}_{\smash{\bm{\mu}}}^{\mathbf{b}}
\otimes
\mathbf{q}_{\smash{\bm{\mu}}}^{\mathbf{a}}
=\mathbf{q}_{\smash{\bm{\mu}}}^{\mathbf{a}}
\otimes
\mathbf{q}_{\smash{\bm{\mu}}}^{\mathbf{b}}
\) 
hold when 
\(
q_{\smash{\smash{\mu_{r}}}}^{a_{r}\ast}
=-q_{\smash{\smash{\mu_{r}}}}^{a_{r}}
\) 
(e.g., \(a_{r}=\mathrm{ACD},\mathrm{AhC}\)), and so 
\(
\check{\mathbf{G}}_{\smash{\mathrm{H}}}^{-1}
(\mathbf{q}_{\smash{\bm{\mu}}}^{\mathbf{a}},
\mathbf{q}_{\smash{\bm{\mu}}}^{\mathbf{b}})
\), is symmetric. 
For other cases (e.g., AFB/R; see below), this is not true in general. 
Restriction to \(\lambda_{\mathrm{H}}=\mu_{\mathrm{H}}\) 
\cite[e.g.,][]{Willot2015}, however, does result in symmetric 
\(
\check{\mathbf{G}}_{\smash{\mathrm{H}}}^{-1}
(\mathbf{q}_{\smash{\bm{\mu}}}^{\mathbf{a}},
\mathbf{q}_{\smash{\bm{\mu}}}^{\mathbf{b}})
\) 
for all \(\mathbf{q}_{\bm{\mu}}^{\mathbf{a}}\) and 
\(\mathbf{q}_{\bm{\mu}}^{\mathbf{b}}\). 

\subsection{FDDs in 3D}

As a first step toward 3D generalization of the 1D FDDs, consider first the 
2D case. To this end, it is useful to work with the difference operators 
\begin{equation}
\begin{array}{rcl}
\delta_{\smash{h_{i}}}^{\mathrm{FD}}
f(\ldots,x_{i},\ldots)
&:=&
h_{i}^{-1}
\lbrack 
f(\ldots,x_{i}+h_{i},\ldots)-f(\ldots,x_{i},\ldots)
\rbrack
\,,\\
\delta_{\smash{h_{i}}}^{\mathrm{BD}}
f(\ldots,x_{i},\ldots)
&:=&
h_{i}^{-1}
\lbrack
f(\ldots,x_{i},\ldots)-f(\ldots,x_{i}-h_{i},\ldots)
\rbrack
\,,\\
\delta_{\smash{h_{i}}}^{\mathrm{CD}}
f(\ldots,x_{i},\ldots)
&:=&
\tfrac{1}{2}
h_{i}^{-1}
\lbrack
f(\ldots,x_{i}+h_{i},\ldots)
-f(\ldots,x_{i}-h_{i},\ldots)
\rbrack
\,,\\
\delta_{\smash{h_{i}}}^{\mathrm{hC}}
f(\ldots,x_{i},\ldots)
&:=&
h_{i}^{-1}
\lbrack
f(\ldots,x_{i}+\tfrac{1}{2}h_{i},\ldots)
-f(\ldots,x_{i}-\tfrac{1}{2}h_{i},\ldots)
\rbrack
\,.
\end{array}
\end{equation}
Given these, consider the 2D grid "cell" with corners at 
\((x_{1},x_{2})\), 
\((x_{1},x_{2}+h_{2})\), 
\((x_{1}+h_{1},x_{2}+h_{2})\), and 
\((x_{1}+h_{1},x_{2})\). With respect to this cell, 
average forward differencing (AFD) is defined as 
\begin{equation}
\begin{array}{rcl}
\nabla_{\!\bm{i}_{1}}^{\mathrm{AFD}}
f(x_{1},x_{2})
&:=&
\tfrac{1}{2}
\lbrack
\delta_{\smash{h_{1}}}^{\mathrm{FD}}
f(x_{1},x_{2})
+\delta_{\smash{h_{1}}}^{\mathrm{FD}}
f(x_{1},x_{2}+h_{2})
\rbrack
\,,\\
\nabla_{\!\bm{i}_{2}}^{\mathrm{AFD}}
f(x_{1},x_{2})
&:=& 
\tfrac{1}{2}
\lbrack
\delta_{\smash{h_{2}}}^{\mathrm{FD}}
f(x_{1},x_{2})
+\delta_{\smash{h_{2}}}^{\mathrm{FD}}
f(x_{1}+h_{1},x_{2})
\rbrack
\,.
\end{array}
\label{equ:FBDiff2D}
\end{equation}
Analogously, average backward differencing (ABD) is defined as 
\begin{equation}
\begin{array}{rcl}
\nabla_{\!\bm{i}_{1}}^{\mathrm{ABD}}
f(x_{1},x_{2})
&:=&
\tfrac{1}{2}
\lbrack
\delta_{\smash{h_{1}}}^{\mathrm{BD}}
f(x_{1},x_{2})
+\delta_{\smash{h_{1}}}^{\mathrm{BD}}
f(x_{1},x_{2}-h_{2})
\rbrack
\,,\\
\nabla_{\!\bm{i}_{2}}^{\mathrm{ABD}}
f(x_{1},x_{2})
&:=& 
\tfrac{1}{2}
\lbrack
\delta_{\smash{h_{2}}}^{\mathrm{BD}}
f(x_{1},x_{2})
+\delta_{\smash{h_{2}}}^{\mathrm{BD}}
f(x_{1}-h_{1},x_{2})
\rbrack
\,,
\end{array}
\label{equ:BDDiff2D}
\end{equation}
with respect to the grid cell with corners at 
\((x_{1},x_{2})\), 
\((x_{1},x_{2}-h_{2})\), 
\((x_{1}-h_{1},x_{2}-h_{2})\), and 
\((x_{1}-h_{1},x_{2})\). 
In the same fashion, we have 
\begin{equation}
\begin{array}{rcl}
\nabla_{\!\bm{i}_{1}}^{\mathrm{ACD}}
f(x_{1},x_{2})
&:=&
\tfrac{1}{2}
\lbrack
\delta_{\smash{h_{1}}}^{\mathrm{CD}}
f(x_{1},x_{2}-h_{2})
+\delta_{\smash{h_{1}}}^{\mathrm{CD}}
f(x_{1},x_{2}+h_{2})
\rbrack
\,,\\
\nabla_{\!\bm{i}_{2}}^{\mathrm{ACD}}
f(x_{1},x_{2})
&:=& 
\tfrac{1}{2}
\lbrack
\delta_{\smash{h_{2}}}^{\mathrm{CD}}
f(x_{1}-h_{1},x_{2})
+\delta_{\smash{h_{2}}}^{\mathrm{CD}}
f(x_{1}+h_{1},x_{2})
\rbrack
\,,
\end{array}
\label{equ:CDDiff2D}
\end{equation}
for average central differencing (ACD), and 
\begin{equation}
\begin{array}{rcl}
\nabla_{\!\bm{i}_{1}}^{\mathrm{AhC}}
f(x_{1},x_{2})
&:=&
\tfrac{1}{2}
\lbrack
\delta_{\smash{h_{1}}}^{\mathrm{hC}}
f(x_{1},x_{2}-\tfrac{1}{2}h_{2})
+\delta_{\smash{h_{1}}}^{\mathrm{hC}}
f(x_{1},x_{2}+\tfrac{1}{2}h_{2})
\rbrack
\,,\\
\nabla_{\!\bm{i}_{2}}^{\mathrm{AhC}}
f(x_{1},x_{2})
&:=& 
\tfrac{1}{2}
\lbrack
\delta_{\smash{h_{2}}}^{\mathrm{hC}}
f(x_{1}-\tfrac{1}{2}h_{1},x_{2})
+\delta_{\smash{h_{2}}}^{\mathrm{hC}}
f(x_{1}+\tfrac{1}{2}h_{1},x_{2})
\rbrack
\,,
\end{array}
\label{equ:hCDDiff2D}
\end{equation}
for average half-central differencing (AhC). 
From \eqref{equ:BDDiff2D} and \eqref{equ:hCDDiff2D} follow 
the direct generalization 
\begin{equation}
\nabla_{\!\bm{i}_{1,2}}^{\mathrm{ABD}}
f(x_{1}+\tfrac{1}{2}h_{1},
x_{2}+\tfrac{1}{2}h_{2})
=\nabla_{\!\bm{i}_{1,2}}^{\mathrm{AhC}}
f(x_{1},x_{2})
\label{equ:BDhCIde2D}
\end{equation}
of \eqref{equ:BDhCIde1D}. 
To obtain the modal forms of these, the notation 
\begin{equation}
f(x_{1}+c_{1}h_{1},\ldots,x_{d}+c_{d}h_{d})
=e^{c_{1}q_{\smash{\mu_{1}}}h_{1}}
\cdots\,
e^{c_{d}q_{\smash{\mu_{d}}}h_{d}}f(x_{1},\ldots,x_{d})
\end{equation}
based on 
\(
f(x_{1},\ldots,x_{d})
=\check{f}_{\smash{\mu_{1}\!\cdots\,\mu_{d}}}
e^{q_{\smash{\mu_{1}}}x_{1}}
\cdots
\,e^{q_{\smash{\mu_{d}}}x_{d}}
\) 
is useful. 
In terms of this, 
\begin{equation}
\begin{array}{rclclcl}
\check{\nabla}_{\!\bm{i}_{r}}^{\mathrm{AFD}}
&:=&
\tfrac{1}{2}h_{r}^{-1}
(e^{q_{\smash{\mu_{r}}}h_{r}}-1)
\,(e^{q_{\smash{\mu_{s}}}h_{s}}+1)
&=&
\tfrac{1}{2}
\,q_{\smash{\mu_{r}}}^{\mathrm{FD}}
\,(e^{q_{\smash{\mu_{s}}}h_{s}}+1)
&=:&
q_{\smash{\mu_{r}}}^{\mathrm{AFD}}
\,,\\
\check{\nabla}_{\!\bm{i}_{r}}^{\mathrm{ABD}}
&:=&
\tfrac{1}{2}h_{r}^{-1}
(1-e^{-q_{\smash{\mu_{r}}}h_{r}})
\,(e^{-q_{\smash{\mu_{s}}}h_{s}}+1)
&=&
\tfrac{1}{2}
\,q_{\smash{\mu_{r}}}^{\mathrm{BD}}
\,(e^{-q_{\smash{\mu_{s}}}h_{s}}+1)
&=:&
q_{\smash{\mu_{r}}}^{\mathrm{ABD}}
\,,\\
\check{\nabla}_{\!\bm{i}_{r}\,}^{\mathrm{ACD}}
&:=&
\tfrac{1}{2}h_{r}^{-1}
(e^{q_{\smash{\mu_{r}}}h_{r}}
-e^{-q_{\smash{\mu_{r}}}h_{r}})
\,\tfrac{1}{2}(e^{q_{\smash{\mu_{s}}}h_{s}}
+e^{-q_{\smash{\mu_{s}}}h_{s}})
&=&
q_{\smash{\mu_{r}}}^{\mathrm{CD}}
\cosh(q_{\smash{\mu_{s}}}h_{s})
&=:&
q_{\smash{\mu_{r}}}^{\mathrm{ACD}}
\,,\\
\check{\nabla}_{\!\bm{i}_{r}\,}^{\mathrm{AhC}}
&:=&
\tfrac{1}{2}h_{r}^{-1}
(e^{\frac{1}{2}q_{\smash{\mu_{r}}}h_{r}}
-e^{-\frac{1}{2}q_{\smash{\mu_{r}}}h_{r}})
\,(e^{\frac{1}{2}q_{\smash{\mu_{s}}}h_{s}}
+e^{-\frac{1}{2}q_{\smash{\mu_{s}}}h_{s}})
&=&
q_{\smash{\mu_{r}}}^{\mathrm{hC}}
\cosh(\frac{1}{2}q_{\smash{\mu_{s}}}h_{s})
&=:&
q_{\smash{\mu_{r}}}^{\mathrm{AhC}}
\,,
\end{array}
\label{equ:FreDifFinFBDF2D}
\end{equation} 
via \eqref{equ:FreDifFinFBDF1D}-\eqref{equ:FreDifFinCDDF1D} 
with \(s=2,1\) for \(r=1,2\). 
Direct generalization of \eqref{equ:FreDifFinFBDF2D} to 3D yields 
\begin{equation}
\begin{array}{rclcl}
\check{\nabla}_{\!\bm{i}_{r}\,}^{\mathrm{AFD}}
&:=&
\frac{1}{4}
\,q_{\smash{\mu_{r}}}^{\mathrm{FD}}
\,(e^{q_{\smash{\mu_{s}}}h_{s}}+1)
\,(e^{q_{\smash{\mu_{t}}}h_{t}}+1)
&=:&
q_{\smash{\mu_{r}}}^{\mathrm{AFD}}
\,,\\
\check{\nabla}_{\!\bm{i}_{r}}^{\mathrm{ABD}}
&:=&
\frac{1}{4}
\,q_{\smash{\mu_{r}}}^{\mathrm{BD}}
\,(e^{-q_{\smash{\mu_{s}}}h_{s}}+1)
\,(e^{-q_{\smash{\mu_{t}}}h_{t}}+1)
&=:&
q_{\smash{\mu_{r}}}^{\mathrm{ABD}}
\,,\\
\check{\nabla}_{\!\bm{i}_{r}}^{\mathrm{ACD}}
&:=&
q_{\smash{\mu_{r}}}^{\mathrm{CD}}
\cosh(q_{\smash{\mu_{s}}}h_{s})
\cosh(q_{\smash{\mu_{t}}}h_{t})
&=:&
q_{\smash{\mu_{r}}}^{\mathrm{ACD}}
\,,\\
\check{\nabla}_{\!\bm{i}_{r}}^{\mathrm{AhC}}
&:=&
q_{\smash{\mu_{r}}}^{\mathrm{hC}}
\cosh(\frac{1}{2}q_{\smash{\mu_{s}}}h_{s})
\cosh(\frac{1}{2}q_{\smash{\mu_{t}}}h_{t})
&=:&
q_{\smash{\mu_{r}}}^{\mathrm{AhC}}
\,.
\end{array}
\label{equ:FreDifFinFBDF3D}
\end{equation} 
with \((s,t)=(2,3),(3,1),(1,2)\) for \(r=1,2,3\). 
Yet another effective wavenumber  
\begin{equation}
q_{\smash{\mu_{r}}}^{\mathrm{R}}
:=\tfrac{1}{4}
\,q_{\smash{\mu_{r}}}^{\mathrm{W}}
(e^{q_{\smash{\mu_{s}}}h_{s}}+1)
(e^{q_{\smash{\mu_{t}}}h_{t}}+1)
\,,\quad 
q_{\mu}^{\mathrm{W}}
:=h^{-1}
\tanh\left(\frac{q_{\mu}h}{2}\right)(e^{q_{\mu}h}+1)
\,,
\label{equ:FreDifFinFBDFR3D}
\end{equation} 
(in the current notation) was obtained by \citet[][Equation (36)]{Willot2015} 
in the context of his "rotated scheme" (R). 
As it turns out, 
\(
q_{\mu}^{\mathrm{W}}
=q_{\mu}^{\mathrm{FD}}
\) 
for \(\mu=2,\ldots,n\) (recall that 
\(
q_{\mu}h
=\imath k_{\mu-1-m}h
=2\pi\imath(\mu-1-m)/n
\)). 
On the other hand, for \(\mu=1\), 
\(
q_{\mu}^{\mathrm{W}}
\neq 
q_{\mu}^{\mathrm{FD}}
\) 
since \(q_{\mu=1}^{\mathrm{W}}\) is indeterminate 
(note \(q_{\mu=1}h=-\pi\imath\) for \(m=n/2\)). At 
least numerically, then, \(\mathrm{AFB}\) and \(\mathrm{R}\) 
are distinct FDDs. On the other hand, note that 
\(
\lim_{\mu\to 1}q_{\mu}^{\mathrm{W}}=q_{\mu=1}^{\mathrm{FD}}
\) 
does hold. Consequently, AFB- and R-based algorithms will be treated as 
as equivalent in what follows. 

Analogous to the 1D case, these results can be employed to 
formulate two types of FDDs. Again following \cite{Willot2015}, the first 
type is based on the direct generalization 
\begin{equation}
q_{\smash{\mu_{r}}}^{b_{r}}
=q_{\smash{\mu_{r}}}^{a_{r}\ast}
\label{equ:EffNumWavConWil3D}
\end{equation} 
of \eqref{equ:EffNumWavConWil1D} to 3D. In particular, note that 
\(
q_{\smash{\mu_{r}}}^{b_{r}}
=-q_{\smash{\mu_{r}}}^{\mathrm{ABD}}
\) 
for \(a_{r}=\mathrm{AFD}\) and 
\(
q_{\smash{\mu_{r}}}^{b_{r}}
=-q_{\smash{\mu_{r}}}^{a_{r}}
\) 
for \(a_{r}=\mathrm{ACD},\mathrm{AhC}\). The second type generalizes 
the choice \((a,b)=(\mathrm{FD},\mathrm{hC})\) in 1D to 
\((a_{r},b_{r})=(\mathrm{AFD},\mathrm{AhC})\) in 3D, and is referred to 
as the "average forward-backward/rotated" (AFB/R) FDD in what follows. 

\subsection{Algorithms based on piecewise-constant discretization} 
\label{sec:DisCenIntSub3D} 

Direct tensor-product-based generalization of \eqref{equ:DisFouPC1D} 
to 3D yields  
\begin{equation}
f_{\mathrm{PC}}(\mathbf{x})
=\varphi_{\smash{\mu_{1}}}^{\mathrm{c}}(x_{1})
\,\varphi_{\smash{\mu_{2}}}^{\mathrm{c}}(x_{2})
\,\varphi_{\smash{\mu_{3}}}^{\mathrm{c}}(x_{3})
\,\check{f}_{\smash{\mu_{1}\mu_{2}\mu_{3}}}^{\mathrm{c}}
=:\varphi_{\smash{\bm{\mu}}}^{\mathrm{c}}(\mathbf{x})
\star
\check{f}_{\smash{\bm{\mu}}}^{\mathrm{c}}
\label{equ:DisFouPC3D}
\end{equation}
again in terms of Rayleigh product notation with  
\begin{equation}
\begin{array}{rclclcl}
f_{\smash{\mathbf{i}}}^{\mathrm{c}}
&:=&
f_{\smash{i_{1}i_{2}i_{3}}}^{\mathrm{c}}
&=&
F_{\!\smash{\mu_{1}i_{1}}\,}^{\mathrm{c}+}
F_{\!\smash{\mu_{2}i_{2}}\,}^{\mathrm{c}+}
F_{\!\smash{\mu_{3}i_{3}}\,}^{\mathrm{c}+}
\,\check{f}_{\smash{\mu_{1}\mu_{2}\mu_{3}}}^{\mathrm{c}}
&=:&
F_{\!\smash{\bm{\mu}\mathbf{i}}}^{\mathrm{c}+}
\star
\check{f}_{\smash{\bm{\mu}}}^{\mathrm{c}}
\,,\\
\check{f}_{\smash{\bm{\mu}}}^{\mathrm{c}}
&:=&
\check{f}_{\smash{\mu_{1}\mu_{2}\mu_{3}}}^{\mathrm{c}}
&=&
F_{\!\smash{\mu_{1}i_{1}}\,}^{\mathrm{c}-}
F_{\!\smash{\mu_{2}i_{2}}\,}^{\mathrm{c}-}
F_{\!\smash{\mu_{3}i_{3}}\,}^{\mathrm{c}-}
\,f_{\smash{i_{1}i_{2}i_{3}}}^{\mathrm{c}}
&=:&
F_{\!\bm{\mu}\mathbf{i}\,}^{\mathrm{c}-}
\star 
f_{\smash{\mathbf{i}}}^{\mathrm{c}}
\,,
\end{array}
\label{equ:DisFouMatPC3D}
\end{equation} 
(again sum on repeated indices) via \eqref{equ:DisFouPC1D}${}_{3,4}$ 
and analogous to \eqref{equ:DisFouMat3D}. On this basis, 
\begin{algorithmP}[H]
\begin{enumerate}
\setlength{\itemsep}{-0.1mm}
\item given:~\(n_{1},\ldots,n_{3}\), 
\(\ssans{C}(\bm{x})\), 
\(\bar{\bm{E}}\), 
\(\ssans{C}_{\mathrm{H}}\), 
\(
\check{\mathbf{G}}_{\smash{\mathrm{H}\bm{\mu}}}^{\mathbf{ab}}
=\hat{\mathbf{G}}_{\mathrm{H}}
(\mathbf{q}_{\smash{\bm{\mu}}}^{\mathbf{a}},
\mathbf{q}_{\smash{\bm{\mu}}}^{\mathbf{b}})
\)
\item initialization \(\iota=0\)
\vspace{-3mm}
\begin{itemize}
\setlength{\itemsep}{-0.5mm}
\item for \(i_{1}=1,\ldots,n_{1}\), \(\ldots\), \(i_{3}=1,\ldots,n_{3}\) 
\newline 
\(
{}\ 
\textsf{C}_{\smash{\mathbf{i}}}^{\mathrm{c}}
=\textsf{C}(\mathbf{x}_{\smash{\mathbf{i}}}^{\mathrm{c}})
\);
\(
{}\ 
\mathbf{T}_{\!\smash{\mathbf{i}}}^{\mathrm{c}(\iota)}
=\textsf{C}_{\smash{\mathbf{i}}}^{\mathrm{c}}\bar{\mathbf{E}}
\);  
\item for \(\mu_{1}=1,\ldots,n_{1}\), \(\ldots\), \(\mu_{3}=1,\ldots,n_{3}\)
\newline
\(
{}\ 
\check{\mathbf{u}}_{\smash{\bm{\mu}}}^{\mathrm{pc}(\iota)}
=\bm{0}
\); 
\(
{}\ 
\check{\mathbf{T}}_{\!\smash{\bm{\mu}}}^{\mathrm{c}(\iota)}
=F_{\!\smash{\bm{\mu}\mathbf{i}}}^{\mathrm{c}-}
\star
\mathbf{T}_{\!\smash{\mathbf{i}}}^{\mathrm{c}(\iota)}
\); 
\(
{}\ 
\Delta\check{\mathbf{u}}_{\smash{\bm{\mu}}}^{\mathrm{pc}(\iota)}
=\check{\mathbf{G}}_{\smash{\mathrm{H}\bm{\mu}}}^{\mathbf{ab}}
\check{\mathbf{T}}_{\!\smash{\bm{\mu}}}^{\mathrm{c}(\iota)}
\mathbf{q}_{\smash{\bm{\mu}}}^{\mathbf{b}}
\); 
\item \(\iota\!+\!\!+\);
\end{itemize}
\vspace{-2mm}
\item while
\(
\sum_{\mu_{1}=1}^{n_{1}}
\cdots
\sum_{\mu_{3}=1}^{n_{3}}
|\Delta\check{\mathbf{u}}_{\smash{\bm{\mu}}}^{\mathrm{pc}(\iota)}
-\Delta\check{\mathbf{u}}_{\smash{\bm{\mu}}}^{\mathrm{pc}(\iota-1)}|
\geqslant\mathrm{tol}
\) 
\& \(\iota\leqslant\mathrm{maxit}\) 
\vspace{-2mm}
\begin{itemize}
\setlength{\itemsep}{-0.5mm}
\item for \(i_{1}=1,\ldots,n_{1}\), \(\ldots\), \(i_{3}=1,\ldots,n_{3}\) 
\newline 
\(
{}\ 
\tilde{\mathbf{E}}_{\smash{\mathbf{i}}}^{\mathrm{c}(\iota)}
=F_{\!\smash{\bm{\mu}\mathbf{i}}}^{\mathrm{c}+}
\star
\mathop{\mathrm{sym}}
(\check{\mathbf{u}}_{\smash{\bm{\mu}}}^{\mathrm{pc}(\iota)}
\otimes
\mathbf{q}_{\smash{\bm{\mu}}}^{\mathbf{a}})
\);
\(
{}\ 
\mathbf{T}_{\!\smash{\mathbf{i}}}^{\mathrm{c}(\iota)}
=\mathbf{T}_{\!\smash{\mathbf{i}}}^{\mathrm{c}(0)}
+\textsf{C}_{\smash{\mathbf{i}}}^{\mathrm{c}}
\tilde{\mathbf{E}}_{\smash{\mathbf{i}}}^{\mathrm{c}(\iota)}
\);
\item for \(\mu_{1}=1,\ldots,n_{1}\), \(\ldots\), \(\mu_{3}=1,\ldots,n_{3}\)
\newline
\(
{}\ 
\check{\mathbf{T}}_{\!\smash{\bm{\mu}}}^{\mathrm{c}(\iota)}
=F_{\!\smash{\bm{\mu}\mathbf{i}}}^{\mathrm{c}-}
\star
\mathbf{T}_{\!\smash{\mathbf{i}}}^{\mathrm{c}(\iota)}
\); 
\(
{}\ 
\Delta\check{\mathbf{u}}_{\smash{\bm{\mu}}}^{\mathrm{pc}(\iota)}
=\check{\mathbf{G}}_{\smash{\mathrm{H}\bm{\mu}}}^{\mathbf{ab}}
\check{\mathbf{T}}_{\!\smash{\bm{\mu}}}^{\mathrm{c}(\iota)}
\mathbf{q}_{\smash{\bm{\mu}}}^{\mathbf{b}}
\); 
\(
{}\ 
\check{\mathbf{u}}_{\bm{\mu}}^{\smash{\mathrm{pc}}(\iota)}
+\!=
\Delta\check{\mathbf{u}}_{\bm{\mu}}^{\smash{\mathrm{pc}}(\iota)}
\);
\item \(\iota\!+\!\!+\);
\end{itemize}
\end{enumerate}
\vspace{-3mm}
\caption{}
\label{alg:AlgDisIntFouDifFinFunGrePC3D}
\end{algorithmP} 
\begin{algorithmD}[H]
\begin{enumerate}
\setlength{\itemsep}{-0.1mm}
\item given:~\(n_{1},\ldots,n_{3}\), 
\(\ssans{C}(\bm{x})\), 
\(\bar{\bm{E}}\), 
\(\ssans{C}_{\mathrm{H}}\),
\(
\check{\mathbf{\Gamma}}
_{\!\smash{\mathrm{H}\,\bm{0}}}^{\mathrm{c}}
=\bm{0}
\),
\newline
\(
\check{\mathbf{\Gamma}}
_{\!\smash{\mathrm{H}\,\bm{\omega}\neq\bm{0}}}^{\mathrm{c}}
=\sum_{\nu_{1}=-m_{1}}^{m_{1}-1}
s_{\smash{\nu_{1}n_{1}+\omega_{1}}}
\cdots
\sum_{\nu_{3}=-m_{3}}^{m_{3}-1}
s_{\smash{\nu_{3}n_{3}+\omega_{3}}}
\ \hat{\mathbf{\Gamma}}_{\!\mathrm{H}}
(k_{\nu_{1}n_{1}+\omega_{1}},
k_{\nu_{2}n_{2}+\omega_{2}},
k_{\nu_{3}n_{3}+\omega_{3}})
\)
\item initialization \(\iota=0\) 
\vspace{-3mm}
\begin{itemize}
\setlength{\itemsep}{-0.5mm}
\item for \(i_{1}=1,\ldots,n_{1}\), \(\ldots\), \(i_{3}=1,\ldots,n_{3}\)
\newline
\(
{}\ 
\textsf{C}_{\smash{\mathbf{i}}}^{\mathrm{c}}
=\textsf{C}(\mathbf{x}_{\smash{\mathbf{i}}}^{\mathrm{c}})
\); 
\(
{}\ 
\mathbf{T}_{\!\smash{\mathbf{i}}}^{\mathrm{c}(\iota)}
=\textsf{C}_{\smash{\mathbf{i}}}^{\mathrm{c}}\bar{\mathbf{E}}
\);
\item for \(\omega_{1}=0,\ldots,n_{1}-1\), \(\ldots\), 
\(\omega_{3}=0,\ldots,n_{3}-1\)
\newline
\(
{}\ 
\check{\mathbf{E}}_{\smash{\bm{\omega}}}^{\mathrm{c}(\iota)}=\bm{0}
\); 
\(
{}\ 
\check{\mathbf{T}}_{\!\smash{\bm{\omega}}}^{\mathrm{c}(\iota)}
=F_{\!\smash{\bm{\omega}\mathbf{i}}\,}^{\mathrm{c}-}
\star
\mathbf{T}_{\!\smash{\mathbf{i}}}^{\mathrm{c}(\iota)}
\);
\(
{}\ 
\Delta\check{\mathbf{E}}_{\smash{\bm{\omega}}}^{\mathrm{c}(\iota)}
=\check{\mathbf{\Gamma}}_{\!\smash{\mathrm{H}\bm{\omega}}}^{\mathrm{c}}
\check{\mathbf{T}}_{\!\smash{\bm{\omega}}}^{\mathrm{c}(\iota)}
\);
\item \(\iota\!+\!\!+\);
\end{itemize}
\vspace{-2mm}
\item while
\(
\sum_{\omega_{1}=0}^{n_{1}-1}
\cdots
\sum_{\omega_{3}=0}^{n_{3}-1}
|\Delta\check{\mathbf{E}}_{\smash{\bm{\omega}}}^{\mathrm{c}(\iota)}
-\Delta\check{\mathbf{E}}_{\smash{\bm{\omega}}}^{\mathrm{c}(\iota-1)}|
\geqslant\mathrm{tol}
\) 
\& \(\iota\leqslant\mathrm{maxit}\) 
\vspace{-2mm}
\begin{itemize}
\setlength{\itemsep}{-0.5mm}
\item for \(i_{1}=1,\ldots,n_{1}\), \(\ldots\), \(i_{3}=1,\ldots,n_{3}\)
\newline
\(
{}\ \ 
\tilde{\mathbf{E}}_{\smash{\mathbf{i}}}^{\mathrm{c}(\iota)}
=F_{\!\smash{\bm{\omega}\mathbf{i}}}^{\mathrm{c}+}
\star
\skew4\check{\mathbf{E}}_{\smash{\bm{\omega}}}^{\mathrm{c}(\iota)}
\);
\(
{}\ 
\mathbf{T}_{\!\smash{\mathbf{i}}}^{\mathrm{c}(\iota)}
=\mathbf{T}_{\!\smash{\mathbf{i}}}^{\mathrm{c}(0)}
+\textsf{C}_{\smash{\mathbf{i}}}^{\mathrm{c}}
\tilde{\mathbf{E}}_{\smash{\mathbf{i}}}^{\mathrm{c}(\iota)}
\);
\item for \(\omega_{1}=0,\ldots,n_{1}-1\), 
\(\ldots\),
\(\omega_{3}=0,\ldots,n_{3}-1\) 
\newline
\(
{}\ \ 
\check{\mathbf{T}}_{\!\smash{\bm{\omega}}}^{\mathrm{c}(\iota)}
=F_{\!\smash{\bm{\omega}\mathbf{i}}\,}^{\mathrm{c}-}
\star
\mathbf{T}_{\!\smash{\mathbf{i}}}^{\mathrm{c}(\iota)}
\);
\(
{}\ 
\Delta\check{\mathbf{E}}_{\smash{\bm{\omega}}}^{\mathrm{c}(\iota)}
=\check{\mathbf{\Gamma}}_{\!\smash{\mathrm{H}\bm{\omega}}}^{\mathrm{c}}
\check{\mathbf{T}}_{\!\smash{\bm{\omega}}}^{\mathrm{c}(\iota)}
\);
\(
{}\ 
\check{\mathbf{E}}_{\smash{\bm{\omega}}}^{\mathrm{c}(\iota)}
+\!=
\Delta\check{\mathbf{E}}_{\smash{\bm{\omega}}}^{\mathrm{c}(\iota)}
\);
\item \(\iota\!+\!\!+\);
\end{itemize}
\vspace{-1mm}
\end{enumerate}
\vspace{-3mm}
\caption{}
\label{alg:AlgDisConPie3D}
\end{algorithmD}
\vspace{-3mm}
are obtained in 3D via direct componentwise generalization of Algorithm 
\ref{alg:AlgDisIntFouDifFinFunGrePC1D} and Algorithm \ref{alg:AlgDisConPie1D}, 
respectively. In particular, both algorithms are based on the Cartesian components 
of the Fourier transform 
\begin{equation}
\hat{\bm{G}}_{\mathrm{H}}(\bm{k})
=\frac{1}{\mu_{\mathrm{H}}|\bm{k}|^{2}}
\left\lbrack
\bm{I}-\frac{1+\lambda_{\mathrm{H}}/\mu_{\mathrm{H}}}
{1+(1+\lambda_{\mathrm{H}}/\mu_{\mathrm{H}})}
\,\frac{\bm{k}}{|\bm{k}|}
\otimes
\frac{\bm{k}}{|\bm{k}|}
\right\rbrack
\,,\quad
\bm{k}\neq\bm{0}
\,,
\end{equation} 
of the isotropic Green tensor. 
In addition, Algorithm \ref{alg:AlgDisConPie3D} utilizes the Cartesian components 
of the Fourier transform \(\hat{\bm{\Gamma}}_{\!\mathrm{H}}\) of the 
3D Lippmann-Schwinger operator \(\bm{\Gamma}_{\!\mathrm{H}}\), defined by 
\(
\hat{\bm{\Gamma}}_{\!\mathrm{H}}(\bm{k})\,\bm{A}
:=-\mathop{\mathrm{sym}}
\lbrack
\hat{\bm{G}}_{\mathrm{H}}(\bm{k})
\,\bm{A}
(\bm{k}\otimes\bm{k})
\rbrack
\) 
for all \(\bm{A}\). 

\section{Computational comparisons in 3D}
\label{sec:Res3D}

In this section, the algorithms formulated in the last section for solution of 
the strong BVP are compared in the 3D matrix-inclusion (MI) case. 
As in the 1D case above, both sharp and smooth MI interfaces are considered. 
More specifically, in the context of TD and 
Algorithm \ref{alg:AlgDisIntFouDifFinFunGre3D}, results are compared 
for the choices \(a_{r}=\mathrm{F},\mathrm{CD},\mathrm{ACD}\) 
in the context of \eqref{equ:EffNumWavConWil3D}, as well as for 
\((a_{r},b_{r})=(\mathrm{AFD},\mathrm{AhC})\) (i.e., AFB/R). 
As well, in the context of PCD, Algorithm \ref{alg:AlgDisIntFouDifFinFunGrePC3D} 
for \((a_{r},b_{r})=(\mathrm{AFD},\mathrm{AhC})\) is compared with 
Algorithm \ref{alg:AlgDisConPie3D}. 
For reference, the results from these in the context of the strong 
BVP are compared with analogous results from the numerical solution of 
the corresponding weak BVP via standard finite element (SFE) discretization. 
In this latter case, note that the sharp MI interface is discretized by element 
boundaries. 

For comparability with the literature, the algorithmic comparisons just 
discussed are carried out in the sequel employing the MI benchmark 
example from \cite{Willot2015}. In particular, this is based on the choices 
\(\lambda_{\mathrm{M}}=\mu_{\mathrm{M}}=0.6\), 
\(\lambda_{\mathrm{H}}=\lambda_{\mathrm{I}}\), and 
\(\mu_{\mathrm{H}}=\mu_{\mathrm{I}}\). 
The 3D computational domain \(\Omega\) of \cite{Willot2015} 
is discretized uniformly based on \(L^{3}\) nodes and unit grid 
spacing \(h=1\). Since his \(L\) then corresponds to \(n_{r}\) 
here, \(n=n_{1}=n_{2}=n_{3}\) and \(h=h_{1}=h_{2}=h_{3}=1\) 
hold in what follows. 
Results are presented for the non-dimensional stress field 
\(\bm{T}/\mu_{\mathrm{M}}\) in what follows for different phase 
contrasts 
\begin{equation}
\chi
=\lambda_{\mathrm{I}}/\lambda_{\mathrm{M}}
=\mu_{\mathrm{I}}/\mu_{\mathrm{M}}
\,.
\label{equ:ConStiPha}
\end{equation}
In terms of \(\chi\), note that 
\(
\lambda(\bm{x})/\mu_{\mathrm{M}}
=(\lambda_{\mathrm{M}}/\mu_{\mathrm{M}})
\,f(\bm{x},\chi)
\) 
and 
\(
\mu(\bm{x})/\mu_{\mathrm{M}}=f(\bm{x},\chi)
\) 
hold, with 
\(
f(\bm{x},\chi)
:=1+\nu_{\!\epsilon}(x_{1})
\,\nu_{\!\epsilon}(x_{2})
\,\nu_{\!\epsilon}(x_{3})
\,(\chi-1)
\). 
The following results are based on control 
of \(\bar{\bm{E}}\), with \(\bar{E}_{xy}=1\), and all other components zero. 

\subsection{Results based on trapezoidal discretization}
\label{sec:3DResTra}

\begin{figure}[H]
\centering
\includegraphics[width=0.8\textwidth]{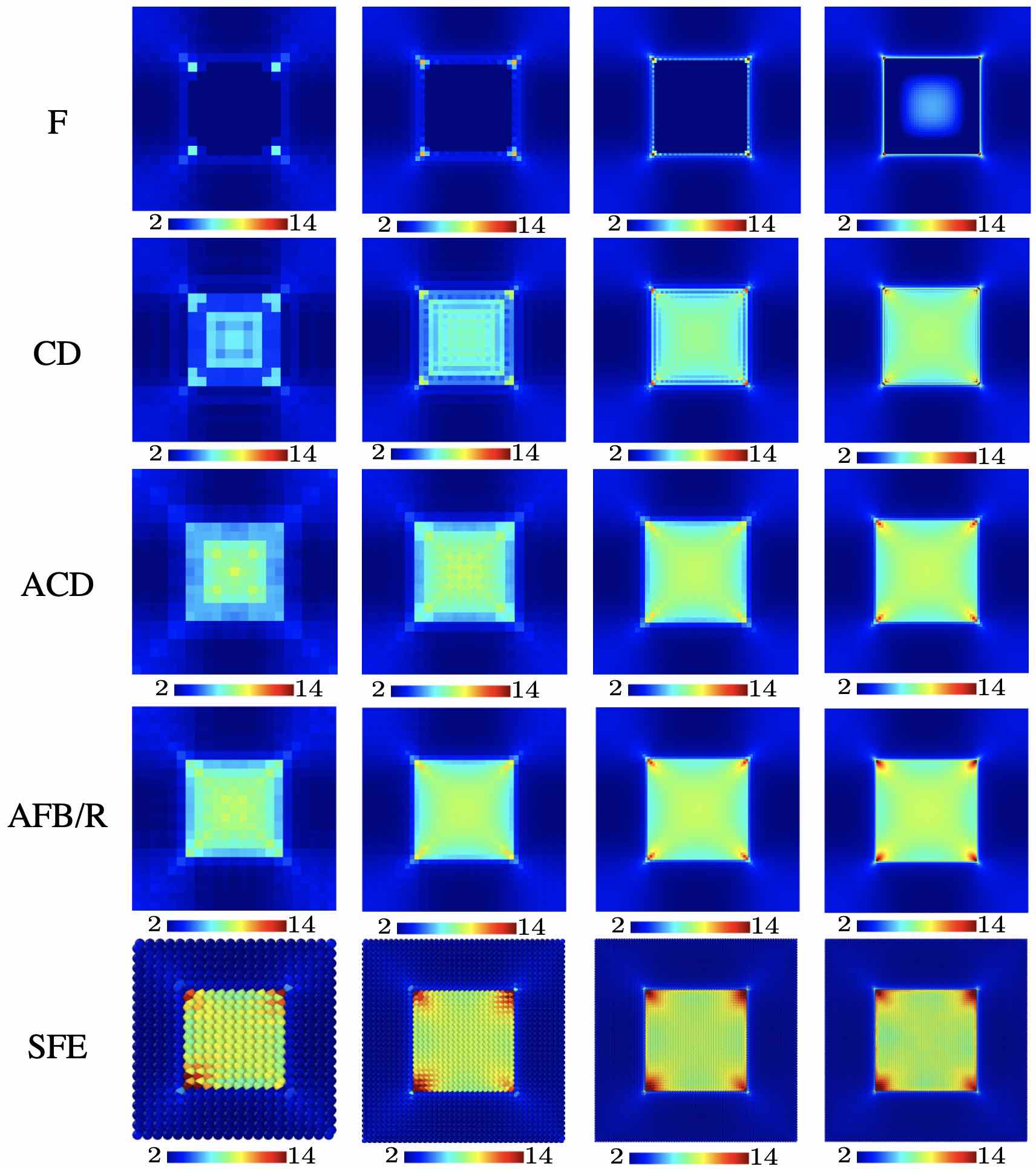}
\caption{Comparison of discrete results for 
\(
T_{\!\smash{xy}}
(x_{i_{1}},x_{i_{2}},z_{\mathrm{I}})/\mu_{\mathrm{M}}
\) 
for a sharp MI interface with \(\chi=1000\) for different choices 
of \(a_{r},b_{r}\) in Algorithm \ref{alg:AlgDisIntFouDifFinFunGre3D} 
and at different numerical resolutions \(n\). In particular, \(n=21\) (left), 
\(n=41\) (middle left),
\(n=81\) (middle right), 
and \(n=161\) (right). 
The same solutions are obtained 
for the even choices \(n=22\) (left), 
\(n=42\) (middle left),
\(n=82\) (middle right), 
and \(n=162\) (right). 
The discrete results shown are from the grid points in the \((x_{1},x_{2})\) 
plane at \(x_{i_{3}}=z_{\mathrm{I}}\) inside the inclusion and 
adjacent to the MI interface.  
\newline 
\color{magenta}
\color{black}
} 
\label{fig:FieStrIntSha}
\end{figure}

\begin{figure}[H]
\centering
\includegraphics[width=0.8\textwidth]{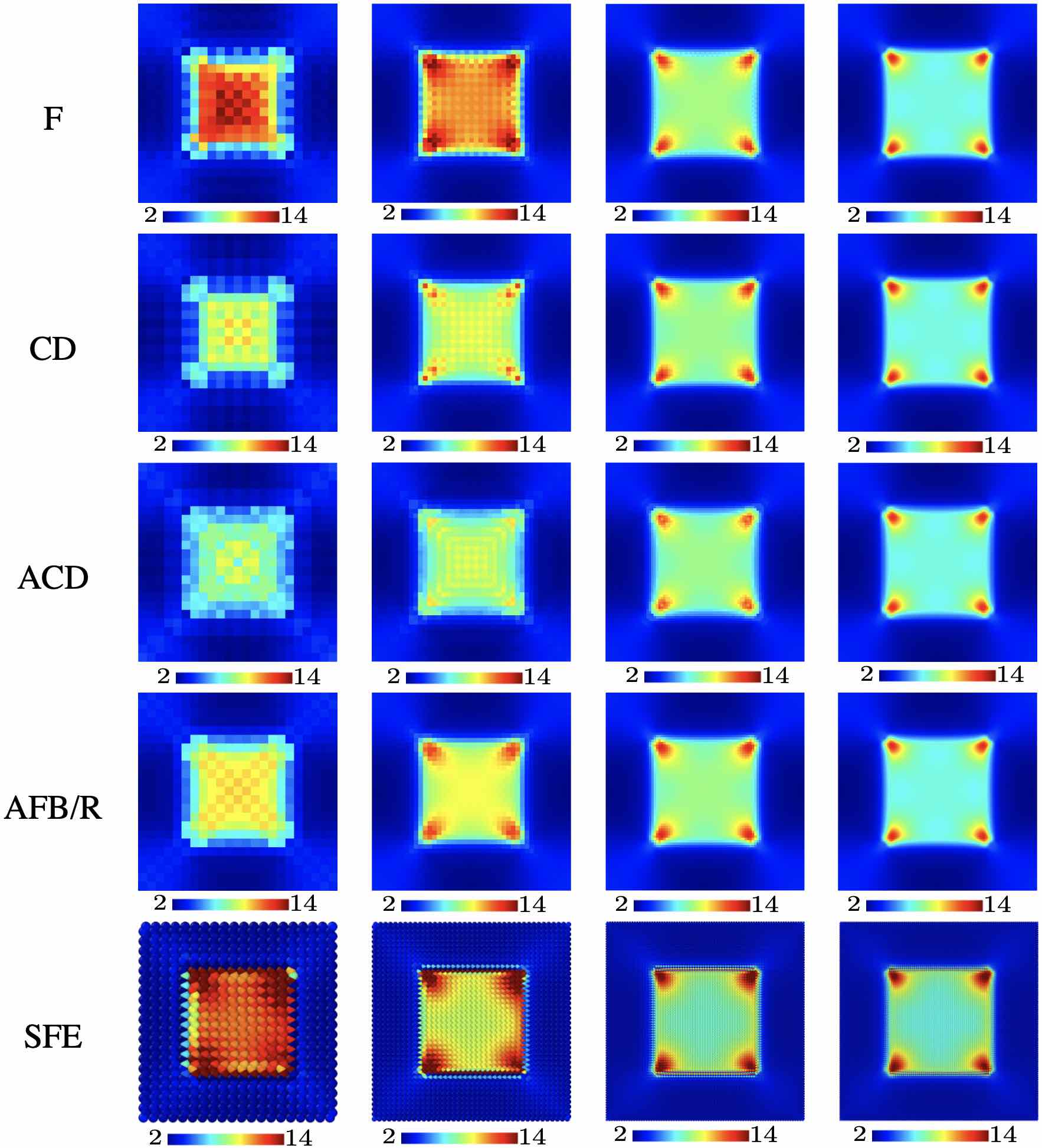}
\caption{Same as Figure \ref{fig:FieStrIntSha} for a smooth MI interface. Here, 
\(
T_{\!\smash{xy}}
(x_{i_{1}},x_{i_{2}},z_{\mathrm{MI}})/\mu_{\mathrm{M}}
\) 
is shown from the grid points in the \((x_{1},x_{2})\) plane on the MI 
interface at \(x_{i_{3}}=z_{\mathrm{MI}}\).  
As in the 1D case and Figure \ref{fig:StnShaSmoStr1D}, the smooth transition in 
stiffness between matrix and inclusion is based on \(\epsilon/l=1/100\) in 
each direction. 
} 
\label{fig:FieStrIntSmo}
\end{figure}

Results for the stress field at a sharp MI interface with a phase contrast 
of \(\chi=1000\) shown in Figure \ref{fig:FieStrIntSha}. 
The analogous results for \(T_{\!xy}\) at a smooth MI interface 
are shown in Figure \ref{fig:FieStrIntSmo}. 
\vspace{-3mm}
As seen in Figure \ref{fig:FieStrIntSha} (top row), for the discontinuous MI 
interface case in the context of \eqref{equ:EffNumWavConWil3D}, 
the choice \(a_{r}=\mathrm{F}\)  
yields qualitatively incorrect results at the resolutions considered. Indeed, 
as shown by \citet[][Figure 4]{Willot2015}, who worked with much finer 
discretizations of \(L=256, 512, 1024\), this choice yields correct 
results for this benchmark case only at much higher resolution. In contrast, the 
central difference choice \(a_{r}=\mathrm{CD}\) 
(Figure \ref{fig:FieStrIntSha}, second row from the top) yields a stress field which 
is qualitatively correct and nearly converges at the highest resolution investigated 
here. Among the choices based on conjugacy \eqref{equ:EffNumWavConWil3D}, 
\(a_{r}=\mathrm{ACD}\) 
(Figure \ref{fig:FieStrIntSha}, middle row) 
converges almost as quickly as the non-conjugate 
AFB/R choice \((a_{r},b_{r})=(\mathrm{AFD},\mathrm{AhC})\) 
(Figure \ref{fig:FieStrIntSha}, second row from the bottom), which 
is closest to the SFE-based reference results 
(Figure \ref{fig:FieStrIntSha}, bottom row). 

Analogous to the 1D case (Figure \ref{fig:StnShaSmoStr1D}), and as expected, 
a quite different picture emerges for the convergence behavior and stress field 
at a smooth MI interface in 3D. Indeed, as for the 1D results in 
Figure \ref{fig:StnShaSmoStr1D} (right), the convergence behavior displayed 
in Figure \ref{fig:FieStrIntSmo} in the 3D case for all choices of \((a_{r},b_{r})\) 
is affected predominantly by numerical resolution. In particular, 
as in the discontinuous MI interface case, 
among the choices based on conjugacy \eqref{equ:EffNumWavConWil3D}, 
\(a_{r}=\mathrm{ACD}\) converges almost as quickly as the non-conjugate 
AFB/R choice \((a_{r},b_{r})=(\mathrm{AFD},\mathrm{AhC})\), which again 
is closest to the SFE-based reference case (bottom row).

\begin{figure}[H]
\centering
\includegraphics[width=0.45\textwidth]{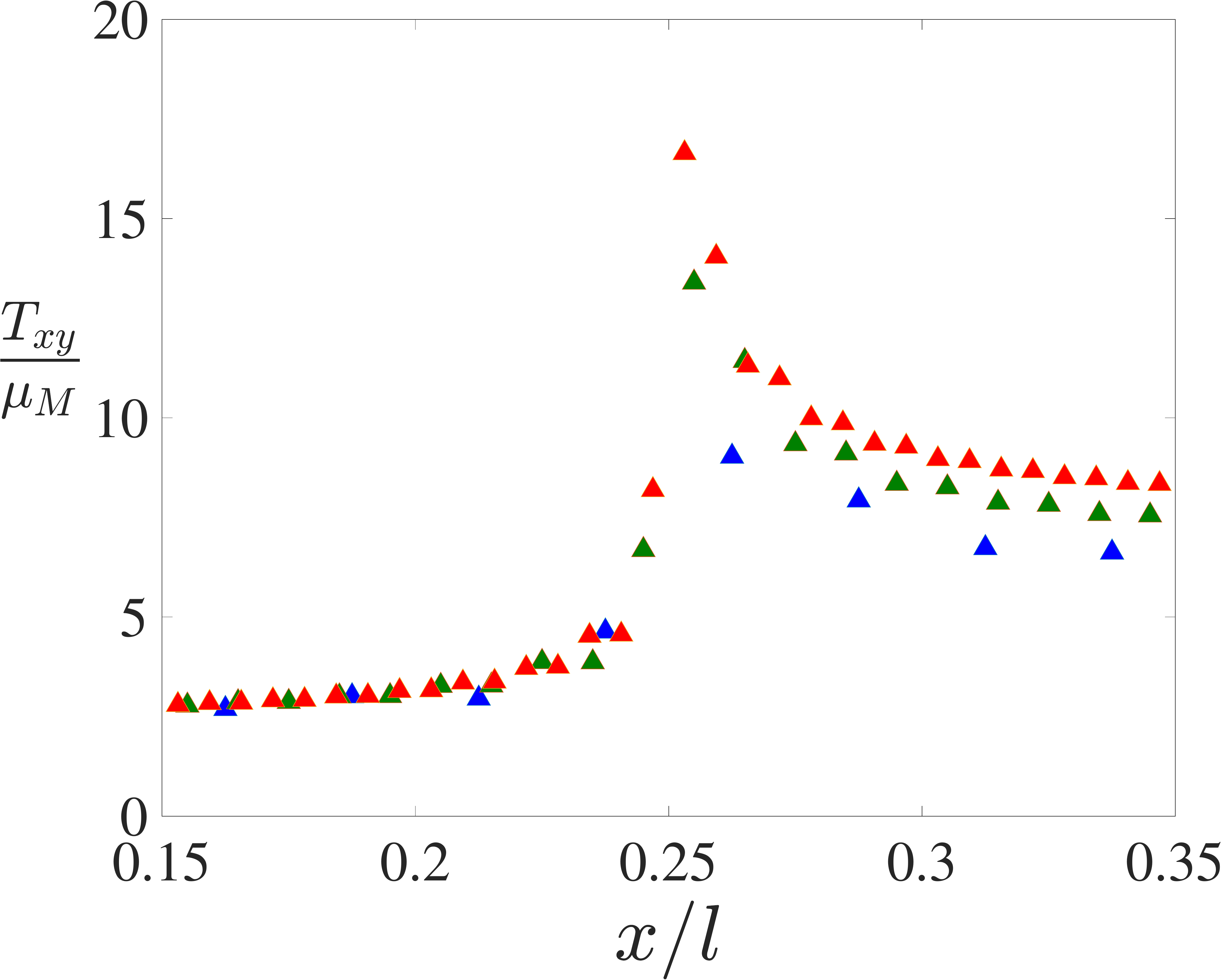}
\hspace{2mm}
\includegraphics[width=0.45\textwidth]{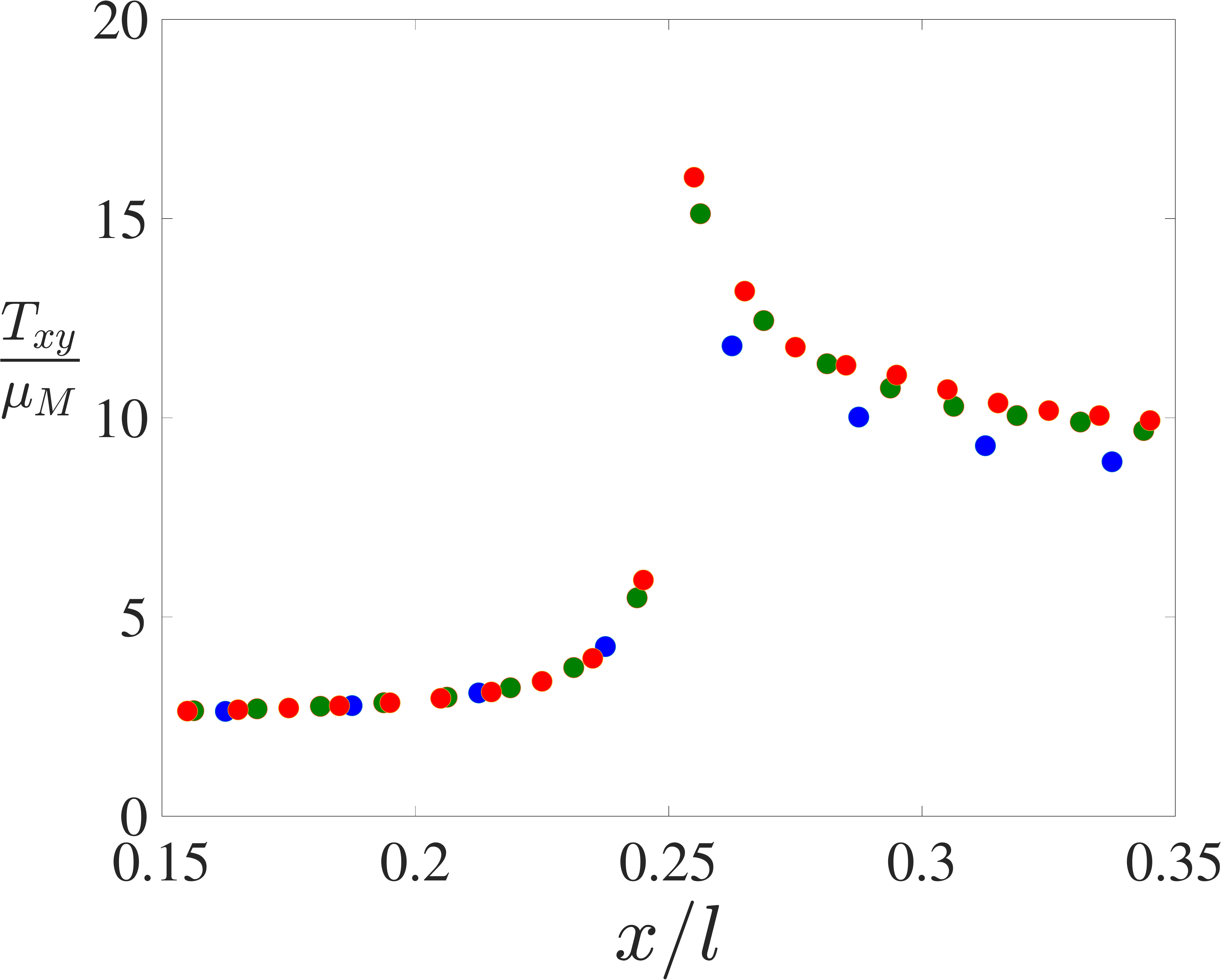}
\caption{Discrete AFB/R (left, triangles) and SFE (right, circles) results for 
\(T_{\!xy}(x_{i_{1}},y_{\mathrm{I}},z_{\mathrm{I}})\) 
across a MI corner for the sharp (left) MI interface. 
Results shown are at grid points in the \(x_{1}\) direction for fixed 
\(x_{i_{2}}=y_{\mathrm{I}},x_{i_{3}}=z_{\mathrm{I}}\) just inside 
the inclusion and for the numerical resolutions of 
\(n_{r}=42\) (blue), \(n_{r}=82\) (green), \(n_{r}=162\) (red). 
} 
\label{fig:ConSpaSha3D}
\end{figure}

\begin{figure}[H]
\centering
\includegraphics[width=0.45\textwidth]{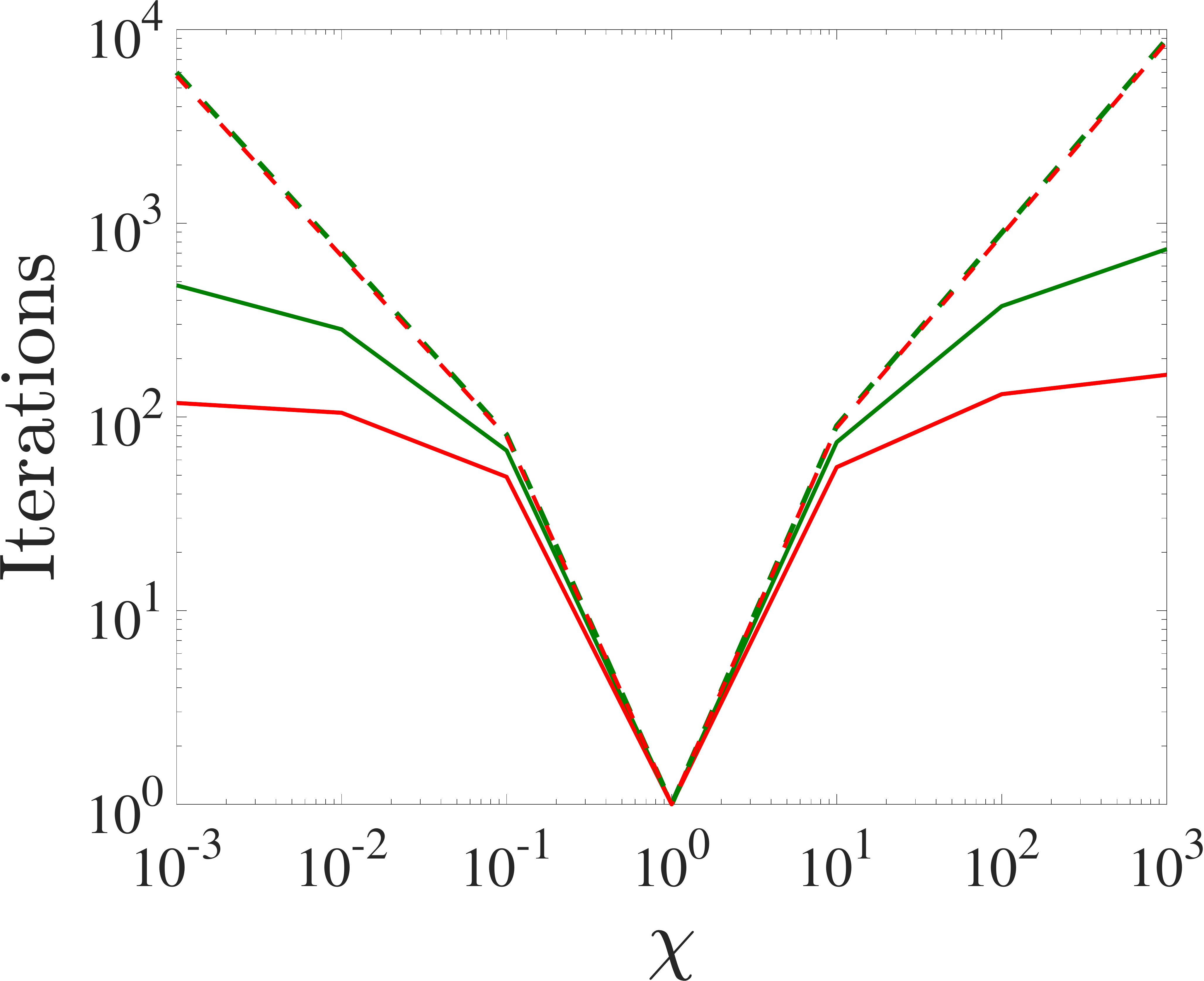}
\caption{Convergence behavior of selected algorithms as a function of phase 
contrast at sharp (solid curves) and smooth (dashed curves) MI interfaces 
for \(n_{r}=161,162\). 
Red:~AFB/R. Green:~\(a_{r}=\mathrm{ACD}\). 
No convergence results for \(a_{r}=\mathrm{F}\) are shown because 
this case does not converge in less than \(10^{4}\) iterations at the highest 
contrasts. 
} 
\label{fig:Conv3D}
\end{figure}

As evident in the results from Figures \ref{fig:FieStrIntSha} and 
\ref{fig:FieStrIntSmo}, the stress field at the MI corners is most sensitive 
to numerical resolution. To look at this in more detail, consider the results 
in Figure \ref{fig:ConSpaSha3D}. 
\vspace{-3mm}
As expected, in the sharp interface case, the solution does not converge 
with increasing resolution, i.e., is mesh-dependent, in contrast to the smooth 
interface case. 
The convergence behavior of the ACD (\(a_{r}=\mathrm{ACD}\)) 
algorithm based on conjugacy \eqref{equ:EffNumWavConWil3D}, 
as well the AFB/R (\((a_{r},b_{r})=(\mathrm{AFD},\mathrm{AhC})\)), 
are compared for both the sharp and smooth MI interfaces are compared 
in Figure \ref{fig:Conv3D}. 
\vspace{-3mm}
As shown, for phase contrasts up to a factor of 10 
(i.e., \(\chi=10^{-1},10^{1}\)), little 
difference in convergence rate is apparent. For higher contrasts, however, 
the effect of oscillations due to Gibbs and aliasing on the convergence rate 
for the sharp interface cases becomes apparent. Indeed, 
as expected from the stress results in Figure \ref{fig:FieStrIntSha}, 
the effective low-pass filtering effect of finite-difference discretization of 
differential operators for \(a_{r}=\mathrm{ACD}\) and AFB/R reduces the 
effect of oscillations due to Gibbs and aliasing on convergence rate. In particular, 
for high phase contrasts, the convergence rate for AFB/R at \(n_{r}=161,162\) 
is about \(10^{2}\) times faster than for \(a_{r}=\mathrm{F}\), 
and about 5 times faster than for \(a_{r}=\mathrm{ACD}\), for the current 
benchmark case. 

\subsection{Results based on piecewise-constant discretization}
\label{sec:3DResPC}

\begin{figure}[H]
\centering
\includegraphics[width=0.9\textwidth]{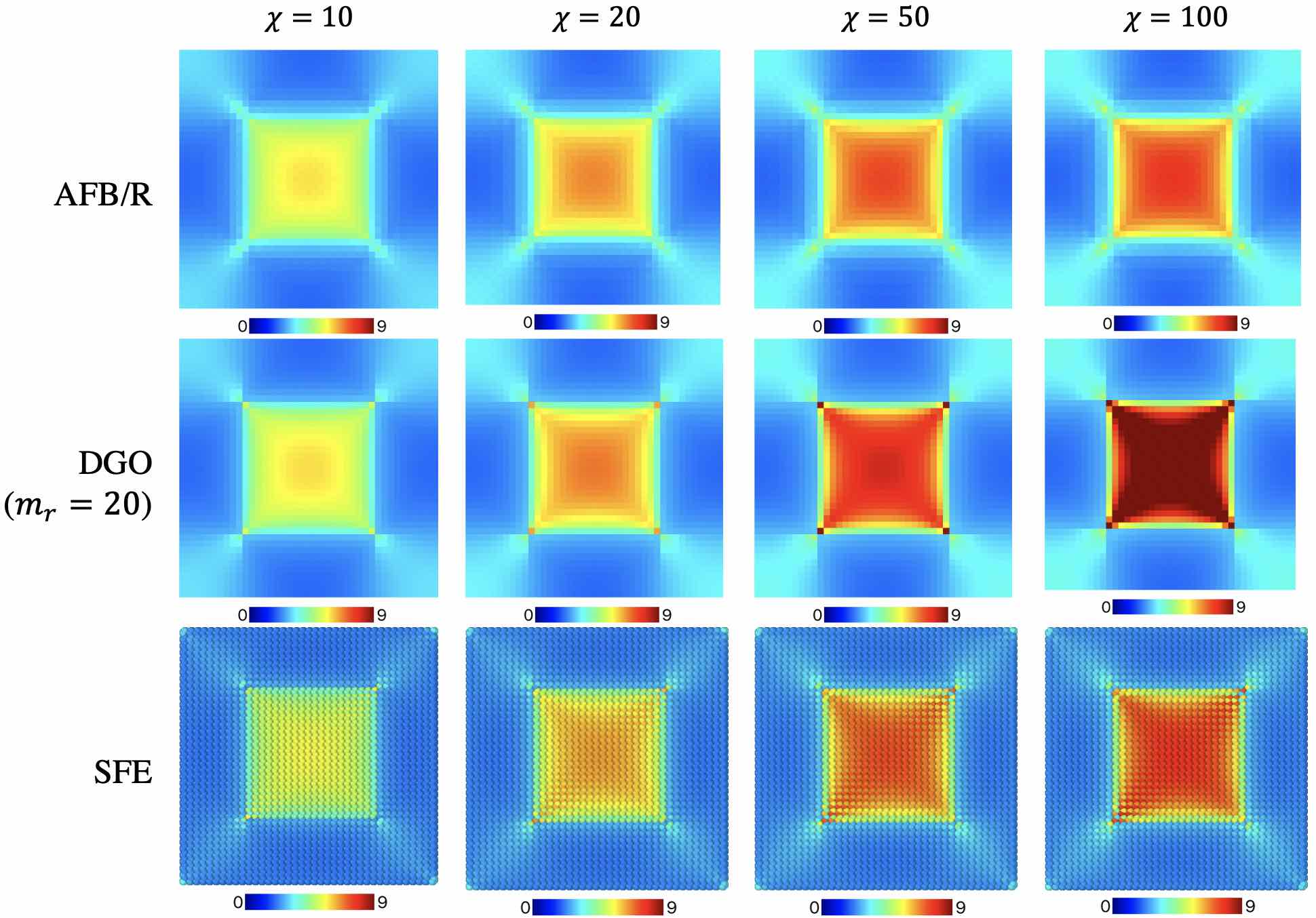}
\caption{PCD-based results from AFB/R (above), the DGO (middle) for 
\(
T_{\!\smash{xy}}^{\mathrm{c}}
(x_{\smash{i_{1}}}^{\mathrm{c}},
x_{\smash{i_{2}}}^{\mathrm{c}},
z_{\mathrm{I}}^{\mathrm{c}})
/\mu_{\mathrm{M}}
\) 
and SFE (below), 
in the sharp MI interface case for different phase contrasts 
\(\chi\) at a resolution of \(n_{r}=40\). 
The discrete results shown are from the grid points in the inclusion adjacent 
(\(x_{\smash{i_{3}}}^{\mathrm{c}}=z_{\mathrm{I}}^{\mathrm{c}}\)) 
to the MI interface. See text for discussion. 
} 
\label{fig:FieStrIntShaDGO}
\end{figure}

Lastly, results from Algorithm \ref{alg:AlgDisIntFouDifFinFunGrePC3D} for 
\((a_{r},b_{r})=(\mathrm{AFD},\mathrm{AhC})\) (i.e., AFB/R) are compared 
with corresponding ones from Algorithm \ref{alg:AlgDisConPie3D} in Figure 
\ref{fig:FieStrIntShaDGO}. 
Except for the smaller phase contrasts and consequently lower stress 
levels, the PCD-based results for AFB/R in Figure \ref{fig:FieStrIntShaDGO} 
(upper row) 
agree with the TD-based results in Figure \ref{fig:FieStrIntSha} 
(middle left row). 
For the chosen nodal resolution of \(n_{r}=40\), recall that 
\(m_{r}=\frac{1}{2}n_{r}=20\) in AFB/R. For comparability, then, 
\(m_{r}=20\) was chosen to obtain the DGO-based results. 
Except for the much higher stress at the MI corners obtained with 
the DGO, the results from both algorithms agree (even quantitatively) 
well for \(\chi=10\). With increasing \(\chi\), however, the solution 
based on the DGO is much stiffer than than based on AFB/R, indicating 
lack of convergence at this resolution. Consequently, the DGO converges 
more slowly than AFB/R in the context of PCD. 

As mentioned in the introduction, in contrast to the current case of cubic 
inclusions, \citet[][\S\S 4.3-4.4]{Eloh2019} consider spherical inclusions 
for which analytic solutions exist \citep[e.g.,][]{Mur87}. In addition, following 
\cite{Anglin2014}, they work with phase contrasts \(\chi\) of \(0.5\) and \(2\) 
in the bulk modulus \(\kappa:=(\lambda+2\mu/3)\). 
Recalling that \(\lambda_{\mathrm{I}}=\chi\lambda_{\mathrm{M}}\) and 
\(\mu_{\mathrm{I}}=\chi\mu_{\mathrm{M}}\) from \eqref{equ:ConStiPha}, 
note that \(\kappa_{\mathrm{I}}=\chi\,\kappa_{\mathrm{M}}\) also holds. 
Consequently, they worked with smaller phase contrasts than \(\chi=10\) 
upon which the results in Figure \ref{fig:FieStrIntShaDGO} are based. 
Further, \cite{Eloh2019} assume 
\(
\kappa_{\mathrm{H}}
=\frac{1}{2}
(\kappa_{\mathrm{M}}+\kappa_{\mathrm{I}})
\), 
\(
\nu_{\mathrm{H}}
=\nu_{\mathrm{M}}
=\nu_{\mathrm{I}}
\), 
and \(\lambda_{\mathrm{M}}/\mu_{\mathrm{M}}=5.2\), 
the latter in contrast to \(\lambda_{\mathrm{M}}/\mu_{\mathrm{M}}=1\) 
assumed by \cite{Willot2015} and here. For isotropic elastic behavior, 
this plays a role in normal strain control cases, but not in the current 
case of shear strain control.  



\section{Summary and discussion} 
\label{sec:DisSum}

In the current work, a number of algorithms have been developed and 
compared for the numerical solution of periodic boundary-value problems 
(BVPs) for the quasi-static mechanical behavior of heterogeneous linear 
elastic materials based on Fourier series discretization. 
In particular, two such discretizations are considered and compared here. 
The first is based on trapezoidal approximation of the integral  
\eqref{equ:RepFieSerFor1D}${}_{2}$ over the unit cell determining the 
Fourier modes \(\hat{f}_{\kappa}\) in the truncated Fourier 
series \eqref{equ:RepFieSerFor1D}${}_{1}$, resulting in the trapezoidal 
discretization \eqref{equ:FieFouAppTra1D} of this series. The second is 
based on approximation of the integrand in \eqref{equ:RepFieSerFor1D}${}_{2}$ 
as piecewise-constant, yielding the piecewise-constant discretization 
\eqref{equ:DisFouPC1D} of \eqref{equ:RepFieSerFor1D}${}_{1}$. 
A comparison of these two implies that a basic difference between 
them lies in the dependence of \eqref{equ:DisFouPC1D} on 
\(s_{\kappa}=\mathop{\mathrm{sinc}}(\pi\kappa/n)\). 
As shown by Algorithms \ref{alg:AlgDisIntFouDifFinFunGre1D} and 
\ref{alg:AlgDisIntFouDifFinFunGrePC1D}, despite this difference, 
equivalent algorithms based on TD and PCD can nevertheless be formulated 
via the fact that \(s_{\kappa}\) is non-zero (and so can be eliminated from 
the algorithm). This is in contrast to the PCD 
\eqref{equ:ConPieSerForTru} of \eqref{equ:RepFieSerFor1D}${}_{1}$ 
employed by \cite{Eloh2019} and resulting in Algorithm \ref{alg:AlgDisConPie1D}. 
Indeed, the DGO 
\(\skew4\check{\Gamma}_{\!\!\smash{\mathrm{H}\,\omega}}^{\mathrm{c}}\) 
from \eqref{equ:SchLipOpeEllConPieApp1D} in the latter algorithm depends 
explicitly on \(s_{\nu n+\omega}\). 

All three algorithms \ref{alg:AlgDisIntFouDifFinFunGre1D}, 
\ref{alg:AlgDisIntFouDifFinFunGrePC1D} and \ref{alg:AlgDisConPie1D} 
are based on Green function preconditioning (GFP). In addition, the first 
two exploit finite difference discretization (FDD) of differential operators, 
in particular of \(\nabla\) and \(\mathop{\mathrm{div}}\). In the context 
of TD \eqref{equ:FieFouAppTra1D} and PCD \eqref{equ:DisFouPC1D} 
of fields, different FDDs for these two operators are obtained in modal form 
(e.g., \eqref{equ:DisFouAppTra1D}-\eqref{equ:AppDifFinDFStn1D} 
for \(\nabla\)) in terms of effective wavenumbers \(q_{\mu}^{a}\) 
(for \(\nabla^{a}\)) and \(q_{\mu}^{b}\) (for \(\mathrm{div}^{b}\)). 
Specific FDDs consider include 
forward (\(a,b=\mathrm{FD}\)), backward (\(a,b=\mathrm{BD}\)), 
central (\(a,b=\mathrm{CD}\)), and half central (\(a,b=\mathrm{hC}\)), 
difference, respectively. Following in particular \cite{Willot2015}, given \(a\), 
one choice for \(b\) is based on conjugacy \eqref{equ:EffNumWavConWil1D}. 
A second one is based on 
choosing \(b\) in such a way that the discretized stress divergence is determined 
in the algorithm at the (displacement) nodes. For example, for the 
choice \(a=\mathrm{FD}\), this results in \(b=\mathrm{hC}\), and so 
the (non-conjugate) combination \((a,b)=(\mathrm{FD},\mathrm{hC})\). 

Computational comparison of these algorithms is carried out in Section 
\ref{sec:Res1D} for the matrix-inclusion benchmark case in 1D for 
material heterogeneity in the form of both 
discontinuous and smooth compliance / stiffness distributions. 
In the former case, as expected from Fourier series approximation  
of the analytic solution and series truncation 
(Figure \ref{fig:SerFouStnDisFluDsc}, left), \(\tilde{E}_{a}(x)\) 
at the sharp MI interface in Figure \ref{fig:StnShaSmoStr1D} 
(left) exhibits Gibbs error independent of, and no convergence with 
increasing, numerical resolution for all choices of 
\(\nabla^{a}\) (\(q_{\mu}^{a}\)) 
and \(\mathrm{div}^{b}\) (\(q_{\mu}^{b}\)) in the context of 
Algorithms \ref{alg:AlgDisIntFouDifFinFunGre1D} 
and \ref{alg:AlgDisIntFouDifFinFunGrePC1D}. On the other hand, 
\(\tilde{E}_{a}(x)\) does converge with increasing numerical resolution 
at the smooth MI interface (Figure \ref{fig:StnShaSmoStr1D}, right). 
In addition, for the sharp interface case, comparison of strain results 
from Algorithms \ref{alg:AlgDisIntFouDifFinFunGre1D}, 
\ref{alg:AlgDisIntFouDifFinFunGrePC1D} 
and \ref{alg:AlgDisConPie1D} in Figure \ref{fig:StnFBDGO1D} 
show that the nodal results from \ref{alg:AlgDisIntFouDifFinFunGrePC1D} 
deviate slightly from those of the anayltic solution and the first two 
algorithms in the (relatively soft) matrix, 
with maximum deviation in the matrix next to the MI interface. 

Multidimensional versions of the 1D algorithms are formulated 
in Section \ref{sec:AlgStr3D} via direct componentwise- and 
tensor-product-based generalization. In this fashion, 
3D versions \ref{alg:AlgDisIntFouDifFinFunGre3D}, 
\ref{alg:AlgDisIntFouDifFinFunGrePC3D} and \ref{alg:AlgDisConPie3D}, 
respectively, of the 1D algorithms \ref{alg:AlgDisIntFouDifFinFunGre1D}, 
\ref{alg:AlgDisIntFouDifFinFunGrePC1D} and \ref{alg:AlgDisConPie1D}, 
respectively, follow directly. As in the 1D case, all three are based on GFP, 
and the first two on FDDs of \(\nabla\) and \(\mathop{\mathrm{div}}\), 
now in 3D. In the context of either TD \eqref{equ:DisFouSca3D} or PCD 
\eqref{equ:DisFouPC3D} of 3D fields, the latter are formulated in 3D 
via componentwise application of the 1D approaches, resulting in the FDDs 
\(\nabla^{\mathbf{a}}\) and \(\mathrm{div}^{\mathbf{b}}\) 
with \(\mathbf{a}=(a_{1},a_{2},a_{3})\) and 
\(\mathbf{b}=(b_{1},b_{2},b_{3})\). 
As in 1D, choices for \(b_{r}\) given \(a_{r}\) are based on 
conjugacy \eqref{equ:EffNumWavConWil3D} and the non-conjugate stress 
divergence criteria, the latter yielding in particular \(b_{r}=\mathrm{AhC}\) 
for \(a_{r}=\mathrm{AFD}\) and so the "average forward-backward/rotated" 
FDD, i.e., AFB/R. 
Computational comparsions of these 
are presented in Section \ref{sec:Res3D}. 
Generally speaking, an increase in phase contrast results in an increase in the 
condition number of the algorithmic equation system being solved, resulting in 
slower convergence. As such, preconditioning is clearly essential to 
improved convergence. As shown for example in \cite{Willot2015} and the 
current work (e.g., Figure \ref{fig:Conv3D}), the combination of GFP and 
FDD results in significant further improvement in convergence rate and 
behavior over approaches based on GFP alone such as the DGO of \cite{Eloh2019}. 

In the current work, the focus has been on material inhomogeneity 
with respect to elastic stiffness. Generalization of the current algorithms 
to the case of such heterogeneity with respect to both elastic stiffness 
and residual strain (e.g., due to lattice mismatch between phases) is 
straightforward.  In the 3D case, the corresponding generalization 
\(
\bm{T}=\ssans{C}\lbrack\bm{E}-\bm{E}_{\ast}\rbrack
\) 
of the stress-strain relation 
(with \(\ssans{C}(\bm{x})\) and \(\bm{E}_{\ast}(\bm{x})\) known)  
induces changes in the algorithms. For example, this results in the change of 
\(
\mathbf{T}_{\smash{i}}^{(0)}
=\textsf{C}_{\smash{i}}\bar{\mathbf{E}}
\) 
to 
\(
\mathbf{T}_{\smash{i}}^{(0)}
=\textsf{C}_{\smash{i}}
\lbrack
\bar{\mathbf{E}}-\mathbf{E}_{\ast\smash{i}}
\rbrack
\) 
in the TD-based algorithm \ref{alg:AlgDisIntFouDifFinFunGre3D}. Likewise, 
\(
\mathbf{T}_{\smash{i}}^{\mathrm{c}(0)}
=\textsf{C}_{\smash{i}}^{\mathrm{c}}
\bar{\mathbf{E}}
\) 
generalizes to 
\(
\mathbf{T}_{\smash{i}}^{\mathrm{c}(0)}
=\textsf{C}_{\smash{i}}^{\mathrm{c}}
\lbrack
\bar{\mathbf{E}}-\mathbf{E}_{\ast\smash{i}}^{\mathrm{c}}
\rbrack
\) 
in the PCD-based algorithms \ref{alg:AlgDisIntFouDifFinFunGrePC3D} 
and \ref{alg:AlgDisConPie3D}. 

In the context of strong mechanical equilbrium, a displacement-based 
approach related to the current one has quite recently been developed by 
\cite{Lucarini2019}. They refer to this approach as displacement-based 
FFT (DBFFT). As the name implies, they also work with the displacement 
as the primary discretant and the split \eqref{equ:DisFluParHom3D}. In 
particular, their algorithm is based on quasi-static mechanical equilibrium 
in Fourier space in the form 
\(
\hat{\bm{A}}(\bm{q})\,\hat{\bm{u}}^{\mathrm{p}}
=\hat{\ssans{C}}\lbrack\bar{\bm{E}}\rbrack\bm{q}
\) 
for \(\bm{q}\neq\bm{0}\) in terms of the generalized acoustic tensor 
\(
\hat{\bm{A}}(\bm{q})\,\hat{\bm{v}}
:=-\mathcal{F}
\lbrace
\ssans{C}
\lbrack
\mathcal{F}^{-1}
\lbrace
\hat{\bm{v}}\otimes\bm{q}
\rbrace
\rbrack
\rbrace
\bm{q}
\). 
As noted by them, Fourier discretization and real-function-based reduction 
yields a discrete system for displacement based on a full-rank associated 
Hermitian matrix. In turn, this facilitates use of preconditioners. In the 
current context, their choice corresponds in particular to those 
\(\textsf{C}_{\mathrm{H}}=\bar{\textsf{C}}\) and 
\((a_{r},b_{r})=(\mathrm{F},\mathrm{F})\) in 
\eqref{equ:OpeFunGreAlgMod3D}. In terms of Fourier transforms, this 
results in 
\(
\hat{\bm{G}}_{\smash{\mathrm{H}}}^{-1}(\bm{q})
\,\bm{a}
=-\bar{\ssans{C}}
\lbrack
\bm{a}\otimes\bm{q}
\rbrack
\bm{q}
\) 
for \(\bm{q}\neq\bm{0}\). Algorithmic extension of this to 
\(
\bm{q}=\bm{0}
\) 
then results in a preconditioning operator \(\hat{\bm{M}}(\bm{q})\) 
which approximately inverts \(\hat{\bm{A}}(\bm{q})\) 
for use in iterative numerical solution. 

Traditional (i.e., linear elastic) micromechanics based on a 
(linear) stress-strain relation \(\bm{T}(\bm{E})\) treats the (linear) 
strain \(\bm{E}\) (and not displacement) as the primary unknown. 
In the computational context, this translates into the treatment of 
\(\bm{E}\), or more recently in the geometrically non-linear case, the  
deformation gradient \(\bm{F}\), as the primary discretant. 
In this case, compatibility needs to be enforced via 
additional algorithmic constraints. Alternatively, as discussed in the current 
work, one can work directly with the displacement or deformation field as 
the primary discretant. This is true in both the current case of strong 
mechanical equilibrium as well as in the case of weak mechanical equilibrium, 
as shown by the recent work of \cite{deGeus2017}. To discuss the latter 
briefly here, consider weak equilibrium 
\(
\int
\bm{T}(\bm{x})
\cdot
\bm{E}_{\mathrm{v}}(\bm{x})
\ dv(\bm{x})
=\int
\skew3\hat{\bm{T}}(\bm{k})
\cdot
\skew3\hat{\bm{E}}_{\smash{\mathrm{v}}}^{\ast}(\bm{k})
\ dv(\bm{k})
\) 
\cite[i.e., via the Rayleigh-Plancherel or power 
theorem:~e.g.,][pp.~119-120]{Bra00} for any "virtual" or "test" strain 
field \(\bm{E}_{\mathrm{v}}\). Given \(\bm{H}\) compatible, 
\(
\skew3\hat{\bm{H}}_{\smash{\mathrm{v}}}^{\ast}
=\hat{\bm{u}}_{\mathrm{v}}\otimes\bm{q}^{\ast}
\) 
holds, again with \(\bm{q}=\imath\bm{k}\). Then 
\(
\hat{\bm{u}}_{\mathrm{v}}
=\skew3\hat{\bm{H}}_{\smash{\mathrm{v}}}^{\ast}
\bm{q}/|\bm{q}|^{2}
\) 
for \(\bm{q}\neq\bm{0}\), inducing in turn 
\(
\skew3\hat{\bm{T}}
\cdot
\hat{\bm{E}}_{\smash{\mathrm{v}}}^{\ast}
=\ssans{P}_{\bm{H}}\,\skew3\hat{\bm{T}}
\cdot
\skew3\hat{\bm{H}}_{\smash{\mathrm{v}}}^{\ast}
\) 
with 
\(
\ssans{P}_{\bm{H}}
:=\bm{I}\,\square\,(\bm{q}\otimes\bm{q}^{\ast})/|\bm{q}|^{2}
\). 
Consequently, the projection \(\ssans{P}_{\bm{H}}\) enforces 
compatibility when \(\bm{H}\) is the primary discretant and unknown.  
Alternatively, as done in the current work, one can simply work directly with 
the displacement field or the deformation field as the primary discretant. 

\bigskip

\textit{Acknowledgements}. Financial support of Subproject M03 of the 
Transregional Collaborative Research Center SFB/TRR 136 by the German 
Science Foundation (DFG) is gratefully acknowledged. 

\bibliographystyle{elsarticle-harv}



\begin{appendix} 

\renewcommand{\thesection}{\Alph{section}}
\setcounter{section}{0}
\renewcommand{\theequation}{\thesection.\arabic{equation}}
\setcounter{equation}{0}
\renewcommand{\thefigure}{\thesection.\arabic{figure}}
\setcounter{figure}{0}

\section{Trapezoidal approximation / discretization} 
\label{app:IntForAppTra1D}

Note that the trapezoidal discretization 
\(\hat{f}_{\smash{\kappa}}^{\mathrm{t}}\) 
of \(\hat{f}_{\smash{\kappa}}\) in \eqref{equ:RepFieSerForTra1D} 
satisfies cardinality 
\begin{equation}
\begin{array}{rclclclcl}
\displaystyle
\sum\nolimits_{\kappa=-m}^{m}
e^{\imath k_{\kappa}x_{i}}
\hat{f}_{\kappa}^{\mathrm{t}}
&=&
\displaystyle
\sum\nolimits_{j=1}^{n}
\sum\nolimits_{\kappa=-(n-1)/2}^{(n-1)/2}
n^{-1}e^{2\pi\imath\kappa (i-j)/n}f(x_{\!j})
&=&
\displaystyle
\sum\nolimits_{j=1}^{n}
\delta_{ij}f(x_{\!j})
&=&
f(x_{i})
\,,\\[2mm]
\displaystyle
\sum\nolimits_{\kappa=-m+1}^{m}
e^{\imath k_{\kappa}x_{i}}
\hat{f}_{\kappa}^{\mathrm{t}}
&=&
\displaystyle
\sum\nolimits_{j=1}^{n}
\sum\nolimits_{\kappa=-n/2+1}^{n/2}
n^{-1}e^{2\pi\imath\kappa (i-j)/n}
\,f(x_{\!j})
&=&
\displaystyle
\sum\nolimits_{j=1}^{n}
\delta_{ij}f(x_{\!j})
&=&
f(x_{i})
\,,\\[2mm]
\displaystyle
\sum\nolimits_{\kappa=-m}^{m-1}
e^{\imath k_{\kappa}x_{i}}
\hat{f}_{\kappa}^{\mathrm{t}}
&=&
\displaystyle
\sum\nolimits_{j=1}^{n}
\sum\nolimits_{\kappa=-n/2}^{n/2-1}
n^{-1}e^{2\pi\imath\kappa (i-j)/n}
\,f(x_{\!j})
&=&
\displaystyle
\sum\nolimits_{j=1}^{n}
\delta_{ij}f(x_{\!j})
&=&
f(x_{i})
\,,
\end{array}
\label{equ:ConIntFor1D}
\end{equation} 
(i.e., the interpolation condition) for both \(n\) odd (\(m=(n-1)/2\)) and 
\(n\) even (\(m=n/2\)). 
Given these, 
\begin{equation}
\mathcal{I}_{\!n\,}^{\mathrm{o}}f(x)
:=\sum\nolimits_{\kappa=-(n-1)/2}^{(n-1)/2}
e^{\imath k_{\kappa}x}
\hat{f}_{\kappa}^{\mathrm{t}}
\label{equ:IntFouModOdd1D}
\end{equation} 
for \(n\) odd, and 
\begin{equation}
\mathcal{I}_{\!n\,}^{\mathrm{e}}f(x)
:=\sum\nolimits_{\kappa=-n/2}^{n/2-1}
e^{\imath k_{\kappa}x}
\hat{f}_{\kappa}^{\mathrm{t}}
=\sum\nolimits_{\kappa=-n/2+1}^{n/2}
e^{\imath k_{\kappa}x}
\hat{f}_{\kappa}^{\mathrm{t}}
\label{equ:IntFouModEve1D}
\end{equation} 
for \(n\) even, interpolate \(f(x)\). 
Note that the second form of \eqref{equ:IntFouModEve1D} 
is employed for example by \citet[][Equations (10)-(12)]{Willot2015}. 


For Fourier discretization based on \eqref{equ:RepFieSerForTra1D}, 
the index transformations 
\begin{equation}
\begin{array}{rclcrclcl}
\kappa
&\mapsto&
\mu=\kappa+1+m
&\colon&
\lbrace-m,\ldots,m\rbrace
&\to&
\lbrace 1,\ldots,n\rbrace
\,,
&&
\hbox{\(m=\frac{1}{2}(n-1)\), \(n\) odd}
\,,\\
\kappa
&\mapsto&
\mu=\kappa+1+m
&\colon&
\lbrace-m,\ldots,m-1\rbrace
&\to&
\lbrace 1,\ldots,n\rbrace
\,,
&&
\hbox{\(m=\frac{1}{2}n\), \(n\) even}
\,,\\
\kappa
&\mapsto&
\mu=\kappa+m
&\colon&
\lbrace-m-1,\ldots,m\rbrace
&\to&
\lbrace 1,\ldots,n\rbrace
\,,
&&
\hbox{\(m=\frac{1}{2}n\), \(n\) even}
\,,
\end{array}
\label{equ:TraIndSerFor}
\end{equation}
are useful. 
Restricting attention to the first two, define 
\begin{equation}
q_{\mu}
:=\imath k_{\mu-1-m}
\,,\quad
\varphi_{\mu}(x)
:=n^{-1/2}e^{q_{\mu}x}
\,,\quad
\check{f}_{\mu}^{\mathrm{t}}
:=n^{-1/2}
\sum\nolimits_{i=1}^{n}
e^{-q_{\mu}x_{i}}
f(x_{i})
\,.
\label{equ:IntTraCoeSerFouShi}
\end{equation}
Based on these, \eqref{equ:IntFouModOdd1D} and 
\eqref{equ:IntFouModEve1D}${}_{2}$ take the common form 
\begin{equation}
\mathcal{I}_{\!n\,}f(x)
=\sum\nolimits_{\mu=1}^{n}
\varphi_{\mu}(x)
\,\check{f}_{\mu}^{\mathrm{t}}
\,,
\label{equ:IntFouApp1D}
\end{equation} 
with 
\(
\check{f}_{\smash{\mu}}^{\mathrm{t}}
=\sum\nolimits_{i=1}^{n}
F_{\!\smash{\mu i}\,}^{\mathrm{t}-}f(x_{i})
\), 
\(
f(x_{i})
=\sum\nolimits_{\mu=1}^{n}
F_{\!\smash{\mu i}\,}^{\mathrm{t}+}
\check{f}_{\smash{\mu}}^{\mathrm{t}}
\), 
and 
\begin{equation}
F_{\!\smash{\mu i}\,}^{\mathrm{t}\pm}
:=n^{-1/2}e^{\pm q_{\mu}x_{i}}
=n^{-1/2}e^{\pm 2\pi\imath(\mu-1-m)(i-1)/n}
\,,\quad 
m=\left\lbrace
\begin{array}{lcl}
\frac{1}{2}(n-1)
&&
\hbox{\(n\) odd}
\\
\frac{1}{2}n
&&
\hbox{\(n\) even}
\end{array}
\right.
\,.
\label{equ:TraFouDisMat}
\end{equation}
Note that 
\(
\varphi_{\mu}(x_{i})=F_{\!\smash{\mu i}\,}^{+}
\) 
and 
\(
\sum_{\mu=1}^{n}
\!F_{\!\smash{\mu i}\,}^{\mathrm{t}+}F_{\!\smash{\mu\!j}\,}^{\mathrm{t}-}
=\delta_{i\!j}
\), 
consistent with cardinality. 



\section{Piecewise-constant approximation / discretization} 
\label{app:DisCenIntSub1DApp}

The form \(\hat{f}_{\kappa}^{\mathrm{c}}\) for 
\(\hat{f}_{\kappa}\) from \eqref{equ:RepFieSerForPieCon1D} 
in the case of piecewise-constant \(f(x)\) determines the corresponding 
Fourier series discretization   
\begin{equation}
\mathcal{F}_{\!\!n}^{\mathrm{c}}f(x)
:=\sum\nolimits_{\mu=1}^{n}
\varphi_{\mu}^{\mathrm{c}}(x)
\,\check{f}_{\mu}^{\mathrm{c}}
\,,\quad
f(x_{i}^{\mathrm{c}})
=\sum\nolimits_{\mu=1}^{n}
F_{\!\smash{\mu i}\,}^{\mathrm{c}+}
\check{f}_{\mu}^{\mathrm{c}}
\,,\quad 
\check{f}_{\mu}^{\mathrm{c}}
=\sum\nolimits_{i=1}^{n}
F_{\!\smash{\mu i}\,}^{\mathrm{c}-}
f(x_{i}^{\mathrm{c}})
\,,
\label{equ:SerForPCTra1D} 
\end{equation} 
via \eqref{equ:TraIndSerFor}${}_{1,2}$. Here, 
\(
\varphi_{\smash{\mu}}^{\mathrm{c}}(x)
:=n^{-1/2}s_{\smash{\mu}}^{\mathrm{c}}
\,e^{q_{\smash{\mu}}x}
\), 
\(
s_{\smash{\mu}}^{\mathrm{c}}
:=s_{\smash{\mu-1-m}}
\)
and 
\(
F_{\!\smash{\mu i}\,}^{\mathrm{c}\pm}
:=n^{-1/2}e^{\pm q_{\mu}x_{i}^{\mathrm{c}}}
\). 
Since \(s_{\smash{\mu}}^{\mathrm{c}}\neq 1\) for \(\mu\neq m+1\), 
note that 
\begin{equation}
\mathcal{F}_{\!\!n}^{\mathrm{c}}f(x_{i}^{\mathrm{c}})
=\sum\nolimits_{\mu=1}^{n}
\varphi_{\mu}^{\mathrm{c}}(x_{i}^{\mathrm{c}})
\,\check{f}_{\mu}^{\mathrm{c}}
=\sum\nolimits_{j=1}^{n}
\sum\nolimits_{\mu=1}^{n}
s_{\mu}^{\mathrm{c}}
F_{\!\smash{\mu i}\,}^{\mathrm{c}+}
F_{\!\mu\!j\,}^{\mathrm{c}-}
f(x_{\!j}^{\mathrm{c}})
\approx
f(x_{i}^{\mathrm{c}})
\label{equ:CarAprPC}
\end{equation}
is only approximately cardinal. 
In the approach of \cite{Eloh2019} also based on \eqref{equ:RepFieSerForPieCon1D}, 
approximate cardinality takes the form 
\begin{equation}
f(x_{i}^{\mathrm{c}})
\approx
\mathcal{F}_{\!\!p\,}^{\mathrm{c}}f(x_{i}^{\mathrm{c}})
=n^{-1/2}\sum\nolimits_{\omega=0}^{n-1}
e^{\imath k_{\omega}x_{i}^{\mathrm{c}}}
\sum\nolimits_{\nu=-m}^{m-1}
(-1)^{\nu}
s_{\nu n+\omega}
\,\check{f}_{\nu n+\omega}^{\mathrm{c}}
\label{equ:CenTraForInfFor}
\end{equation} 
via \eqref{equ:ConPieSerForTru} and 
\(
e^{-\imath k_{\nu n}x_{i}^{\mathrm{c}}}
=e^{-2\pi\imath\nu(i-\frac{1}{2})}
=e^{\pi\imath\nu}e^{-2\pi\imath\nu i}
=(-1)^{\nu}
\) 
(\(p=nm\) even). Comparison of \eqref{equ:CenTraForInfFor} and 
\eqref{equ:CenTraForDisConBac}${}_{2}$ results in  
\begin{equation}
\check{f}_{\omega}^{\mathrm{c}}
\approx
\sum\nolimits_{\nu=-m}^{m-1}
(-1)^{\nu}
s_{\nu n+\omega}
\,\check{f}_{\nu n+\omega}^{\mathrm{c}}
\,,\quad
\sum\nolimits_{\nu=-m\atop\nu\neq 0}^{m-1}
(-1)^{\nu}
s_{\nu n}
\,\check{f}_{\nu n}^{\mathrm{c}}
\approx 
0
\,,
\label{equ:CenTraForDisConFor}
\end{equation}
via the fact that 
\(
\bar{f}
=n^{-1/2}\check{f}_{0}^{\mathrm{c}}
=n^{-1}
\sum\nolimits_{i=1}^{n}
f(x_{i}^{\mathrm{c}})
\) 
from \eqref{equ:FieMeaCelUni1D}${}_{2}$, 
\eqref{equ:RepFieSerFor1D}${}_{2}$ and 
\eqref{equ:RepFieSerForPieCon1D}${}_{3}$. 
Given \eqref{equ:CenTraForDisConFor}, note that 
\eqref{equ:CenTraForInfFor} reduces to 
\(
\tilde{f}(x_{i}^{\mathrm{c}})
\approx
\sum\nolimits_{\omega=1}^{n-1}
e^{\imath k_{\omega}x_{i}^{\mathrm{c}}}
\sum\nolimits_{\nu=-m}^{m-1}
(-1)^{\nu}
s_{\nu n+\omega}
\,\check{f}_{\nu n+\omega}
\) 
for the fluctuation part \(\tilde{f}(x_{i}^{\mathrm{c}})\) of 
\(
f(x_{i}^{\mathrm{c}})=\bar{f}+\tilde{f}(x_{i}^{\mathrm{c}})
\) 
in the context of \eqref{equ:FieMeaCelUni1D}. 

As done by \cite{Eloh2019}, the principle application of these relations 
and in particular of \eqref{equ:CenTraForDisConFor}${}_{1}$ is to 
obtain the discretization 
\begin{equation}
\skew4\check{\Gamma}_{\!\!\mathrm{H}\omega}^{\mathrm{c}}
\check{T}_{\!\smash{\omega}}^{\mathrm{c}}
=0
\,,\quad
\omega=1,\ldots,n-1
\,,
\label{equ:EleOpeAlgStnConPie1D}
\end{equation}
of \eqref{equ:SchLipEquMecFouTra1D} in terms of the "discrete Green operator" 
(DGO) 
\begin{equation}
\skew4\check{\Gamma}_{\!\!\mathrm{H}\omega}^{\mathrm{c}}
:=\sum\nolimits_{\nu=-m}^{m-1}
s_{\nu n+\omega}
\ \skew4\hat{\Gamma}_{\!\!\mathrm{H}}(k_{\nu n+\omega})
\,,\quad
\omega=1,\ldots,n-1
\,.
\label{equ:SchLipOpeEllConPieApp1D}
\end{equation} 
Both \eqref{equ:EleOpeAlgStnConPie1D} and \eqref{equ:SchLipOpeEllConPieApp1D} 
are employed in Algorithm \ref{alg:AlgDisConPie1D}, and via (Cartesian) 
tensor product generalization in Algorithm \ref{alg:AlgDisConPie3D}. 

\end{appendix}

\end{document}